\documentclass[12pt]{article}
\pdfoutput=1
\usepackage{amsmath,amsfonts,amssymb,amsthm,amssymb,bbm,bm}
\usepackage{pdfsync}
\usepackage{graphicx,import}
\usepackage[utf8]{inputenc}
\usepackage[pagebackref]{hyperref}
\renewcommand*{\backref}[1]{}
\renewcommand*{\backrefalt}[4]{\small{
    \ifcase #1%
          \or (Cited on page~#2.)%
          \else (Cited on pages~#2.)%
    \fi%
    }}
\hypersetup{colorlinks,linkcolor={RedViolet},citecolor={Blue},urlcolor={Blue}}
\usepackage{empheq}
\usepackage[all]{xy}
\usepackage{stmaryrd}
\usepackage{rotating}
\usepackage[dvipsnames]{xcolor}  
\usepackage{slashed}
\usepackage{cancel}
\usepackage{array, makecell} %
\usepackage{comment}
\usepackage{tikz-cd}
\usepackage{hhline}
\usepackage{cancel}
\usepackage{dsfont}
\usepackage{stackengine}
\usepackage[breakable]{tcolorbox}
\usepackage{mathrsfs}
\usepackage{booktabs}
\usepackage{float}
\usepackage{changepage}
\usepackage[normalem]{ulem}
\usepackage{abraces}
\usepackage{multicol}
\usepackage{orcidlink}

\allowdisplaybreaks

\DeclareMathAlphabet\mathrsfso{U}{rsfso}{m}{n}

\usepackage[scr=boondoxo,scrscaled=1.1]{mathalpha}

\makeatletter

\DeclareFontFamily{U}{matha}{\hyphenchar\font45}
\DeclareFontShape{U}{matha}{m}{n}{
      <5> <6> <7> <8> <9> <10> gen * matha
      <10.95> matha10 <12> <14.4> <17.28> <20.74> <24.88> matha12
      }{}
\DeclareSymbolFont{matha}{U}{matha}{m}{n}
\DeclareMathSymbol{\oright}       {2}{matha}{"69}

\usepackage[top=2cm, bottom=2.5cm, left=2.5cm, right=2.5cm]{geometry}

\newcommand{\doublehat}[1]{%
\begingroup%
  \let\macc@kerna\z@%
  \let\macc@kernb\z@%
  \let\macc@nucleus\@empty%
  \hat{\raisebox{.55ex}{\vphantom{\ensuremath{#1}}}\smash{\hat{#1}}}%
\endgroup%
}

\renewcommand{\ni}{\noindent}

\newcommand{\bit}{\begin{itemize}}
\newcommand{\eit}{\end{itemize}}
\newcommand{\bd}{\begin{description}}
\newcommand{\ed}{\end{description}}

\newcommand{\bc}{\begin{center}}
\newcommand{\ec}{\end{center}}

\newcommand{\C}{{\mathbb C}}
\newcommand{\N}{{\mathbb N}}

\newcommand{\E}{\mathsf{E}}

\newcommand{\gbms}{\mathfrak{gbms}}

\newcommand{\cT}{{\mathcal T}}
\newcommand{\cA}{{\mathcal A}}

\def\be#1\ee{\begin{align}#1\end{align}}
\newcommand{\bea}{\begin{eqnarray}}
\newcommand{\eea}{\end{eqnarray}}
\newcommand{\bs}{\begin{subequations}}
\newcommand{\es}{\end{subequations}}
\newcommand{\nn}{\nonumber}

\newcommand{\mat}[4]{\left(\begin{matrix}{#1}&{#2}\\{#3}&{#4}\end{matrix}\right)}

\usepackage[numbers,sort&compress]{natbib}

\makeatletter
\newcommand{\oset}[3][0ex]{%
  \mathrel{\mathop{#3}\limits^{
    \vbox to#1{\kern-2\ex@
    \hbox{$\scriptstyle#2$}}}}}
\makeatother

\newcommand{\rd}{\mathrm{d}}
\newcommand{\pa}{\partial}
\newcommand{\Dcal}{\mathcal{D}}
\newcommand{\DGR}{\overset{\mkern-1mu\sgr\mkern1mu}{\mathcal{D}}}
\newcommand{\DYM}{D_{\mkern-3mu A}}
\newcommand{\DEYM}{\overset{\mkern-0mu\scriptscriptstyle{\textsc{eym}}\mkern0mu}{\mathcal{D}}}
\newcommand{\DTOT}{\overset{\mkern-1mu\sym\mkern1mu}{\mathcal{D}}}
\newcommand{\DYMc}{D_{\Ac0}}

\newcommand{\nablaYM}{\nabla_{\mkern-4mu A}}
\newcommand{\nablaGR}{\overset{\sgr}{\nabla}}
\newcommand{\nablaEYM}{\overset{\scriptscriptstyle{\textsc{eym}}}{\nabla}}

\newcommand{\dt}{\delta_\tau}
\newcommand{\dtp}{\delta_{\tau'}}

\newcommand{\da}{\delta_\alpha}

\newcommand{\dta}{\delta_{(\tau,\alpha)}}
\newcommand{\dtap}{\delta_{(\tau'\!,\alpha')}}

\newcommand{\dbt}{\delta_{\bar{\tau}}}
\newcommand{\dbtp}{\delta_{\bar{\tau}'}}
\newcommand{\dbtpp}{\delta_{\bar{\tau}''}}

\newcommand{\odt}{\delta_{\text{\scalebox{1.1}{$\ensurestackMath{\stackon[0.9pt]{\tau}{\mkern 1mu\scriptscriptstyle{\circ}}}$}}}}

\newcommand{\odtp}{\delta_{\text{\scalebox{1.1}{$\ensurestackMath{\stackon[-1.9pt]{\tau'}{\mkern-4mu\scriptscriptstyle{\circ}}}$}}}}

\newcommand{\odtpp}{\delta_{\text{\scalebox{1.1}{$\ensurestackMath{\stackon[-2pt]{\tau''}{\mkern-9mu\scriptscriptstyle{\circ}}}$}}}}

\newcommand{\hda}{\hat{\delta}_\alpha}
\newcommand{\hdap}{\hat{\delta}_{\alpha'}}
\newcommand{\hdapp}{\hat{\delta}_{\alpha''}}

\newcommand{\dT}{\delta_T}

\newcommand{\dPta}{\delta_\gamma}
\newcommand{\dPtap}{\delta_{\gamma'}}
\newcommand{\dPtapp}{\delta_{\gamma''}}

\newcommand{\dbPta}{\delta_{\bar{\gamma}}}
\newcommand{\dbPtap}{\delta_{\bar{\gamma}'}}
\newcommand{\dbPtapp}{\delta_{\bar{\gamma}''}}

\newcommand{\sI}{\mathsf{I}}

\newcommand{\Ll}{\mathbb{L}}

\newcommand{\overbar}[1]{\mkern 3mu\overline{\mkern-3.5mu#1\mkern-2mu}\mkern 2mu}

\newcommand{\overbarT}[1]{\mkern 3mu\overline{\mkern-2.5mu#1\mkern-1.5mu}\mkern 2mu}

\newcommand{\overbarF}[1]{\mkern 2mu\overline{\mkern-4mu#1\mkern-1mu}\mkern 2mu}

\newcommand{\bnews}{\overbarF{F}}

\newcommand{\bC}{\overbar{C}}
\newcommand{\bN}{\overbar{N}}
\newcommand{\bz}{\bar{z}}
\renewcommand{\bm}{\overbar{m}}
\newcommand{\bY}{\overbar{Y}}
\newcommand{\bD}{\overbar{D}}

\newcommand{\bcT}{\overbarT{\mathcal{T}}}
\newcommand{\bsfT}{\overbarT{\mathsf{T}}}

\newcommand{\bt}{\bar{\tau}}
\newcommand{\btp}{\bar{\tau}'}
\newcommand{\btpp}{\bar{\tau}''}

\newcommand{\bPta}{\bar{\gamma}}
\newcommand{\bPtap}{\bar{\gamma}'}
\newcommand{\bPtapp}{\bar{\gamma}''}

\newcommand{\ft}{\hat{\tau}}
\newcommand{\fa}{\hat{\alpha}}

\newcommand{\fPta}{\hat{\Pta}}

\makeatletter
\newcommand{\hhat}[1]{%
\begingroup%
  \let\macc@kerna\z@%
  \let\macc@kernb\z@%
  \let\macc@nucleus\@empty%
  \widehat{\mathchoice%
    {\raisebox{.2ex}{\vphantom{\ensuremath{\displaystyle #1}}}}%
    {\raisebox{.4ex}{\vphantom{\ensuremath{\textstyle #1}}}}%
    {\raisebox{.16ex}{\vphantom{\ensuremath{\scriptstyle #1}}}}%
    {\raisebox{.14ex}{\vphantom{\ensuremath{\scriptscriptstyle #1}}}}%
    \smash{\widehat{#1}}}%
\endgroup%
}
\makeatother

\makeatletter
\DeclareRobustCommand\widecheck[1]{{\mathpalette\@widecheck{#1}}}
\def\@widecheck#1#2{%
    \setbox\z@\hbox{\m@th$#1#2$}%
    \setbox\tw@\hbox{\m@th$#1%
       \widehat{%
          \vrule\@width\z@\@height\ht\z@
          \vrule\@height\z@\@width\wd\z@}$}%
    \dp\tw@-\ht\z@
    \@tempdima\ht\z@ \advance\@tempdima2\ht\tw@ \divide\@tempdima\thr@@
    \setbox\tw@\hbox{%
       \raise\@tempdima\hbox{\scalebox{1}[-1]{\lower\@tempdima\box
\tw@}}}%
    {\ooalign{\box\tw@ \cr \box\z@}}}
\makeatother

\newcommand{\tQ}{\widetilde{Q}}

\renewcommand{\tt}[1]{\tilde{\tau}_{#1}}
\newcommand{\ttp}[1]{\tilde{\tau}'_{#1}}

\newcommand{\Tr}{\mathrm{Tr}}

\newcommand{\gYM}{g_\textsc{ym}}
\newcommand{\GN}{G_{\mkern-2mu N}}
\newcommand{\X}{\mathfrak{X}}
\newcommand{\g}{\mathfrak{g}}

\newcommand{\Ocal}{\mathcal{O}}

\newcommand{\cTo}{\ocirc{\mathcal{T}}}

\newcommand{\Cc}[1]{\sigma_{#1}}
\newcommand{\Ac}[1]{\beta_{#1}}

\newcommand{\sn}{\mathsf{SN}}
\newcommand{\PS}{\mathcal{P}}

\newcommand{\cS}{\mathcal{S}}

\newcommand{\CP}[1]{\C\mathbb{P}^{#1}}

\newcommand{\ym}{\textsc{ym}}
\newcommand{\gr}{\textsc{gr}}
\newcommand{\sym}{\scriptscriptstyle{\textsc{ym}}}
\newcommand{\sgr}{\scriptscriptstyle{\textsc{gr}}}
\newcommand{\QYM}[1]{\overset{\mkern-4mu\sym\mkern4mu}{Q^u}_{\mkern-11mu#1}}
\newcommand{\QGR}[1]{\overset{\mkern-4mu\sgr\mkern4mu}{Q^u}_{\mkern-11mu#1}}

\newcommand{\sfST}{\mathsf{ST}}
\newcommand{\sfSK}{\mathsf{S}_{\mkern-1mu K}}
\newcommand{\sfTK}{\mathsf{T}_{\mkern-3mu K}}
\newcommand{\sfSTK}{\mathsf{ST}_{\mkern-3mu K}}
\newcommand{\sfK}{\mathsf{K}}

\newcommand{\cST}{\mathcal{ST}}
\newcommand{\cSK}{\mathcal{S}_K}
\newcommand{\cTK}{\mathcal{T}_K}
\newcommand{\cSTK}{\mathcal{ST}_{\mkern-5mu K}}

\newcommand{\W}{\mathsf{W}}
\newcommand{\Wcal}{\mathcal{W}}

\newcommand{\sfT}{{\mathsf{T}}}
\newcommand{\sfS}{\mathsf{S}}

\newcommand{\TN}{\mathscr{N}}
\newcommand{\ocT}{\ocirc{\mathcal{T}}}
\newcommand{\osfT}{\ocirc{\mathsf{T}}}

\newcommand{\Ccel}[1]{\mathcal{C}^\textsf{cel}_{(#1)}(S)}
\newcommand{\Ccar}[1]{\mathcal{C}^\textsf{car}_{(#1)}(\scri)}

\newcommand{\Ccarg}[1]{\mathcal{C}^\textsf{car}_{(#1)}(\scri,\g)}
\newcommand{\dCar}{\delta}

\newcommand{\Wt}{\bt\vartriangleright}
\newcommand{\Wtp}{\btp\vartriangleright}
\newcommand{\Wtpp}{\btpp\vartriangleright}

\newcommand{\dotvartriangleright}{\ensurestackMath{\stackon[-6.6pt]{\vartriangleright}{\mkern-3mu\cdot}}}

\newcommand{\dotblacktriangleright}{\ensurestackMath{\stackon[-6.6pt]{\blacktriangleright}{\mkern-3mu\textcolor{white}{\cdot}}}}

\newcommand{\scriptdotblacktriangleright}{\text{\scalebox{1.1}{$\ensurestackMath{\stackon[-4.4pt]{\blacktriangleright}{\mkern-3mu\textcolor{white}{\cdot}}}$}}}

\newcommand{\dotblacktriangleleft}{\ensurestackMath{\stackon[-6.6pt]{\blacktriangleleft}{\mkern3mu\textcolor{white}{\cdot}}}}

\newcommand{\scriptdotblacktriangleleft}{\text{\scalebox{1.1}{$\ensurestackMath{\stackon[-4.3pt]{\blacktriangleleft}{\mkern3mu\textcolor{white}{\cdot}}}$}}}

\newcommand{\dWt}{\bt\mkern5mu\dotvartriangleright\mkern5mu}
\newcommand{\dWtp}{\btp\mkern5mu\dotvartriangleright\mkern5mu}

\newcommand{\blackwhitetriangle}{\ensurestackMath{\stackon[-1pt]{\,}{\mkern5mu\begin{tikzpicture}
\draw[fill=black] (0,-0.04) -- (0,0.075) -- (0.26,0.075) -- (0,-0.04);
\draw (0,0.075) -- (0,0.19) -- (0.26,0.075);
\end{tikzpicture}\mkern5mu}}}

\newcommand{\BWt}{\tau\blackwhitetriangle}
\newcommand{\BWtp}{\tau'\blackwhitetriangle}

\newcommand{\barWhiteTriangle}{\ensurestackMath{\mkern1mu\stackon[-1pt]{\,}{\mkern4mu\begin{tikzpicture}
\draw (0,0) -- (0,0.25) -- (0.28,0.125) -- (0,0);
\draw (0.05,0.03) -- (0.05,0.23);
\end{tikzpicture}}\mkern4mu}}

\newcommand{\bWt}{\tau\barWhiteTriangle}
\newcommand{\bWtp}{\tau'\barWhiteTriangle}

\newcommand{\ocirc}[1]{\ensurestackMath{\stackon[1.5pt]{#1}{\mkern 1mu\scriptscriptstyle{\circ}}}}

\newcommand{\ocircp}[1]{\ensurestackMath{\stackon[-2pt]{#1}{\mkern-3mu\scriptscriptstyle{\circ}}}}

\newcommand{\ocircpp}[1]{\ensurestackMath{\stackon[-2pt]{#1}{\mkern-6mu\scriptscriptstyle{\circ}}}}

\newcommand{\Bt}{\ocirc{\tau}\blacktriangleright}
\newcommand{\Btp}{\ocircp{\tau'}\blacktriangleright}
\newcommand{\Btpp}{\ocircpp{\tau''}\blacktriangleright}

\newcommand{\dBt}{\ocirc{\tau}\mkern5mu \dotblacktriangleright\mkern5mu}
\newcommand{\dBtp}{\ocircp{\tau'}\mkern5mu \dotblacktriangleright\mkern5mu}
\newcommand{\dBtpp}{\ocircpp{\tau''}\mkern5mu \dotblacktriangleright\mkern5mu}

\newcommand{\rBt}{\ocirc{\tau}\blacktriangleleft}
\newcommand{\rBtp}{\ocircp{\tau'}\!\blacktriangleleft}
\newcommand{\rBtpp}{\ocircpp{\tau''}\!\blacktriangleleft}

\newcommand{\drBt}{\ocirc{\tau}\mkern5mu \dotblacktriangleleft\mkern5mu}
\newcommand{\drBtp}{\ocircp{\tau'}\mkern3mu \dotblacktriangleleft\mkern5mu}
\newcommand{\drBtpp}{\ocircpp{\tau''}\mkern2mu \dotblacktriangleleft\mkern5mu}

\newcommand{\Pairg}[1]{\big\langle #1\big\rangle_\g}

\newcommand{\Bprod}{\mkern5mu\begin{tikzpicture}
\draw[fill=black] (0,0) -- (0,0.23) -- (0.175,0.115) -- (0,0);
\draw[fill=black] (0.175,0.115) -- (0.35,0.23) -- (0.35,0) -- (0.175,0.115);
\end{tikzpicture}\mkern5mu}

\newcommand{\Wprod}{\mkern5mu\begin{tikzpicture}
\draw (0,0) -- (0,0.23) -- (0.35,0);
\draw (0,0) -- (0.35,0.23);
\end{tikzpicture}\mkern5mu}

\newcommand{\bWprod}{\mkern5mu\begin{tikzpicture}
\draw (0,0) -- (0,0.25) -- (0.40,0);
\draw (0,0) -- (0.40,0.25);
\draw (0.05,0.03) -- (0.05,0.22);
\end{tikzpicture}\mkern6mu}

\newcommand{\BWprod}{\mkern5mu\begin{tikzpicture}
\draw (0,0) -- (0,0.23) -- (0.35,0);
\draw (0,0) -- (0.35,0.23);
\draw[fill=black] (0,0) -- (0,0.115) -- (0.175,0.115) -- (0,0);
\end{tikzpicture}\mkern5mu}

\newcommand{\ot}{\ocirc{\tau}}
\newcommand{\otp}{\ocircp{\tau'}}
\newcommand{\otpp}{\ocircpp{\tau''}}

\newcommand{\Pta}{\gamma}
\newcommand{\Ptap}{\gamma'}
\newcommand{\Ptapp}{\gamma''}

\newcommand{\ada}{\mathrm{ad}_A}

\renewcommand{\t}[1]{\tau_{#1}}
\newcommand{\T}[1]{T_{#1}}
\newcommand{\tp}[1]{\tau'_{#1}}
\newcommand{\Tp}[1]{T'_{#1}}
\newcommand{\tpp}[1]{\tau''_{#1}}

\renewcommand{\a}[1]{\alpha_{#1}}
\newcommand{\Aa}[1]{\mathsf{a}_{#1}}
\newcommand{\ap}[1]{\alpha'_{#1}}
\newcommand{\Aap}[1]{\mathsf{a}'_{#1}}
\newcommand{\app}[1]{\alpha''_{#1}}

\newcommand{\scri}{\mathrsfso{I}}

\newcommand{\Qt}{Q_\tau}
\newcommand{\Qa}{Q_\alpha}
\newcommand{\Qta}{Q_{(\tau,\alpha)}}
\newcommand{\QPta}{Q_\Pta}

\newcommand{\lbr}{\llbracket}
\newcommand{\rbr}{\rrbracket}
\newcommand{\poisson}[1]{\big\{#1\big\}}
\newcommand{\paren}[1]{\,\Lbag #1\Rbag\,}

\newcommand{\cyc}{\overset{\circlearrowleft}{=}}

\newcommand{\Ham}{\eta}

\newcommand{\bfm}[1]{\boldsymbol{#1}}

\newcommand{\bfAc}{\bfm{\beta}}
\newcommand{\bfCc}{\bfm{\sigma}}

\usepackage{tikz}

\definecolor{myorange}{RGB}{223, 109, 20}

\definecolor{argile}{RGB}{239, 239, 239}
\definecolor{beige}{RGB}{254, 253, 240}

\begin{document}

\title{\Large{\bf Asymptotic Higher Spin Symmetries IV:\\
Einstein-Yang-Mills Theory
}}

\author{Nicolas Cresto\,\orcidlink{0009-0006-7263-8777}$^{1,2}$\thanks{ncresto@perimeterinstitute.ca}\,,
Laurent Freidel\,\orcidlink{0000-0001-6964-0100}$^1$\thanks{lfreidel@perimeterinstitute.ca} 
}
\date{\small{\textit{
$^1$Perimeter Institute for Theoretical Physics,\\ 31 Caroline Street North, Waterloo, Ontario, N2L 2Y5, Canada\\ \smallskip
$^2$Department of Physics \& Astronomy, University of Waterloo,\\Waterloo, Ontario, N2L 3G1, Canada
}}}

\maketitle
\begin{abstract}
We generalize the analysis of the asymptotic higher spin symmetries developed in the first three parts of this series by considering the minimal coupling of Einstein Gravity and Yang-Mills theory.
We show that there exist symmetry parameters that satisfy a collection of dual equations of motion, which allow the construction of an infinite collection of charges that are conserved in the absence of radiation.
These Noether charges act on the Einstein Yang-Mills phase space canonically and non-linearly. Their action defines a symmetry algebroid 
which reduces to a symmetry algebra at non-radiative cuts and generalizes the celestial $sw_{1+\infty}$ algebra. 
The corresponding symmetry bracket is shown to satisfy the Jacobi identity and an interesting cross-product structure, which is analyzed in details.
\end{abstract}

\newpage
\tableofcontents
\newpage

\section{Introduction}

In this letter, we piece together our analysis on asymptotic higher spin symmetries in General Relativity (GR) \cite{Cresto:2024fhd, Cresto:2024mne} and Yang-Mills (YM) theory \cite{Cresto:2025bfo}, and study the symmetry algebra (and symmetry algebroid) resulting from the minimal coupling of gravity and non-abelian gauge theory, namely Einstein-Yang-Mills (EYM) theory.
We refer the reader to the \emph{Introduction} and \emph{Preliminaries} sections of our companion papers, respectively for the list of relevant references on the topic and for some extra details about the notation that we shall use throughout this work.
In brief, see \cite{Bondi:1962px, Sachs:1962wk, Barnich:2011mi, Barnich:2016lyg, He:2014laa, Campiglia:2014yka, Compere:2018ylh, Campiglia:2020qvc, Grant:2021sxk, Freidel:2021fxf, Freidel:2021qpz, Freidel:2024jyf} for references about the BMS, eBMS, GBMS and BMSW algebras.
See \cite{Guevara:2021abz, Strominger:2021mtt, Himwich:2021dau} for the relevance of the $Lw_{1+\infty}$ algebra in celestial holography and the connection with scattering of soft gravitons, together with \cite{Ball:2021tmb, Penrose:1976js, Adamo:2021lrv, Adamo:2021zpw, Donnay:2024qwq, Kmec:2024nmu} for its appearance in self-dual gravity and its twistor space formulation.
We refer the reader to \cite{Strominger:2013jfa, ashtekar_angular_1979, Prabhu:2019fsp, Prabhu:2021cgk, Strominger:2017zoo, Hamada:2018vrw, Raclariu:2021zjz} for the interpretation of soft theorems as conservation laws and the relation to the $\cS$-matrix.
The phase space description of the gravitational higher spin charges was developed in \cite{Freidel:2021dfs, Freidel:2021ytz, Freidel:2023gue} and \cite{Geiller:2024bgf}, as a generalization to higher spin of the original spin 0 and spin 1 construction of \cite{Strominger:2013jfa, Strominger:2014pwa, Campiglia:2014yka, Campiglia:2015yka, Kapec:2014opa, Pasterski:2015tva}.
Concerning non-abelian gauge theory, the equivalence between soft gluon theorems \cite{Casali:2014xpa, Broedel:2014fsa, SabioVera:2014mkb} and Ward identities of asymptotic charges was developed in 
\cite{Strominger:2013lka, He:2015zea, Adamo:2015fwa}.
Importantly, \cite{Hamada:2018vrw, Li:2018gnc} discovered a whole tower of sub${}^s$-leading energetically soft theorems in gravity and Yang-Mills theory, then reformulated as a tower of sub${}^s$-leading conformally soft currents \cite{Guevara:2021abz, Strominger:2021mtt}, whose operator product expansions satisfy the $s$-algebra, $w_{1+\infty}$ algebra and $sw_{1+\infty}$ algebra for pure YM, pure GR and EYM theory respectively.
The main input of the current manuscript, namely the EYM evolution equations for the charge aspects, comes from the paper of Agrawal,  Charalambous and Donnay \cite{Agrawal:2024sju}.
\medskip

In this paper, we follow the notation and logic of \cite{Cresto:2024fhd, Cresto:2024mne, Cresto:2025bfo}. We derive in section \ref{secEYM:DualEOM+Noether} the set of dual equations of motion (EOM) that the Carrollian (field dependent) symmetry parameters $\t{}$ and $\a{}$, respectively dual to the gravitational and Yang-Mills charge aspects have to satisfy in order for the higher spin symmetries to be canonically realized via Noether charges.
These evolution equations take the form\footnote{$[\cdot\,,\cdot]_\g$ denotes the YM Lie algebra commutator, $C$ is the gravitational shear, $A$ the asymptotic gauge connection and $F=\dot{A}$ its curvature.}
\bs
\begin{align}
    \pa_u\t{s}&=D\t{s+1}-(s+3)C\t{s+2}, \\
    \pa_u\a{s}&=D\a{s+1}+[A,\a{s+1}]_\g-(s+2)C\a{s+2}-(s+2)F\t{s+1}.
\end{align}
\es
Moreover, the infinitesimal symmetry transformation on the asymptotic phase space variables, namely the shear $C$ and the transverse gauge field $A$, are given by\footnote{$N=\pa_uC$ is the news.}
\bs
\begin{align}
    \dta C &\equiv\dt C=N\t0-D^2\t0+2DC\t1+3CD\t1-3C^2\t2, \\
    \dta A &=F\t0-D\a0-[A,\a0]_\g+C\a1.
\end{align}
\es
We show that these transformations are generated by a Noether charge $\Qta$, which is given by the sum of the Noether charges $\Qt$ and $\Qa$ introduced in \cite{Cresto:2024mne} and \cite{Cresto:2025bfo} plus cross terms.
Next, in section \ref{secEYM:algebroid}, we define the `$\sfST$-algebroid' bracket $\lbr\cdot\,,\cdot\rbr$ which encodes the commutator of Noether charges and symmetry transformations on the phase space of EYM,
\begin{equation}
    \big[\dta,\dtap\big]\cdot=-\delta_{\lbr (\tau,\a{}),(\tau'\mkern-3mu,\ap{})\rbr}\cdot.
\end{equation}
The $\sfST$-bracket reproduces the pure GR $\sfT$-bracket \cite{Cresto:2024mne} and pure YM $\sfS$-bracket \cite{Cresto:2025bfo}, while incorporating the effect of the coupling in terms of a bicrossed product structure.

We first report in section \ref{secEYM:T-bracket} a definition of the $\sfST$-bracket. 
Then, in section  \ref{secEYM:YMBracket} we introduce a kinematical Lie algebra bracket defined \emph{off-shell} of the dual EOM. 
We prove that the $\sfST$-algebroid bracket is obtained from this kinematical Lie algebra bracket  after imposition of the dual EOM. 
This off-shell algebra has the structure of a semi-direct product. 

Imposing the dual EOM destroys the semi-direct structure. 
We prove in section \ref{secEYM:BicrossedJacobi} that the $\cST$-algebroid is realized through a combination of a bicrossed product and a semi-direct action of $\cT$ on $\cS$. 
This means that
$\cST\simeq\cTo\Bprod\big(\bcT \Wprod\cS\big)$, where $\cTo$ is the sub-algebroid of super-translations with the center, while $\bcT$ represents the sub-algebroid of sphere diffeomorphisms and higher-spin parameters. 
$\cS$ is the YM algebroid and an ideal.
This structure allows us to prove in subsec.\,\ref{secEYM:JacobiIdentity} that the $\sfST$-bracket satisfies the Jacobi identity.

Besides, we study in section \ref{secEYM:CovWedge} the covariant wedge algebra $\Wcal_{CA}(\scri)$ associated to the $\cST$-algebroid and how it represents a deformation of the celestial $sw_{1+\infty}$ bracket, cf.\,\ref{secEYM:celestialBracket}, while on the other hand it can also be recast as a shifted Schouten-Nijenhuis bracket.
Interestingly, we find that the 3 actions forming the bicrossed/semi-direct products of Lie algebroids recombine into a single action $\blackwhitetriangle$ such that at a non-radiative cut $S$ of $\scri$, the EYM wedge algebra takes the form of the semi-direct product of the gravitational wedge algebra with the Yang-Mills one. 
Finally we point out in section \ref{secEYM:twistor} the relationship with twistor theory and its Poisson bracket.
On top of the main text, a large collection of appendices detail most demonstrations.\\

\ni\textbf{Summary of notations: }
To help the reader follow the paper we summarize here some of the main notations, introduced with a reference to their definitions.

\begin{multicols}{2}

\ni\textbf{Vector spaces}
\begin{itemize}
    \item $\sfSK,\sfTK,\sfTK^+,\sfSTK$: Kinematical spaces \eqref{DefSTspace}.
    \item $\sfSK=\bigoplus_{s=0}^\infty\Ccar{0,-s}$.
    \item $\sfTK=\bigoplus_{s=-1}^\infty\Ccar{-1,-s}$.
    \item $\sfTK^+=\bigoplus_{s=0}^\infty\Ccar{-1,-s}$.
    \item $\sfSTK=\sfSK\oplus\sfTK$.
    \item $\sfS,\sfT,\sfST,\sfST^+$: On-shell spaces \eqref{DefSTspace}. Same vector space structure as their kinematical pendant, but with their respective dual EOM imposed. 
    \item $\osfT,\bsfT$: sub-spaces of $\sfST$ \eqref{defSpaceobar}.
    \item $\W_{CA}(\scri), \W_{\bfCc\bfAc}(S)$: Wedge spaces (\ref{CovWedgeScri}-\ref{CovWedgeEYMdefBis}).
\end{itemize}

\ni\textbf{Algebras and algebroids}
\begin{itemize}
    \item $\cS$: YM algebroid \cite{Cresto:2025bfo} and \eqref{hatdaA}.
    \item $\cT$: GR algebroid \cite{Cresto:2024mne} and \eqref{defdeltaEYM}.
    \item $\cST$: EYM algebroid \eqref{defdeltaEYM}.
    \item $\ocT, \bcT$: sub-algebroids of $\cST$ \eqref{defBracketobar}.
    \item $\cSK,\cTK,\cSTK$: kinematical algebras \eqref{Kbracket}.
    \item $\Wcal_{CA}(\scri), \Wcal_{\bfCc\bfAc}(S), \Wcal^\gr_{\bfCc}(S), \Wcal^\ym_{\bfCc\bfAc}(S)$: covariant wedges (\ref{CovWedgeScri}-\ref{CovWedgeEYMdefBis}-\ref{secEYM:SemiDirectWedge}).
\end{itemize}

\ni\textbf{Brackets}
\begin{itemize}
    \item $[\cdot\,,\cdot]_\g$: YM Lie algebra bracket.
    \item $[\a{},\ap{}]^\g, \lbr\a{},\ap{}\rbr^\g$: $\sfS$-bracket \eqref{SbracketEYM} and \eqref{hatdaA}.
    \item $\{\tau,\tau'\}$: GR kinematical bracket \eqref{CbracketScridef}.
    \item $\{\Pta,\Ptap\}^\ym$: YM kinematical bracket \eqref{Kbracket}.
    \item $\{\Pta,\Ptap\}$: EYM kinematical bracket \eqref{Kbracket}.
    \item $[\tau,\tau']^C$: GR $C$-bracket \eqref{TbracketEYM}.
    \item $[\Pta,\Ptap]^\ym$: YM-bracket \eqref{YMbracketCFdef} or \eqref{YMbracketDTOTdef}.
    \item $\lbr\tau,\tp{}\rbr$: GR $\sfT$-bracket \eqref{STbracket}.
    \item $\lbr\Pta,\Ptap\rbr$: EYM $\sfST$-bracket \eqref{STbracket}.
    \item $\big\lbr\ot,\otp\big\rbr^\circ$: $\circ$-bracket \eqref{defBracketobar}.
    \item $\overline{\lbr\bt,\btp\rbr}$: bar-bracket \eqref{defBracketobar}.
\end{itemize}

\ni\textbf{Covariant derivatives}
\begin{itemize}
    \item $\Dcal,\DGR,\DYM,\DTOT$: \eqref{covd}.
    \item $\DEYM$: \eqref{defDEYM}.
    \item $\nabla,\nablaGR,\nablaYM,\nablaEYM$: \eqref{defNabla}.
\end{itemize}

\ni\textbf{Charges}
\begin{itemize}
    \item $\Qt^u,\Qa^u,\QPta^u$: master charges (\ref{defMasterChargeEYM}-\ref{defMasterChargeEYM2}).
    \item $\widehat{Q}_\Pta,\QPta$: Noether charges (\ref{NoetherChargeEYM}-\ref{NoetherChargeEYM3}) and \eqref{NoetherChargeEYM2}.
\end{itemize}

\ni\textbf{Actions}
\begin{itemize}
    \item $\barWhiteTriangle$: \eqref{KinematicalAction}.
    \item $\vartriangleright,\blacktriangleright, \blacktriangleleft$: \eqref{defAction}.
    \item $\dotvartriangleright, \dotblacktriangleright,\dotblacktriangleleft$: \eqref{defActionDot2}. 
    \item $\blackwhitetriangle$: \eqref{BlackWhiteAction}.
\end{itemize}
\end{multicols}

\section{Dual EOM, infinitesimal action and Noether charge \label{secEYM:DualEOM+Noether}}

From the paper of Agrawal,  Charalambous and Donnay \cite{Agrawal:2024sju}, we know that the EYM equations of motion for the higher spin charge aspects are given by\footnote{We use an anti-hermitian basis of generators for $\g$. The pairing $\Pairg{Y,Z}=-\Tr(YZ),~Y,Z\in\g$,  is positive definite. }
\bs\label{Chargeev}
\begin{align}
    \pa_u\tQ_s^\ym &=D\tQ_{s-1}^\ym+\big[A,\tQ_{s-1}^\ym\big]_\g +sC\tQ_{s-2}^\ym, && s\geqslant 0,\\
    \pa_u\tQ_s^\gr &=D\tQ_{s-1}^\gr+(s+1)C\tQ_{s-2}^\gr+(s+1)\Pairg{F,\tQ_{s-1}^\ym}, & & s\geqslant -1,
\end{align}
\es
with $C$ and $N=\pa_u C$ the shear and curvature news as defined in \cite{Cresto:2024mne}, and $A$ and $F=\pa_u A$ the asymptotic gauge field and Yang-Mills curvature \cite{Cresto:2025bfo}. 
The lowest charge aspects encode the radiative data\footnote{Our conventions are such that $\tQ_{-3}^\gr=\tQ_{-2}^\ym=0$.}
\be
\tQ_{-2}^\gr=\frac{\dot{\bN}}{4\pi \GN}, \qquad\tQ_{-1}^\gr=\frac{D\bN}{4\pi \GN}, \qquad 
\tQ_{-1}^\ym=- \frac{ 2\bnews}{\gYM^2}.
\ee
The gravitational charge aspects  $\tQ_{s}^\gr$  have a Carrollian weight\footnote{A Carrollian field $\Phi(u,z,\bz) \in \Ccar{\dCar,s}$ of Carrollian weight $\delta$ and helicity (or spin-weight) $s$ transforms under sphere diffeomorphisms as
\be
\delta_{\mathcal Y}  \Phi 
= \big(YD + \tfrac12 (\delta  + u\pa_u + s) DY   \big)\Phi
+ \left(\bY\bD+\tfrac12 (\delta + u\pa_u  - s)\bD\bY \right) \Phi ,
\label{sDeltaDefCarrollEYM}
\ee
where $Y$ and $\bY$ are the projections of $\mathcal Y\in\mathrm{Diff}(S)$ along the dyad onto the sphere $\bm$ and $m$ respectively, namely $\mathcal Y=\mathcal Y^A\pa_A=(m^A\mathcal Y_A)\bD+(\bm^A\mathcal Y_A)D\equiv \bY\bD+YD$. We have that $A \in \Ccar{1,1}$, $C\in \Ccar{1,2}$ while $\pa_u$ raises the Carrollian weight by $1$.} equal to $3$ and a Carrollian spin equal to $s$ which we denote as $\tQ_s^\gr\in\Ccar{3,s}$. Similarly, the charge aspects  $\tQ_{s}^\ym$ are Lie algebra valued and of Carrollian weight equal to $2$ and spin $s$ which we denote as $\tQ_s^\ym\in\Ccarg{2,s}$.

To define the integrated Noether charge we  introduce the dual transformation parameters
\be 
\t{s}\in\Ccar{-1,-s}, \qquad \a{s}\in\Ccarg{0,-s},
\ee
which can be used to form the integrated  master charges 
\bs
\label{defMasterChargeEYM}
\begin{align}
    \Qt^u &:=\sum_{s=-1}^{\infty}\QGR{s}[\t{s}]\equiv\sum_{s=-1}^{\infty}\int_{S_u}\tQ_s^\gr\t{s}, \\
    \Qa^u &:=\sum_{s=0}^{\infty}\QYM{s}[\a{s}]\equiv\sum_{s=0}^{\infty}\int_{S_u}\Pairg{\tQ_s^\ym,\a{s}}.
\end{align}
\es
Here and in the following we denote the series $\tau=(\t{-1},\t0,\t1,\ldots)$ and $\a{}=(\a0,\a1,\a2,\ldots)$ simply by the letters without the subscript.
The integrals are taken over the sphere $S_u$ at the cut $u=\mathrm{cst}$. We naturally define the EYM master charge $\Qta^u$ as the sum of the respective master charges $\Qt^u$ and $\Qa^u$:
\begin{equation} \label{defMasterChargeEYM2}
    \boxed{\QPta^u\equiv\Qta^u:=\Qt^u+\Qa^u},
\end{equation}
where $\Pta$ stands for the pair of elements $(\tau,\a{})$.
We denote $\Pta_s=(\t{s},\a{s}),\, s\geq 0$ and $\Pta_{-1}=(\t{-1},0)$.
We assume that there is no massive matter so that the master charge at $i^+$ vanishes: $\QPta^{+\infty}=0$.

\subsection{Dual equations of motion}

As in \cite{Cresto:2025bfo}, we find the dual set of EOM for $\gamma$ in order for the  master charge to be  conserved in the absence of left-handed radiation.\footnote{See \cite{Cresto:2024mne} (and what follows) for an extensive discussion about the subtleties of defining the non-radiative condition either as $\bN\equiv 0$ or as $\dot{\bN}\equiv 0$. \label{footNNdot}}

\begin{tcolorbox}[colback=beige, colframe=argile]
\textbf{Theorem [Master charge conservation]}\\
In the absence of left-handed radiation, $\dot\bN=0=\bnews$, the master charge is conserved 
\be 
\pa_u \QPta^u =0,
\ee 
iff the dual EOM  $\E(\Pta)=0$ are satisfied,  where $\E(\Pta)=\big( \E^\gr(\t{}), \E^\ym(\Pta)\big)$ and
\bs \label{DualEOMEYM}
\begin{align} 
    \E_s^\gr(\t{})&:=\pa_u\t{s}-\Big(D\t{s+1}-(s+3)C\t{s+2}\Big), & s\geq -1, \\
    \E_s^\ym(\Pta)&:=\pa_u\a{s}-\Big(\DYM\a{s+1}-(s+2)C\a{s+2}-(s+2)F\t{s+1}\Big), & s\geq 0.
\end{align}
\es
\end{tcolorbox}

\ni We have denoted the asymptotic gauge covariant derivative onto the sphere by
\begin{equation}
    \DYM:=D+\ada=D+[A,\cdot\,]_\g.
\end{equation}

\paragraph{Remark:} In the following it will be useful to note that the equations of motion can be simply written as 
\begin{equation}\label{dEOM}
    \boxed{\E(\Pta)=\pa_u\Pta-\Dcal\Pta},
\end{equation}
where we define a covariant derivative $ \Dcal\Pta :=\big(\DGR\tau,\DTOT\Pta\big)$ with\footnote{$\DGR$ was introduced in \cite{Cresto:2024fhd}.}
\bs \label{covd}
\begin{align}
    \big(\DGR\t{}\big)_s &:=D\t{s+1}-(s+3)C\t{s+2}, \\
    \big(\DTOT\Pta\big)_s &:=\DYM\a{s+1}-(s+2)\big(C\a{s+2}+F\t{s+1}\big).
\end{align}
\es

Notice that the gravitational evolution equations for the Carrollian symmetry parameters $\tau$ are unchanged compare to \cite{Cresto:2024mne}.
However, the dual EOM for $\a{}$ get corrected by interaction terms of the type $C\a{}$ and $F\tau$.
It is worth emphasizing the structure of the couplings by rewriting $\E(\Pta)$ in matrix form as
\begin{equation}
    \pa_u\Pta_s=(D+\cA_s)\Pta_{s+1} -\mathcal{C}_s\Pta_{s+2},
\end{equation}
where
\begin{equation}
    \Pta_s=\binom{\t{s}}{\a{s}},\qquad \cA_s=\mat{0}{0}{-(s+2)F}{\ada},\qquad \mathcal{C}_s=\mat{(s+3)C}{0}{0}{(s+2)C}.
\end{equation}
We see that the gravitational field acts diagonally as $-(\delta+ s+2) C\phi_{s+2}$,  on  a symmetry parameter $\phi_s\in\Ccar{-\delta,-s}$. Note that $\dCar+s+2$ is the homogeneous holomorphic weight \cite{gelfand1966generalized, Cresto:2024mne} of the field $\phi_{s+2}$.\\

We define $\dot\bN=\bnews=0$ as the self-dual non-radiative conditions. 
They are implied by but not equivalent to the self-duality of the gravitational and Yang-Mills fields.\footnote{Self-duality imposes also that $\psi_2=\psi_1=\psi_0=\phi_1=\phi_0=0$, in the Newman-Penrose notation \cite{newman_new_1968}.}
If one relaxes these conditions, we find that the master charge satisfies a simple flux equation.

\begin{tcolorbox}[colback=beige, colframe=argile]
\textbf{Theorem [Charge flux]}\\
Assuming that the dual EOM $\E(\gamma)=0$ are satisfied, in the presence of left-handed radiation the master charge flux is linear in $(\dot\bN,\bnews)$ and given by 
\be \label{Qflux}
\pa_u \QPta^u =
\frac{1}{4\pi\GN}\dot\bN\big[\sI_\tau\big]-\frac{2}{\gYM^2}\bnews\big[\sI_\Pta^\ym\big],
\ee 
where we define the \emph{initial constraint} $\sI_\Pta$ as
\begin{equation}
    \label{defI} \sI_\Pta\equiv\big(\sI_\tau,\sI_\Pta^\ym \big):= \Big(C\t0-D\t{-1}\,,-\DYM\a0+C\a1+F\t0\Big).
\end{equation}
\end{tcolorbox}

\paragraph{Proof of Theorems:}
The proof of both theorems follows from the same calculation. 
Taking the time derivative of \eqref{defMasterChargeEYM2}, we find, using the charge conservation laws \eqref{Chargeev} and the convention $\QYM{-2}=\QGR{-3}=0$, that
\begin{align}
    \pa_u\QPta^u 
&=\sum_{s=0}^\infty \left(\QYM{s}\big[\pa_u\a{s}\big]-\QYM{s-1}\big[\DYM\a{s}\big] + \QYM{s-2}\big[sC\a{s}\big]  \right) \cr  &
+\sum_{s=-1}^\infty \left( \QGR{s}\big[ \pa_u\t{s}\big]- \QGR{s-1} \big[D\t{s}\big] +\QGR{s-2}\big[(s+1)C\t{s}\big]+ \QYM{s-1}\big[(s+1)F\t{s}\big]\right)
\nn\\
&=\sum_{s=0}^\infty \QYM{s}\big[\pa_u\a{s} - \DYM\a{s+1} + (s+2) (C\a{s+2} + F \t{s+1})\big]  \cr  &
+\sum_{s=-1}^\infty \QGR{s}\big[ \pa_u\t{s}- D\t{s+1} + (s+3)C\t{s+2}\big]
\\
&+ \QYM{-1}\big[-\DYM\a0+C\a1+F\t0\big]
+\QGR{-2}\big[C\t0-D\t{-1}\big]. \nn
\end{align}
Therefore we get that
\begin{equation} \label{QuPtaDot}
 \boxed{  \pa_u\QPta^u =\frac1{4\pi \GN}\dot\bN\big[\sI_\tau\big]-\frac2{\gYM^2}\bnews\big[\sI_\Pta^\ym\big] 
    + \sum_{s=-1}^\infty \QGR{s}\big[\E_{s}^\gr\big]
+\sum_{s=0}^\infty\QYM{s}\big[\E_{s}^\ym\big] }
\end{equation}
where we used the definitions (\ref{DualEOMEYM}-\ref{defI}).
Imposing the dual evolution equations 
$\E_s^\gr(\tau)=0$, $\forall\,s\geq -1$ and $\E_s^\ym(\Pta)=0$, $\forall\,s\geq 0$, implies \eqref{Qflux}.
Setting $\dot\bN=\bnews=0$ in addition, we see that $\pa_u\QPta^u=0$, which concludes the proofs.

\subsection{Infinitesimal variations and Noether charge \label{secEYM:NoetherChargeBondi}}

Given the central relevance of the dual EOM we introduce the following vector spaces:

\begin{tcolorbox}[colback=beige, colframe=argile]
\textbf{Definition [$\sfST$-space]} 
\bs \label{DefSTspace}
\begin{align}
\sfST^+ :=&\Big\{\,\Pta\equiv(\tau,\a{})\in\sfTK^+ \oplus\sfSK~\big|~\E_s(\Pta)=0,~  \forall\,s\geqslant 0~ \Big\},\\
\sfST :=& \Big\{\,\Pta\equiv(\tau,\a{})\in\sfTK\oplus\sfSK~\big|~\E_{s-1}^\gr(\tau )=\E_{s}^\ym(\Pta)=0,~  \forall\,s\geqslant 0~ \Big\}.
\end{align}
\es
\end{tcolorbox}

\ni $\sfTK$ and $\sfSK$ are short-hand notations for the spaces 
\be \label{defSTKspace}
\sfTK :=\bigoplus_{s=-1}^\infty\Ccar{-1,-s}, \qquad 
\sfSK:=\bigoplus_{s= 0}^\infty\Ccarg{0,-s}.
\ee 
The subscript $K$ stands for \emph{Kinematical}, i.e. without any dual EOM imposed.
$\sfTK$ was defined in \cite{Cresto:2024mne}.
Similarly we denote $\sfTK^+ :=\bigoplus_{s=0}^\infty\Ccar{-1,-s}$ as the space similar to $\sfTK$ with the $\tau_{-1}$ parameter excluded. 
For shortness, we write $\sfSTK\equiv\sfTK\oplus\sfSK$ while $\sfST$ is the space where the dual EOM are imposed.
The difference between $\sfST$ and $\sfST^+\subset \sfST$ is in the range of parameters and of  dual EOM imposed. 
In $\sfST^+$ only dual EOM for positive $s$ hold.
$\Pta\equiv(\tau,\a{})$ denotes an element of $\sfST$ (or $\sfSTK$ according to the context).
The $\sfS$-space and the $\sfT$-space are just the restrictions of $\sfST$ to the YM, respectively GR sub-spaces.

Another master charge plays a key role in the analysis of gravitational symmetry.
It is defined as 
\be 
\boxed{\widehat{Q}^u_\Pta:= Q^u_\Pta - \frac{1}{4\pi \GN} \bN[\sI_\tau]}. 
\ee 
By construction $\widehat{Q}^u_\Pta $ is independent of $\t{-1}$.
While $\tQ^\gr_0$ is the covariant mass aspect introduced in \cite{Freidel:2021qpz}, $\widehat{Q}^\gr_0 = \tQ^\gr_0-\bN C/{4\pi\GN}$ corresponds to the Bondi mass aspect. 
The other (higher spin) charge aspects are unmodified.

If we assume that $\gamma\in \sfST^+$  we obtain that
this master charge satisfies the conservation law  
\be \label{cons2}
\boxed{\pa_u  \widehat Q^u_\Pta  = 
- \frac{1}{4\pi \GN} \bN\big[\dPta C\big]
- \frac2{\gYM^2} \bnews\big[\dPta A\big]},
\ee 
where we have defined the infinitesimal variations of the shear and gauge potential as

\begin{tcolorbox}[colback=beige, colframe=argile]
\textbf{Definition [Infinitesimal variations]}\\
For $\Pta\in \sfST^+$,
\vspace{-0.6cm}

\bs \label{deltaPtaCA}
\begin{align}
    \dPta C &=N\t0-D^2\t0+2DC\t1+3CD\t1-3C^2\t2, \\
    \dPta A &=F\t0-\DYM\a0+C\a1.
\end{align}
\es
\end{tcolorbox}

\ni The proof of \eqref{cons2} follows from \eqref{QuPtaDot} from which we infer that the modified master charge satisfies the conservation law
\be 
\pa_u  \widehat Q^u_\Pta  = 
- \frac{1}{4\pi \GN} \bN\big[\pa_u \sI_\tau + D \E_{-1}^\gr(\tau)\big]
- \frac2{\gYM^2} \bnews\big[\sI_\Pta^\ym\big].
\ee
Moreover we have that 
\begin{align}
\dPta A  =\sI_\Pta^\ym  \qquad\textrm{and}\qquad
\dPta C\equiv \dt C =\pa_u\sI_\tau+D\E_{-1}^\gr(\t{}). \label{dtCoffShell}
\end{align}

According to \eqref{deltaPtaCA}, note that the action of $\t0$ is the supertranslation action $\t0\pa_u$ on $C$ and $A$ while $\a0$ acts as a gauge transformation on $A$. 
The subleading parameter $\t1$ acts as a holomorphic diffeomorphism on $C$ while $\a1$ and $\t2$ respectively act on $A$ and $C$ via a term proportional to the gravitational shear.
The action on $C$ is unmodified compare to the purely gravitational case while the action on $A$  can be written as 
\be 
\delta_{(\tau,\alpha)}A= \da A + F\t0 + C\a1.
\ee 
The term $\da A = -\DYM\a0$ is the usual gauge action, the extra two terms represent the coupling of Yang-Mills to gravity.
\medskip

The symmetry charge $\widehat{Q}_\Pta:= \widehat{Q}_\Pta^{-\infty}$ is defined as the limit of $\widehat Q^u_\Pta$ to the asymptotic corner $\scri^+_-$, i.e. when $u\to -\infty$.  
In the absence of matter fields, the boundary condition at timelike infinity is such that $\lim_{u\to +\infty} \widehat{Q}^u_\Pta =0$. This means that for $\gamma \in\sfST^+$, we can express the charge as an integral over $\scri$ 
\begin{equation} \label{NoetherChargeEYM}
    \boxed{\widehat{Q}_{(\tau,\alpha)}=\frac{1}{4\pi\GN}\int_\scri\bN\dt C+\frac{2}{\gYM^2}\int_\scri\Pairg{\bnews,\dta A}}.
\end{equation}
Besides, the holomorphic\footnote{The gravitational Ashtekar-Streubel symplectic potential is real and given by 
\be 
\Theta^{\mathsf{AS}}=\frac1{8\pi \GN} \int_{\scri}\big(\bN \delta C +N \delta \bC\big).
\ee 
Adding to it a total variation and boundary term $\frac{1}{8\pi\GN}\big(\delta\left(\int \bN C\right)- \int \pa_u(C\delta\bC)\big)$ transforms it into its holomorphic version.}
Ashtekar-Streubel symplectic potential \cite{ashtekar1981symplectic} for EYM is given by  
\begin{equation} \label{HolAshtekrStreubel}
    \Theta^{\mathsf{HAS}}=\frac1{4\pi \GN} \int_{\scri}\bN \delta C+\frac{2}{\gYM^2}\int_{\scri}\Pairg{\bnews,\delta A}.
\end{equation}
We see that the charge is obtained as the contraction of the symplectic potential with the field space variation:
$ \widehat{Q}_{\Pta}=I_{\dPta} \Theta^{\mathsf{HAS}}$ and therefore, following  the Noether analysis of \cite{Cresto:2024mne, Cresto:2025bfo}, we know that the limit to the corner $\scri^+_-$ of the master charge is the Noether charge.\footnote{Assuming the fields satisfy Schwartzian fall-off conditions \cite{schwartz2008mathematics}. Going beyond is an interesting challenge which is necessary to include hyperbolic scattering data, see \cite{Choi:2024ajz, Choi:2024mac}.}
By construction, $\widehat Q_\Pta$ generates the symmetry transformation, namely
\begin{equation}
    \poisson{\widehat Q_{(\tau,\a{})},C}=\dt C\qquad\textrm{and} \qquad\poisson{\widehat Q_{(\tau,\a{})},A}=\dta A.
\end{equation}

\paragraph{Remark:}
The Noether charge splits as follows
\be \label{NoetherChargeEYM3}
\widehat{Q}_\Pta= \widehat{Q}^\gr_\tau + \Qa^\ym + \frac{2}{\gYM^2}
\int_{\scri} \Pairg{\bnews, F} \t0  +\frac{2}{\gYM^2}\int_{\scri} \Pairg{\bnews C,\a1},
\ee 
where the gravitational and Yang-Mills charges were already constructed in \cite{Cresto:2024mne, Cresto:2025bfo}:
\bs \label{NoetherChargeEYM4}
\begin{align}
\widehat{Q}^\gr_\tau &\equiv 
\frac1{4\pi \GN}\left(
\int_{\scri} \left(\bN N - D^2\bN\right)\t0
-\int_{\scri} \left(3D\bN C + \bN DC\right)\t1
-\int_{\scri} 3\bN C^2\t2
\right),\\
\Qa^\ym &= -\frac2{\gYM^2} \int_{\scri} \Pairg{\bnews, \DYM\a0}.
\end{align}
\es
In particular, the Noether charge associated to a global time translation $\t0=1$, i.e. the total energy, takes the Bondi form (with $\hat\Ham:=(1,0,0,\ldots)\in\sfT^+$)
\begin{equation}
    \widehat{Q}_{\hat\Ham}=\frac{1}{4\pi\GN}\int_\scri\bN N+\frac{2}{\gYM^2}\int_\scri\Pairg{\bnews, F}.
\end{equation}
Although we have a complete Noether understanding for all spin $s$ charges, our work does not provide any spacetime interpretation for the higher spin symmetries.
Connecting our results to the analysis of  \cite{Campiglia:2021oqz, Nagy:2022xxs, Nagy:2024dme, Nagy:2024jua}, which interprets the spin 2 symmetry as over-leading gauge transformations is left for future work.
Note however than the cubic term for the spin 2 charge has recently been understood by \cite{Jorstad:2025qxu} as part of the soft theorem in the $S$-matrix analysis.

\subsection{Explicit action of the low spin symmetry}

In order to write the Noether charge for the spin 0, 1 and 2 transformations, we solve the dual EOM \eqref{DualEOMEYM} for some initial conditions given at the $u=0$ cut for simplicity.
These are the usual celestial symmetry parameters.
We first focus on the gravity side.
The spin 0 symmetry is labeled by a supertranslation parameter $\T0$.
The spin 1 is labeled by the time independent vector $\T1$ while the parameter $\T2\in\Ccel{-1,-2}$ labels the spin 2 transformation.
On the gauge theory side, $\Aa0$ is a super-phase rotation while $\Aa1$ labels the subleading spin 1 transformation.
To proceed, it is convenient to introduce the potentials $(h,a)$, $h\in \Ccar{0,2}$ and $a\in \Ccar{0,1}$, which are such that 
\be 
\pa_u h=C, \qquad \pa_u a= A.
\ee 
We write the solutions of the dual EOM for parameters that vanish at high spin: $\t{s}=0$, $s\geqslant 3$ and $\a{s}=0$ for $s\geqslant 2$.
These solutions are expressed in terms of functions on the sphere $(T_0,T_1,T_2)$ for the gravitational sector and $(\Aa0,\Aa1)$ for YM. 
These encode the first subleading symmetries. 
\bs \label{solEOMEYM}
\begin{align}
    \t2 &=\T2, \\
    \t1 &=uD\T2+\T1, \\
    \t0 &=\frac{u^2}{2}D^2\T2+uD\T1+\T0-3h\T2.
\end{align}
Similarly,
\begin{align}
    \a1 &=\Aa1 -3 \pa_ua \T2, \\
    \a0 &=uD\Aa1+\Aa0+[a,\Aa1]_\g-2 \pa_ua \T1-2u\pa_ua  D\T2-aD\T2-3Da\T2,
\end{align}
\es
The validity of the dual EOM can be checked by direct inspection.
The various Noether charges for $\T0,\T1,\T2,\Aa0,\Aa1$ are then obtained by plugging the solutions \eqref{solEOMEYM} into \eqref{NoetherChargeEYM}.
After some algebra we get\footnote{Since the context is clear, we write $\widehat{Q}_{\T{s}}\equiv\widehat{Q}_{(\tau(\T{s}),\a{}(\T{s}))}$ and $\widehat{Q}_{\text{\scalebox{1.24}{$\Aa{s}$}}}\equiv\widehat{Q}_{(0,\a{}(\Aa{s}))}$ for conciseness.}
\bs
\begin{align}
    \widehat{Q}_{\T0} &= \frac1{4\pi \GN}
    \int_{\scri} \bN\Big(\!- D^2\T0+\Ll_{\T0} h\Big)+\frac{2}{\gYM^2}\int_{\scri} \Pairg{\bnews,\Ll_{\T0} a}, \\
    \widehat{Q}_{\T1} &=\frac1{4\pi \GN}    \int_{\scri} \bN\Big(\!-u D^3\T1+\Ll_{\T1}h\Big)+\frac{2}{\gYM^2}\int_{\scri} \Pairg{\bnews,\Ll_{\T1}a}, \\
    \widehat{Q}_{\T2} &=\frac1{4\pi \GN}    \int_{\scri} \bN\Big(\!-\frac{u^2}{2} D^4\T2+\Ll_{\T2}h\Big)+\frac{2}{\gYM^2}\int_{\scri} \Pairg{\bnews,\Ll_{\T2}a} \label{QhatT2EYM}\\
    &+\frac{2}{\gYM^2}\int_{\scri} \Big\langle\bnews,\big[A,(3\T2 D+2D\T2)a\big]_\g\Big\rangle_\g, \nn\\
    \widehat{Q}_{\text{\scalebox{1.24}{$\Aa0$}}} &= \frac{2}{\gYM^2}\int_{\scri}\Big\langle\bnews,-D\Aa0+[\Aa0,\pa_u a]_\g\Big\rangle_\g, \\
    \widehat{Q}_{\text{\scalebox{1.24}{$\Aa1$}}} \!&= \frac{2}{\gYM^2}\int_{\scri}\!\Big\langle\bnews, -uD^2\Aa1+[\Aa1,Da]_\g-\big[[\Aa1,a]_\g,\pa_u a\big]_\g +\big[D\Aa1(1+u\pa_u),a\big]_\g +C\Aa1\Big\rangle_\g,
\end{align}
\es
where the differential operators $\Ll$ are defined as
\bs
\begin{align}
    \Ll_{\T0} &:=\T0\pa_u^2, \\
    \Ll_{\T1} &:=\big(2\T1 D+(s+1+u\pa_u)D\T1 \big)\pa_u, \\
    \Ll_{\T2} &:=3\T2\big(D^2-C\pa_u-h\pa_u^2\big)+2D\T2 (s+1+u\pa_u)D+\frac12 D^2\T2(s+u\pa_u)(s+1+u\pa_u).\!\!
\end{align}
\es
Here $s$ is the spin-weight of the field $\Ll$ acts on, i.e. 1 for YM and 2 for GR.
$\widehat{Q}_{\T1}$ is the Dray-Streubel charge, which is covariant \cite{Dray:1984rfa, Flanagan:2015pxa,Chen:2022fbu, Freidel:2024jyf}. The charges $ \widehat{Q}_{\text{\scalebox{1.24}{$\Aa1$}}}$ and $\widehat{Q}_{\T2}$ contain terms which are cubic in the field variables $(\bN, \bnews, h, a)$.

\paragraph{Remark:} 
The master charge, respectively the Noether charge, is either written as a flux integral over a portion of $\scri$, respectively over full $\scri$, or as a corner integral \eqref{defMasterChargeEYM}.
We refer the reader to \cite{Cresto:2024mne, Cresto:2025bfo} for similar corner expressions involving the renormalized charge aspects and the initial condition $\Pta\big|_{u=0}\equiv(\T{},\Aa{})$ of the dual EOM.

\subsection{Self-dual phase space}

The self dual phase space is defined in terms of a potential $h$ which is related to the shear via 
\be 
C=\pa_u h.
\ee 
The asymptotic self-dual symplectic potential is then given by 
\begin{equation}
    \Theta^{\mathsf{SD}}= -\frac1{4\pi \GN} \int_{\scri} \dot \bN \delta h +\frac{2}{\gYM^2}\int_{\scri}\Pairg{\bnews,\delta A}.
\end{equation} 
It is related to the Holomorphic Ashtekar-Streubel potential \eqref{HolAshtekrStreubel} by a corner term which vanishes under the Schwartzian fall-off conditions.
An analysis similar to the subsection \ref{secEYM:NoetherChargeBondi} lets us conclude that when $\Pta\in\sfST$ the charge 
\be \label{NoetherChargeEYM2}
\boxed{\QPta := \QPta^{-\infty}=-\frac1{4\pi \GN} \int_{\scri} \dot \bN \dt h +\frac{2}{\gYM^2}\int_{\scri}\Pairg{\bnews,\dPta A}}
\ee 
is the Noether charge for the action $\dPta$,
where $\dPta A$ is the same as before while 
\be \label{dtahEYM}
\boxed{\dt h = \sI_\tau= C\t0 -D\t{-1}}.
\ee  
Note that thanks to \eqref{dtCoffShell} and the fact that $E_{-1}^\gr(\tau)=0$ for $\Pta\in \sfST$ we have that 
\be 
\pa_u \dt h = \dt C, \label{dthdtC}
\ee 
so the action of $\sfST$ on $h$ extends the action of $\sfST^+$ on $C=\pa_u h$.
Notice that the analog of the Bondi mass aspect involves now the covariant mass introduced in \cite{Freidel:2021qpz}.
Given $\Ham= (V, 1, 0,0,\ldots)\in\sfT$,
\be \label{Qham}
Q_{\Ham}= 
\frac{1}{4\pi\GN} \int_\scri  \dot \bN \big(DV-C\big) +\frac{2}{\gYM^2}\int_\scri\Pairg{\bnews, F},
\ee 
where $V=\tau_{-1}$ such that $\pa_u V=0$ labels the value of the central element.
We see that the central element evaluates to a vanishing boundary term under the boundary conditions we work with.

\section{$\cST$-algebroid \label{secEYM:algebroid}}

In the previous section we have defined the action $\dPta$ for $\Pta \in \sfST^+ $ or $\Pta \in \sfST$.
We intend to show in this section that this action closes under the commutator, namely
\begin{equation}
\big[\dPta,\dPtap\big]\subset\mathrm{Im}(\delta).
\end{equation}
The fact that this action closes means that there exists a bracket 
on $\sfST$, denoted $\lbr\cdot\,,\cdot\rbr$, which is such that 
\be \label{morphismEYM1}
\big[\dPta,\dPtap\big]=-\delta_{\lbr\Pta,\Ptap\rbr}
\ee 
when acting on functionals of $C$ and $A$, for  $\Pta,\Ptap\in\sfST$. 
In this section we first define the bracket, then prove that it preserves the dual EOM and that it satisfies \eqref{morphismEYM1} and the Jacobi identity.

\subsection{$C$-, YM- and $\sfST$-brackets}\label{secEYM:T-bracket}

For the reader's convenience, we recall the expressions of $[\t{},\tp{}]^C$ and $\paren{\t{},\tp{}}^\gr$, respectively called the $C$-bracket and GR Dali-bracket, both defined in \cite{Cresto:2024mne}.

\begin{tcolorbox}[colback=beige, colframe=argile]
\textbf{Definition [$C$-bracket]}\\
The $C$-bracket $[\cdot\,,\cdot]^C:\sfTK \times\sfTK\to\sfTK$ is given by 
\bs \label{TbracketEYM}
\begin{align}
    [\tau,\tau']_s^C &:=\sum_{n=0}^{s+1}(n+1)\big(\t{n}D\tp{s+1-n}-\tp{n}D\t{s+1-n}\big)-C\paren{\t{},\tp{}}_s^\gr, \\
    \paren{\t{},\tp{}}_s^\gr &:=(s+3)\big(\t0\tp{s+2}-\tp0\t{s+2}\big).
\end{align}
\es
\end{tcolorbox}

\ni We then introduce the YM-bracket. 
It contains a piece  $[\a{},\ap{}]^\g$ which denotes the algebra part of the $\sfS$-bracket \cite{Cresto:2025bfo},
\begin{equation}
\big[\a{},\ap{}\big]^\g_s:=\sum_{n=0}^s\big[\a{n} ,\ap{s-n}\big]_\g. \label{SbracketEYM}
\end{equation}

\begin{tcolorbox}[colback=beige, colframe=argile]
\textbf{Definition [YM-bracket]}\\
The YM-bracket $[\cdot\,,\cdot]^\ym:\sfSTK\times\sfSTK\to\sfSK$  between $\Pta,\Ptap\in\sfSTK$ is defined as
\begin{align}
    \big[\Pta,\Ptap\big]^\ym_s &:= \big[\a{},\ap{}\big]^\g_s+ \bigg\{ \sum_{n=0}^s(n+1) \Big(\t{n}\DYM\ap{s+1-n}-\ap{n+1}D\t{s-n}\Big)-\Pta\leftrightarrow\Ptap\bigg\} \label{YMbracketCFdef}\\
    &\quad -(s+2)C \Big(\t0\ap{s+2}-\tp0\a{s+2}+\a1\tp{s+1}-\ap1\t{s+1}\Big)-F\paren{\t{},\tp{}}^\ym_s, \nn
\end{align}
where the YM Dali-bracket takes the form
\begin{equation}
\paren{\t{},\tp{}}^\ym_s :=(s+2)\big(\t0\tp{s+1}-\tp0\t{s+1}\big) \quad \in~\Ccar{-2,-s-1}.
\end{equation}
\end{tcolorbox}

These brackets are antisymmetric pairings defined off-shell. 
In order to connect with the Noether charge construction and to prove the Jacobi identity we need to impose the dual EOM. 
The on-shell condition for $(\tau,\alpha)$ means that they are field dependent and therefore the brackets need to be promoted to an algebroid bracket with the anchor map $\delta$. 
See \cite{Cresto:2024mne} for a reminder about algebroids \cite{algebroid, Ciambelli:2021ujl, Barnich:2010eb, Barnich:2010xq}. 

\begin{tcolorbox}[colback=beige, colframe=argile, breakable]
\textbf{Definition [$\sfST$-bracket]}\\
The $\sfST$-bracket $\lbr\cdot\,,\cdot\rbr:\sfSTK\times\sfSTK\to\sfSTK$ between $\Pta,\Ptap\in\sfSTK$ is given by
\begin{align} \label{STbracket}
    \big\lbr\Pta,\Ptap\big\rbr=\big[\Pta, \Ptap\big]+\dPtap\Pta-\dPta\Ptap=
    \Big( \big\lbr\Pta,\Ptap\big\rbr^\gr ,\big\lbr\Pta,\Ptap\big\rbr^\ym\Big),
\end{align}
with the projection onto the GR subspace given by the $\sfT$-bracket \cite{Cresto:2024mne}, namely the sum of the $C$-bracket \eqref{TbracketEYM} with the anchor map,
\begin{equation}
    \big\lbr\Pta,\Ptap \big\rbr^\gr:=\lbr\tau,\tau'\rbr =[\tau,\tau']^C+\dtp\tau-\dt\tau',
\end{equation}
and the projection onto the YM subspace given by the sum of the YM-bracket \eqref{YMbracketCFdef} with the EYM anchor map,
\begin{align}
    \big\lbr\Pta,\Ptap\big\rbr^\ym :=\big[\Pta,\Ptap\big]^\ym+\dPtap\a{}-\dPta\ap{}.
\end{align}
\end{tcolorbox}

\paragraph{Remark:}
From \eqref{YMbracketCFdef} and the results of \cite{Cresto:2024fhd, Cresto:2024mne}, it is clear that $\t{-1}$ spans the center of the $\sfST$-bracket.\\

The first major property of this bracket is its closure in the $\sfST$-space.

\begin{tcolorbox}[colback=beige, colframe=argile] \label{LemmaClosureEYM}
\textbf{Lemma [$\sfST$-bracket closure]}\\
The $\sfST$-bracket closes, i.e. satisfies the dual EOM \eqref{DualEOMEYM}.
If $\Pta,\Ptap\in\sfST$ then $\E\big(\lbr\Pta,\Ptap\rbr\big)=0$. 
Hence $\lbr\Pta,\Ptap\rbr \in \sfST$.
\end{tcolorbox}

\paragraph{Proof of Lemma [$\sfST$-bracket closure]:}
We showed already in \cite{Cresto:2024mne} that if $\tau,\tau'\in\sfT$
\begin{equation} \label{dualEOMGRbracket}
    \pa_u\lbr\tau,\tau'\rbr=\DGR \lbr\tau,\tau'\rbr.
\end{equation}
We report the brute force demonstration of
\begin{equation} \label{dualEOMEYMbracket}
    \pa_u\big\lbr\Pta,\Ptap\big\rbr^\ym= \DTOT\big\lbr\Pta,\Ptap\big\rbr,\qquad\Pta,\Ptap \in\sfST
\end{equation}
in App.\,\ref{App:dualEOMEYM}. 
The covariant derivatives are introduced in \eqref{covd}.
The lengthy part is to compute
\begin{align}
    \Big(\big(\pa_u-\Dcal\big)\big[\Pta,\Ptap\big] \Big)^\ym_s=\Big\{\big[\dPta A,\ap{s+1}\big]_\g-(s+2)\ap{s+2}\dt C-(s+2)\tp{s+1}\dPta F\Big\}-\Pta\leftrightarrow\Ptap.
\end{align}
It is then easy to show that 
\begin{equation}
    \Big(\big(\pa_u-\Dcal\big)\big(\dPtap\Pta-\dPta\Ptap\big)\Big)^\ym=-\Big(\big(\pa_u-\Dcal\big)\big[\Pta,\Ptap\big]\Big)^\ym,
\end{equation}
so that \eqref{dualEOMEYMbracket} holds.\\

One of the main results of this paper is then the following theorem.

\begin{tcolorbox}[colback=beige, colframe=argile, breakable] \label{theoremSTAlgebroid}
\textbf{Theorem [$\cST$-algebroid]}\\
The space $\cST\equiv\big(\sfST,\lbr\cdot\,,\cdot\rbr,\delta\big)$ equipped with the $\sfST$-bracket \eqref{STbracket} and the anchor map $\delta$,
\vspace{-0.3cm}
\begin{align}
    \delta : \,\,&\cST\to \X(\PS) \nn\\
     &(\tau,\alpha)\mapsto\dta, \label{defdeltaEYM}
\end{align}
is a Lie algebroid over the Einstein-Yang-Mills phase space $\PS$. 
This means in particular that the anchor map is an anti-homomorphism of Lie algebroids, i.e.
\begin{equation}
    \big[\dPta,\dPtap\big]\cdot=-\delta_{\lbr\Pta,\Ptap\rbr}\cdot
\end{equation}
and that the $\sfST$-bracket satisfies the Jacobi identity,\footnote{$\cyc$ indicates an equality valid upon cyclic permutation of both the sides.}
\begin{equation}
    \big\lbr\Pta,\lbr\Ptap,\Ptapp\rbr\big\rbr\cyc 0.
\end{equation}
\end{tcolorbox}

\ni The demonstration of this theorem is split into three lemmas.
We already showed the first one, namely \hyperref[LemmaClosureEYM]{[$\sfST$-bracket closure]}.
The second lemma goes as follows:

\begin{tcolorbox}[colback=beige, colframe=argile] \label{lemmaAlgebroidAction}
\textbf{Lemma [Algebroid action]}\\
The action $\delta$ defined in \eqref{deltaPtaCA} and \eqref{defdeltaEYM} is a realization of the $\sfST$-bracket \eqref{STbracket} onto the phase space $\PS$:
\begin{equation} \label{AlgebroidActionEYM}
    \big[\dPta,\dPtap\big]\cdot=-\delta_{\lbr\Pta,\Ptap\rbr}\cdot.
\end{equation}
\end{tcolorbox}

\paragraph{Proof of Lemma [Algebroid action]:}
The fact that the anchor map restricted to the GR subspace is an anti-homomorphism of Lie algebroids, i.e.
\begin{equation}
    \big[\dPta,\dPtap\big]C=\big[\dt,\dtp\big]C =-\delta_{\lbr\tau,\tau'\rbr}C,
\end{equation}
is already one of the main result of \cite{Cresto:2024mne}.
Concerning the YM subspace, we report in appendix \ref{AppEYM:AlgAct} the demonstration of the morphism property 
\begin{equation} \label{AlgebroidActionEYM1}
    \big[\dPta,\dPtap\big]A=-\delta_{\lbr\Pta,\Ptap\rbr}A.
\end{equation}
for $\Pta,\Ptap\in \sfST$.

The third lemma \hyperref[LemmaJacEYM]{[Jacobi identity]} is given is section \ref{secEYM:JacobiIdentity} and necessitates a careful analysis of the algebraic structure of the $\sfST$-bracket.
We dedicate the entire section \ref{secEYM:BicrossedJacobi} to the latter.
\medskip 

We now come back to the YM-bracket \eqref{YMbracketCFdef} and detail one way to construct it, highlighting the complementarity of the Carrollian and celestial perspectives.
Doing so, we also give a complementary proof of the aforementioned lemma \hyperref[LemmaJacEYM]{[Jacobi identity]}.
For a first reading, we recommend the reader to focus on the next subsection and skip section \ref{secEYM:BicrossedJacobi}.

\subsection{Kinematical algebra \label{secEYM:YMBracket}}

In this subsection, we provide a simplified construction of  the YM-bracket \eqref{YMbracketCFdef}, as the on-shell restriction of a simple \emph{kinematical} bracket.

Let us start with the pure gravitational analysis.
In \cite{Cresto:2024mne}, we showed that the $C$-bracket \eqref{TbracketEYM} can equivalently be written as
\begin{equation}\label{CbracketDGRdefEYM}
    [\tau,\tau']_s^C =\sum_{n=0}^{s+1}(n+1)\big(\t{n}(\DGR\tp{})_{s-n}-\tp{n}(\DGR\tau)_{s-n}\big)
\end{equation}
with the covariant derivative defined in \eqref{covd}.
This was a consequence of the identity\footnote{This is obvious upon changing $ n \to s+2-n$.}
\begin{align}
    &\sum_{n=0}^{s+2}(n+1)(s+3-n)\big(\t{n}\tp{s+2-n}-\tp{n}\t{s+2-n}\big)=0 \nn\\
    \Leftrightarrow \quad & \sum_{n=0}^{s+1}(n+1)(s+3-n)\big(\t{n}\tp{s+2-n}-\tp{n}\t{s+2-n}\big)=\paren{\t{},\tp{}}_s^\gr.
\end{align}
The remarkable point is that by working on $\scri$ and using the dual EOM $\pa_u\tau=\DGR\tau$, one can replace the derivative onto the sphere by $\pa_u$ so that the $C$-bracket takes the simple form
$[\t{},\tp{}]^C_s=\{\t{},\tp{}\}_s$ with 
\begin{equation}
    \boxed{\big\{\t{},\tp{}\}_s:=\sum_{n=0}^{s+1}(n+1)\big(\t{n}\pa_u\tp{s-n} -\tp{n}\pa_u\t{s-n}\big) }.\label{CbracketScridef}
\end{equation}
In appendix \ref{AppEYM:JacobiScri} we show that $\{\tau,\tp{}\}$ is a Lie bracket, namely
\begin{equation} \label{JacobiKGRbracket}
    \big\{\tau,\{\tp{},\tpp{}\}\big\}\cyc 0.
\end{equation}
This means that the kinematical algebra 
$\cTK \equiv\big(\sfTK,\{\cdot\,,\cdot\}\big)$ is a Lie algebra which admits a natural action on $\cSK\equiv\big(\sfSK,[\cdot\,,\cdot]^\g\big)$.

\begin{tcolorbox}[colback=beige, colframe=argile]
\textbf{Definition [$\barWhiteTriangle$ action]}\\
We define the action\! $\barWhiteTriangle$ of GR onto YM by
\begin{align}
    \barWhiteTriangle :\sfTK &\to\mathrm{End}\big( \sfSK\big)\\
    \tau&\mapsto\bWt,
\end{align}
such that
\begin{equation} \label{KinematicalAction}
    \big(\bWt\ap{}\big)_s :=\sum_{n=0}^s(n+1) \Big(\t{n}\pa_u\ap{s-n}-\ap{n+1}\pa_u\t{s-n-1}\Big).
\end{equation}
\end{tcolorbox}

\ni We prove in App.\,\ref{AppEYM:tauBarAction} and \ref{AppEYM:HomtauBarAction} that the action $\!\barWhiteTriangle$ is a derivative and a homomorphism of Lie algebras, i.e.
\begin{equation}
    \barWhiteTriangle :\cTK\to\mathrm{Der}\big(\cSK\big),
\end{equation}
which amounts to the two properties
\bs \label{barWhiteProperties}
\begin{align}
    \bWt\big[\ap{},\app{}\big]^\g &=\big[\bWt\ap{},\app{}\big]^\g+\big[\ap{}, \bWt\app{}\big]^\g,\\
    \big\{\t{},\tp{}\big\} \barWhiteTriangle\app{} &=\big[\bWt,\bWtp\!\big]\app{}.
\end{align}
\es
This shows that $\barWhiteTriangle$ is a semi-direct action, which in turns guarantees that the Jacobi identity holds for $\{\cdot\,,\cdot\}^\ym$ defined as follows (so that $\cSTK=\cTK\bWprod\cSK$ is a Lie algebra).

\begin{tcolorbox}[colback=beige, colframe=argile, breakable]
\textbf{Theorem [$\cSTK$-algebra]}\\
$\cTK\equiv\big(\sfTK,\{\cdot\,,\cdot\}\big)$ and $\cSK\equiv\big(\sfSK,[\cdot\,,\cdot]^\g\big)$, respectively equipped with the brackets \eqref{CbracketScridef} and \eqref{SbracketEYM} are Lie algebras.
Moreover, $\cSTK\equiv\big(\sfSTK,\{\cdot\,,\cdot\}\big)$, equipped with the  kinematical $\sfK$-bracket $\{\cdot\, ,\cdot\}:\sfSTK\times\sfSTK\to\sfSTK$  defined as
\begin{equation}
    \{\Pta,\Ptap\}:=\big(\{\t{},\tp{}\},\{\Pta,\Ptap \}^\ym\big), \label{Kbracket}
\end{equation}
where the GR projection is given by \eqref{CbracketScridef} and the YM projection takes the form 
\begin{align}
    \{\Pta,\Ptap \}^\ym:=[\a{},\ap{}]^\g+\bWt\ap{}-\bWtp\a{}, \label{YMbracketScridef}
\end{align}
is the  semi-direct product Lie algebra 
\begin{equation}
    \cSTK=\cTK\bWprod\cSK. \label{Kalgebra}
\end{equation}
\end{tcolorbox}

We now establish that the Carrollian kinematical bracket projected on-shell is the same as the celestial bracket\footnote{Celestial in the sense that it involves differential operators onto the sphere. The relation with the usual celestial $sw_{1+\infty}$ bracket is given in Sec.\,\ref{secEYM:celestialBracket}.} defined earlier.

\begin{tcolorbox}[colback=beige, colframe=argile]
\textbf{Theorem [on-shell $\sfK$-bracket]} \label{KbracketTheorem}\\
The on-shell projection of the $\sfK$-bracket \eqref{Kbracket}, i.e. its restriction to the $\sfST$-space, is the $\sfST$-bracket \eqref{STbracket}:
\begin{equation}
   \big[\Pta, \Ptap\big]= \big\{\Pta,\Ptap\big\}\qquad \textrm{if}\quad\Pta,\Ptap\in \sfST,
\end{equation} and therefore the algebroid brackets coincide on-shell
\begin{equation}
   \big\lbr\Pta, \Ptap\big\rbr= \big\{\Pta,\Ptap\big\} +\dPtap\Pta-\dPta\Ptap \qquad \textrm{if}\quad\Pta,\Ptap\in \sfST.
\end{equation}
\end{tcolorbox}

\paragraph{Proof of Theorem [on-shell $\sfK$-bracket]:}
We start by showing how to get \eqref{YMbracketCFdef} from \eqref{YMbracketScridef} after the  on-shell projection.
For this we use the dual EOM in the form \eqref{dEOM} and the definition \eqref{covd} of the covariant derivatives in order to recast the $\sfK$-bracket as a bracket involving derivative operators over the sphere.
Hence, when $\gamma,\gamma'\in \sfST$ we have 
\begin{equation}
    \boxed{\big\{\Pta,\Ptap\big\}^\ym_s\overset{\sfST}{=} \big[\a{},\ap{}\big]^\g_s+\bigg\{\sum_{n=0}^s(n+1) \Big(\t{n}(\DTOT\Ptap)_{s-n}-\ap{n+1}(\DGR\t{})_{s-n-1}\Big)-\Pta\leftrightarrow\Ptap\bigg\}}. \label{YMbracketDTOTdef}
\end{equation} 
Next we expand $\Dcal\Pta$ and leverage two identities.
The first one is that
\begin{equation}
    \sum_{n=0}^{s+1}(n+1)(s+2-n)\big(\t{n}\tp{s+1-n}-\tp{n}\t{s+1-n}\big)=0,
\end{equation}
or equivalently 
\begin{equation}
    \sum_{n=0}^{s}(n+1)(s+2-n)\big(\t{n}\tp{s+1-n}-\tp{n}\t{s+1-n}\big)=(s+2)\big(\t0\tp{s+1}-\tp0\t{s+1}\big)=\paren{\tau,\tau'}^\ym_s.
\end{equation}
The second one is
\begin{equation}
    \sum_{n=0}^{s+1}(n+1)(s+2-n)\big(\t{n}\ap{s+2-n}-\ap{n+1}\t{s+1-n}\big)=0,
\end{equation}
which amounts to
\begin{equation}
    C\sum_{n=0}^s(n+1)(s+2-n)\big(\t{n}\ap{s+2-n}-\ap{n+1}\t{s+1-n}\big)=(s+2)C \big(\t0\ap{s+2}-\ap1\t{s+1}\big).
\end{equation}
From there we deduce that
\begin{align}
    \big\{\Pta,\Ptap\big\}^\ym_s &\overset{\sfST}{=} \big[\a{},\ap{}\big]_s+ \bigg\{ \sum_{n=0}^s(n+1) \Big(\t{n}\DYM\ap{s+1-n}-\ap{n+1}D\t{s-n}\Big)-\Pta\leftrightarrow\Ptap\bigg\} \\
    &\qquad-(s+2)C \Big(\t0\ap{s+2}-\tp0\a{s+2}+\a1\tp{s+1}-\ap1\t{s+1}\Big)-F\paren{\t{},\tp{}}^\ym_s, \nn
\end{align}
which is indeed the YM-bracket \eqref{YMbracketCFdef}.
Similarly, we already showed in the discussion at the beginning of this subsection that $\{\tau,\tp{}\} \overset{\sfST}{=}[\tau,\tp{}]^C$.
The general structure of Lie algebroid ensures that there is a unique choice of anchor map $\delta_\gamma$ on $(C,A)$ which preserves the dual EOM. 
This phenomenon is generic in the context of gauge fixing \cite{ Barnich:2009se, Barnich:2011mi, Freidel:2021fxf, Riello:2024uvs} where an algebra is promoted after gauge fixing to an algebroid which preserves the gauge fixing in question.
Once this choice is made the bracket is an algebroid bracket by construction.

Let us illustrate how this works  on the gravitational sector.
It is useful to introduce the concept of Leibniz rule anomaly of a bracket $[\cdot\,,\cdot]$ with respect to a differential operator $\mathscr{D}$, encoded in the expression 
\begin{equation} \label{LeibnizRuleNotation}
    \cA\big([\cdot\,,\cdot],\mathscr{D} \big):=\mathscr{D}[\cdot\,,\cdot]-[\mathscr{D}\cdot\,,\cdot]-[\cdot\,,\mathscr{D}\cdot].
\end{equation}
We find that
\begin{equation}
    \cA\Big(\big\{\tau,\tp{}\big\},\pa_u-\DGR\Big)=-\cA\Big(\big\{\tau,\tp{}\big\},\DGR\Big),
\end{equation}
where
\begin{equation}\label{Leibnizgr}
    \cA_s\Big(\big\{\tau,\tp{}\big\},\DGR\Big) =\left((s+3) \t{s+2} \pa_u\big(\DGR\tp{} \big)_{-2}+ \sum_{n=-1}^{s+1} \big(\DGR\tau\big)_n\big(\pa_u\tau'-\DGR\tau'\big)_{s-n}\right)-\tau\leftrightarrow\tau'.
\end{equation}
This clearly shows that even when $\tau,\tp{}\in\sfT$, such that $\pa_u\tau=\DGR\tau$, we are still left with
\begin{equation}
    \Big(\big(\pa_u-\DGR\big)\big\{\tau,\tp{}\big\} \Big)_{\!s} \overset{\sfT}=(s+3) \left[\tp{s+2} \pa_u\big(\DGR\t{} \big)_{-2}-\t{s+2} \pa_u\big(\DGR\tp{} \big)_{-2}\right],
\end{equation}
where $\big(\DGR\t{} \big)_{-2}=D\t{-1}-C\t{0}$.
The action of $\pa_u-\DGR$ is thus anomalous on the \emph{Lie bracket}.
The algebroid correction provides a counter-term. 
Indeed it is direct to check that 
\begin{equation}
    \Big(\big(\pa_u-\DGR\big)\big(\dtp\tau-\dt\tp{}\big)\Big)_{\!s}\overset{\sfT}=(s+3)\left(\tp{s+2}\dt C-\t{s+2}\dtp C\right).
\end{equation}
This means that for $\tau,\tau'\in\sfT$,\footnote{Albeit presented differently, this computation was done already in \cite{Cresto:2024mne}.}
\begin{equation}
    \Big(\big(\pa_u-\DGR\big)\big\lbr\tau, \tp{}\big\rbr\Big)_{\!s}=(s+3)\tp{s+2}\Big(\dt C +\pa_u\big(\DGR\tau\big)_{-2}\Big)-\tau\leftrightarrow\tp{}.
\end{equation}
Requiring that $\dt C=-\pa_u\big(\DGR\tau\big)_{-2}$ in order for the RHS of this last equation to vanish can be viewed as a \emph{definition} of $\dt C$. 
This definition matches the one given in \eqref{deltaPtaCA} since
\be 
\pa_u\big(\DGR\tau\big)_{-2} &=  
D\pa_u\t{-1}-C\pa_u\t0 - N\t0 \nn\\*
&\overset{\sfT}= D^2\t0 - 2D(C\t1)- C D\t1 +3C^2 \t2 -N\t0\equiv -\dt C,
\ee
where we use the dual EOM $\E_{-1}(\tau)= \E_0(\tau)=0$ and the definition of the action \eqref{deltaPtaCA} in the last equality (see also \eqref{dthdtC}). 
The story is similar for the YM projection of the $\sfK$-bracket.
This thus concludes the proof that 
\begin{equation}
    \big\{\Pta,\Ptap\big\}+ \dPtap\Pta-\dPta\Ptap \overset{\sfST}=\big\lbr\Pta, \Ptap\big\rbr.
\end{equation}

\paragraph{Remark:}
The various ways \eqref{TbracketEYM} or  \eqref{CbracketScridef} and \eqref{YMbracketCFdef} or \eqref{YMbracketScridef} of writing the $C$- and YM-brackets each have pros and cons. 
On one hand \eqref{TbracketEYM}, \eqref{YMbracketCFdef} and \eqref{CbracketDGRdefEYM}, \eqref{YMbracketDTOTdef} emphasize the \emph{corner algebra} structure, since only derivative operators onto the sphere appear.
Once projected at a cut of $\scri$, the bracket keeps the same functional form with the Carrollian parameters $(\t{},\a{})$ replaced by the celestial ones $(T,\Aa{})$.
Moreover, as we will see in section \ref{secEYM:celestialBracket}, once we impose the wedge condition, the resulting algebra is a deformation of the $sw_{1+\infty}$ algebra studied in \cite{Agrawal:2024sju}, where the radiative data play the role of deformation parameters.
We expect  the deformed algebra to be isomorphic to the undeformed one, as this was the case in gravity \cite{Cresto:2024fhd}.\footnote{Note that the brackets \eqref{TbracketEYM}, \eqref{YMbracketCFdef} do not satisfy the Jacobi identity outside of the wedge, see \eqref{JacobiEYMCbracket} and \eqref{JacobiEYMYMbracket} below.}
On the other hand, \eqref{CbracketScridef} and \eqref{YMbracketScridef} emphasize the simplicity of the brackets once expressed on $\scri$, fully committing to a Carrollian picture, where this is now the derivative along the null generator which appears explicitly. 
These brackets satisfy the Jacobi identity off-shell and exhibit the semi-direct product structure of the off-shell algebra. 
The algebroid arises naturally by imposing the dual EOM, which mimics a gauge fixing procedure.

\section{Bicrossed product and Jacobi identity \label{secEYM:BicrossedJacobi}}

In the previous section, we showed that the kinematical $\sfK$-bracket $\{\cdot\,,\cdot\}$ satisfies the Jacobi identity and projects to the $\sfST$-bracket on shell of the dual EOM.
We also showed that the $\cSTK$-algebra possesses a semi-direct product structure \eqref{Kalgebra}.
Unfortunately once we impose the dual EOM, the semi-direct product structure no longer holds on the $\sfST$-space. 

In this section, we show that the $\cST$-algebroid is instead built as the combination of a bicrossed product \cite{Majid_2002, michor1992knitproductsgradedlie} and a semi-direct product of Lie sub-algebroids.
Unraveling this algebraic structure also provides a direct proof that the Jacobi identity holds for the $\sfST$-bracket.
This section is technical and can be skipped on the first reading.
If the reader wishes to have a glance at the results, 
the main theorem is \hyperref[STisomorphismBis]{[Bicrossed product structure]}.

\subsection{Bicrossed product structure \label{secEYM:Bicrossed}}

We already mentioned the two vector subspaces $\sfS=\sfST\cap\sfSK$ and $\sfT=\sfST\cap\sfTK$.
Let us introduce the $\sfS$-bracket, which generalizes the one of \cite{Cresto:2025bfo} when the anchor map depends on the shear.

\begin{tcolorbox}[colback=beige, colframe=argile, breakable]
\textbf{Definition [$\sfS$-bracket]}\\
The $\sfS$-bracket $\lbr\a{},\ap{}\rbr^\g =[\a{},\ap{}]^\g+\hdap\a{}-\hda\ap{}$ is simply the $\sfST$-bracket restricted to $\sfS$. 
It naturally generalizes the one introduced in \cite{Cresto:2025bfo}, where the only difference appears in the usage of the anchor $\hda$, satisfying 
\be \label{hatdaA}
\hda A=-\DYM\a0+C\a1.
\ee 
$\cS\equiv\big(\sfS,\lbr\cdot\,,\cdot\rbr^\g,\hat\delta\big)$ is the $\cS$-algebroid in the presence of a non-zero shear. 
\end{tcolorbox}

\ni We can readily check that the $\cS$-algebroid is an ideal. 
Indeed, taking $\Ptap=(\tp{},\ap{})$, we have that
\begin{equation}
    \big\lbr(0,\a{}),\Ptap\big\rbr=\Big(0,\big\lbr (0,\a{}),\Ptap\big\rbr^\ym\Big).
\end{equation}
However the $\cT$-algebroid is \emph{not} a sub-algebroid! 
This is clear since 
\begin{equation} \label{anomalyTsubAlgebroid}
    \big[(\t{},0),(\tp{},0)\big]= \Big([\tau,\tau']^C,-F\paren{\tau,\tau'}^\ym\Big).
\end{equation}
In other words, the Dali-bracket prevents us from having a semi-direct action of $\cT$ onto $\cS$.
The situation is thus subtler.
To address it, we define $\osfT$ and $\bsfT$ as follows.

\begin{tcolorbox}[colback=beige, colframe=argile]
\textbf{Definition [$\osfT$- \& $\bsfT$-subspaces]}

\vspace{-0.6cm}
\bs \label{defSpaceobar}
\begin{align}
    \osfT &:=\sfST\cap\big( \Ccar{-1,1}\oplus\Ccar{-1,0}\big), \\
    \bsfT &:=\sfST\cap\bigg(\bigoplus_{s=1}^\infty \Ccar{-1,-s}\bigg).
\end{align}
\es
\vspace{-0.4cm}

$\osfT$ is the subspace of ``Carrollian super-translations'' together with the center. 
We denote its elements as $\ot\equiv(\t{-1},\t0,0,0,\ldots)$.
$\bsfT$ is the subspace of ``Carrollian sphere diffeomorphisms'' and higher-spin parameters. 
We denote its elements by $\bt\equiv(0,0,\t1,\t2,\t3,\ldots)$.
\end{tcolorbox}

\ni For convenience we denote an element of $\bsfT\oplus\sfS$ as $\bPta=(\bt,\a{})$ and an element of $\sfT$ as $\t{}=(\ot;\bt)$.

\begin{tcolorbox}[colback=beige, colframe=argile]
\textbf{Definition [$\circ$- \& bar-brackets]}\\
We define the $\circ$-bracket $\big\lbr\ot,\otp\big\rbr^\circ=\big[\ot,\otp \big]^\circ+\odtp\ot-\odt\otp$ and the bar-bracket $\overline{\lbr\bt,\btp\rbr}=\overline{[\bt,\btp]} +\dbtp\bt-\dbt\btp$, respectively as the restriction of the $\sfST$-bracket to $\osfT$ and $\bsfT$ :
\vspace{-0.5cm}

\bs \label{defBracketobar}
\begin{align}
    \lbr\cdot\,,\cdot\rbr^\circ&:\osfT\times \osfT\to\osfT \!\!\qquad\textrm{with}\qquad\! \big[\ot,\otp\big]^\circ= \big(\t0D\tp0-\tp0D\t0\,,0,0,\ldots\big), \\
    \overline{\lbr\cdot\,,\cdot\rbr}\,&:\bsfT\times \bsfT\to \bsfT\!\!\qquad\!\textrm{with}\mkern6mu\qquad\! \overline{[\bt,\btp]}_s=\sum_{n=1}^s(n+1)\big(\t{n}D\tp{s+1-n}-\tp{n}D\t{s+1-n}\big).
\end{align}
\es
\end{tcolorbox}

\ni Since $\paren{\bt,\btp}^\ym=0$, we infer that $\big\lbr(\bt,0),(\btp,0)\big\rbr= \big(\overline{\lbr\bt,\btp\rbr},0\big)$, so that $\bcT\equiv\big(\bsfT,\overline{\lbr\cdot\,,\cdot \rbr},\delta\big)$ is a sub-algebroid.
Similarly, since $\paren{\ot,\otp}^\ym=0$, we deduce that $\big\lbr(\ot,0),(\otp,0)\big\rbr= \big(\lbr\ot,\otp\rbr^\circ,0\big)$, so that $\ocT\equiv\big(\osfT,\lbr\cdot\,,\cdot\rbr^\circ, \delta\big)$ is a sub-algebroid of $\cST$ as well.
Notice that $\big\lbr(\bt,0),\Ptap\big\rbr^\ym\neq 0$ and $\big\lbr(\ot,0),\Ptap\big\rbr^\ym\neq 0$, so that neither $\bcT$ nor $\ocT$ are ideals.
Finally, we see that $\big\lbr\bPta,\bPtap \big\rbr^\gr=\big(0;\overline{\lbr\bt,\btp\rbr}\big)\in\bsfT$, which proves that $\big(\bsfT\oplus\sfS,\lbr\cdot\,,\cdot\rbr,\delta \big)$\footnote{Here it is understood that $\lbr\cdot\,,\cdot\rbr$ is the restriction of the $\sfST$-bracket to the subspace $\bsfT\oplus\sfS$.
We do not introduce any new notation for this bracket. 
The context should always make it clear which one we consider.} is itself a sub-algebroid of $\cST$ (though it is not an ideal).

We now define several actions and a co-reaction to describe the algebroid structure. The white action $\vartriangleright$ allows to construct the semi-direct product of algebroids $\bcT\Wprod\cS$.
We also introduce the black action $\blacktriangleright$ and the black back-reaction $\blacktriangleleft$ to form the bicrossed product \cite{michor1992knitproductsgradedlie} $\ocT\Bprod\big(\bcT\Wprod\cS\big)$.

\begin{tcolorbox}[colback=beige, colframe=argile]
\textbf{Definition [$\vartriangleright,\blacktriangleright, \blacktriangleleft$ Actions]}\\
We define 
\begin{align} \label{defAction2}
    \vartriangleright\,&:\bsfT_{\mkern-3mu K}\to\textrm{End} \big( \sfSK\big)&   \blacktriangleright\,&: \osfT_{\mkern-3mu K}\to\textrm{End} \big(\bsfT_{\mkern-3mu K} \oplus\sfSK \big)&  \blacktriangleleft\,&:\bsfT_{\mkern-3mu K}\oplus \sfSK\to\textrm{End}(\osfT_{\mkern-3mu K}) \\
    &:\bt\mapsto\Wt, &  &:\ot\mapsto\Bt, &  &: \bPta\mapsto\,\,\blacktriangleleft\bPta. \nn
\end{align}
such that\footnote{The first 3 lines are valid for $s\geqslant 0, s\geqslant 1$ and $s\geqslant 0$ respectively.}
\vspace{-0.8cm}

\bs \label{defAction}
\begin{align} 
    \big(\Wt\ap{}\big)_s \,&:=\sum_{n=1}^s\Big( (n+1)\t{n}\DYM\ap{s+1-n}-n\ap{n}D\t{s+1-n}\Big)+(s+2)C\ap1\t{s+1}, \label{defWhiteAction} \\
    \big(\Bt\bPtap\big)_s^\gr &:=\t0 D\tp{s+1}-(s+2)\tp{s+1}D\t0-(s+3)C\t0\tp{s+2}, \label{defBlackActionGR} \\
    \big(\Bt\bPtap\big)_s^\ym &:=\t0\DYM\ap{s+1}-(s+1)\ap{s+1}D\t0-(s+2)C\t0\ap{s+2}-(s+2)F\t0\tp{s+1}, \label{defBlackActionYM} \\
    \big(\rBt\bPtap\big)_{-1} &:=-2C\t0\tp1, \label{defReverseBlackAction-1} \\
    \big(\rBt\bPtap\big)_0 \,&:=\t0D\tp1-2\tp1D\t0-3C\t0\tp2. \label{defReverseBlackAction0}
\end{align}
\es
\end{tcolorbox}

\paragraph{Remark:} 
Notice that if we restrict ourselves to $\gbms$ transformations, i.e. spin 0 and 1, then $\Bt\bPtap\equiv 0$ and the entire $\gbms$ bracket comes from $\big(\rBt\bPtap\big)_0$ after setting the center, namely the degree $-1$ element, to 0.\\

One of the main result of this paper is the following theorem, see also \eqref{STisomorphismBis}.\footnote{We could also have written the algebroid as a different cross-product structure, of the type $\bcT\,'\!\Bprod\!'\big(\ocT\,'\!\Wprod\cS\big)$ for some primed actions related to the unprimed ones.}

\begin{tcolorbox}[colback=beige, colframe=argile]
\textbf{Theorem [Bicrossed product structure]}\\
\begin{equation} \label{STisomorphism}
\cST\simeq\ocT\Bprod\big(\bcT\Wprod\cS\big),
\end{equation}
such that in particular
\begin{align}
    \big[\Pta,\Ptap\big] &=\Big(\big[ \ot,\otp\big]^\circ+\rBt\bPtap-\rBtp\bPta\,;\overline{\big[\bt,\btp\big]}+ \big(\Bt\bPtap-\Btp\bPta\big)^\gr, \label{bicrossedBracket}\\
    &\qquad\big[\a{},\ap{}\big]^\g+\Wt\ap{}-\Wtp\a{}+ \big(\Bt\bPtap-\Btp\bPta\big)^\ym\Big).\nn
\end{align}
\end{tcolorbox}

\paragraph{Proof:} 
By definition of a bicrossed product bracket \cite{michor1992knitproductsgradedlie}, we have that
\begin{align}
    \big[(\ot,\bPta),(\otp,\bPtap)\big]=\Big(\big[ \ot,\otp\big]^\circ+\rBt\bPtap-\rBtp\bPta\,, \big[\bPta,\bPtap\big]+\Bt\bPtap-\Btp\bPta\Big). 
\end{align}
Using the semi-direct structure $\bcT\Wprod\cS$, we can further expand the bracket as
\begin{align}
    \big[(\ot,\bPta),(\otp,\bPtap)\big] &= \big[(\ot;\bt,\a{}),(\otp;\btp,\ap{})\big] \nn\\
    &=\Big(\big[ \ot,\otp\big]^\circ+\rBt\bPtap-\rBtp\bPta\,;\overline{\big[\bt,\btp\big]}+ \big(\Bt\bPtap-\Btp\bPta\big)^\gr, \label{bicrossedBracket1}\\
    &\qquad\big[\a{},\ap{}\big]^\g+\Wt\ap{}-\Wtp\a{}+ \big(\Bt\bPtap-\Btp\bPta\big)^\ym\Big). \nn
\end{align}
From the definitions \eqref{defBracketobar} and \eqref{defAction} of the various brackets and actions, we find by inspection that \eqref{bicrossedBracket1} is indeed equal to $\big[\Pta,\Ptap\big]$ as defined in \eqref{STbracket}.

\paragraph{Remark:}
At this stage, \eqref{STisomorphism} is technically valid only if the anchor map $\delta$ is identically vanishing.
Indeed, we only expressed $[\Pta,\Ptap]$, and not $\lbr\Pta,\Ptap\rbr$, in terms of the bicrossed product.
We made that choice to highlight a feature of the proof of the Jacobi identity of the $\sfST$-bracket, that we address in the next subsection. 
For completeness, we report the algebroid version of this discussion in section \ref{secEYM:BicrossedMathy}.
On top of that, all the computations of the next subsection are necessary for \ref{secEYM:BicrossedMathy} anyway.

\subsection{Jacobi identity \label{secEYM:JacobiIdentity}}

We can now come back to the proof of the Jacobi identity for the $\sfST$-bracket.
It is a well-known fact that the Jacobi identity for a bracket constructed via bicrossed (or semi-direct) product amounts to a list of properties that the actions have to satisfy (we recall the proof in what follows).
For instance, the semi-direct action $\vartriangleright$ has to be a derivative operator on the product over the $\sfS$-space, namely the bracket $[\cdot\,,\cdot]^\g$.
$\vartriangleright$ must also be a Lie algebra homomorphism.
As we just mentioned in the previous remark, at this stage we do not include the anchor map part of the brackets. 
We thus proceed as follows: we check all the properties that the actions must satisfy. 
Most of them are ``anomalous'', where the anomaly is the difference between what the property is and what the property should be in a traditional Lie algebra setting \cite{Majid_2002, michor1992knitproductsgradedlie}.
We show that these anomalies can always be expressed in terms of the action of the anchor map onto the deformation parameters, i.e. the infinitesimal variations of $C,A$ and $F$.
We can then easily compute the Jacobi identity for the algebra part of the $\sfST$-bracket, namely $\big[\Pta,\Ptap\big]$ and see how the various anomalies recombine to define a \emph{Jacobi identity anomaly}.
Next, we show that the latter is equal to the \emph{Leibniz anomaly} of the anchor onto $\big[\Pta,\Ptap\big]$. 
This fact then allows us to conclude that the Jacobi identity for the algebroid $\sfST$-bracket $\big\lbr\Pta,\Ptap\big\rbr$ holds.
Recall that the Leibniz anomaly is defined in \eqref{LeibnizRuleNotation}.

\paragraph{Properties of $\vartriangleright$:}
We can show that $\Wt$ is a derivative operator up to an anomaly (see appendix \ref{AppEYM:tauAction}):
\begin{equation} \label{LeibnizAnomalyTauAction}
    \boxed{\cA_s\big([\ap{},\app{}]^\g,\Wt\!\big)
    =-\sum_{n=1}^s (n+1)\t{n}\big[\hdap A,\app{s+1-n}\big]_\g-(\ap{}\leftrightarrow\app{})}.
\end{equation}
Recall that $\hda A=-\DYM\a0+C\a1$.
The action $\vartriangleright$ is also a homomorphism of Lie algebras up to an anomaly (see appendix \ref{AppEYM:tauHom})
\begin{equation}
    \boxed{\Big(\overline{[\bt{},\btp{}]} \vartriangleright \app{}-\big[\Wt,\Wtp\!\big]\app{} \Big)_s=(s+2)\app1\Big(\t{s+1}\dbtp C-\tp{s+1}\dbt C\Big)}, \label{HomAnomalyTauAction}
\end{equation}
where $\dbt C$ is the variation $\dt C$ restricted to the $\bsfT$-subspace, i.e.
\begin{equation}
    \dbt C=2DC\t1+3CD\t1-3C^2\t2.
\end{equation}

\paragraph{Properties of $\blacktriangleright$:}
Similarly for the black action, we get (see App.\,\ref{AppEYM:BlackTauActionLeibnizGR})
\begin{equation} \label{LeibnizAnomalyBlackTauActionGR}
    \boxed{\cA_s^\gr\Big(\big[\bPtap,\bPtapp \big], \Bt\!\Big)=\Big\{ \big((\rBt\bPtap)\blacktriangleright \bPtapp\big)^\gr_s+(s+3)\t0\tpp{s+2} \dbtp C\Big\}-\bPtap\leftrightarrow\bPtapp}
\end{equation}
and (see App.\,\ref{AppEYM:BlackTauActionLeibnizYM})
\begin{empheq}[box=\fbox]{align}
    \cA_s^\ym\Big(\big[\bPtap,\bPtapp \big], \Bt\!\Big) &=\bigg\{\big((\rBt\bPtap)\blacktriangleright \bPtapp\big)^\ym_s+(s+2)\t0\app{s+2}\dbtp C-(s+2)\ap1\tpp{s+1}\odt C \nn\\
    &\quad -\t0 \big[\hdap A,\app{s+1}\big]_\g+(s+2)\t0\tpp{s+1}\dbPtap F \label{LeibnizAnomalyBlackTauActionYM}\\
    &\quad +\sum_{n=1}^{s}(n+1)\tp{n}\big[\odt A, \app{s+1-n}\big]_\g\bigg\}-\bPtap\leftrightarrow\bPtapp, \nn
\end{empheq}
where $\odt C=-D^2\t0+N\t0$ and (using the dual EOM \eqref{DualEOMEYM})
\begin{align}
    \dPta F=\pa_u(\dPta A)&=-\DYM^2\a1+3C\DYM\a2+ 2DC\a2-3C^2\a3-[F,\a0]_\g+2\DYM F\t1 \nn\\
    &+3FD\t1+N\a1-6CF\t2+\dot F\t0.
\end{align}
$\dbPta F$ is just $\dPta F$ with $\t0=0$.
These Leibniz rule anomalies contain 2 different parts.
The term $(\rBt\bPtap)\blacktriangleright \bPtapp$ mixing the black action and co-reaction is a classic feature of a bicrossed product \cite{michor1992knitproductsgradedlie}.
The other part of the anomalies involves the infinitesimal variation of the asymptotic data $C,A$ and $F$.
Concerning the homomorphism anomaly, we show in appendix \ref{AppEYM:BlackTauHom} that
\begin{equation} \label{HomAnomalyBlackTauActionYM}
    \boxed{\Big(\big[\ot,\otp\big]^\circ \blacktriangleright\bPtapp-\big[\Bt,\Btp\!\big]\bPtapp\Big)^\ym_s=(s+2)\app{s+2}\Big(\tp0\odt C-\t0\odtp C\Big)}.
\end{equation}
Similarly, the projection onto the $\bsfT$-subspace is
\begin{equation} \label{HomAnomalyBlackTauActionGR}
    \boxed{\Big(\big[\ot,\otp\big]^\circ \blacktriangleright\bPtapp-\big[\Bt,\Btp\!\big]\bPtapp\Big)^\gr_s=(s+3)\tpp{s+2}\Big(\tp0\odt C-\t0\odtp C\Big)}.
\end{equation}

\paragraph{Properties of $\blacktriangleleft$:}
The Leibniz rule anomaly at degree $-1$ is (proof in App.\,\ref{AppEYM:BackBlackTauLeibniz})
\bs \label{LeibnizAnomalyrBlackTauAction}
\begin{equation}
    \boxed{\cA_{-1}\Big(\big[\otp,\otpp \big]^\circ, \blacktriangleleft\bPta \Big)=\Big\{ \big(\rBtp(\Btpp\bPta)\big)_{-1}+2\t1\tpp0 \odtp C\Big\}-\otp\leftrightarrow\otpp},
\end{equation}
while the degree 0 is anomaly free,
\begin{equation}
     \boxed{\cA_0\Big(\big[\otp,\otpp \big]^\circ, \blacktriangleleft\bPta \Big)=0}.
\end{equation}
\es
Finally, the homomorphism anomaly takes the form (proof in App.\,\ref{AppEYM:BackBlackTauHom})
\begin{equation} \label{HomAnomalyrBlackTauAction}
    \boxed{\Big(\rBtpp[\bPta,\bPtap]-\otpp\big[\!\blacktriangleleft\bPta, \blacktriangleleft\bPtap\big] \Big)_{s}= (s+3)\tpp0\big(\tp{s+2}\dbt C-\t{s+2}\dbtp C\big)},\qquad s=-1,0.
\end{equation}

\paragraph{Jacobi identity anomaly:}
As we had proven in \cite{Cresto:2024fhd,Cresto:2024mne}, the GR projection of the Jacobi anomaly of the bracket $[\Pta,\Ptap]$ is given by
\begin{equation} \label{JacobiEYMCbracket}
    \big[\Pta,[\Ptap,\Ptapp]\big]^\gr_s=\big[\tau,[\tau',\tau'']^C\big]^C_s \cyc-\dt C\paren{\tp{},\tpp{}}^\gr_s.
\end{equation}
Besides, the Jacobi anomaly of the YM-bracket takes the form
\begin{empheq}[box=\fbox]{align}
    \big[\Pta,[\Ptap,\Ptapp]\big]^\ym_s &\cyc \sum_{n=0}^s(n+1)\Big(\tp{n}\big[\dPta A,\app{s+1-n}\big]_\g-\tpp{n}\big[\dPta A,\ap{s+1-n}\big]_\g\Big) \label{JacobiEYMYMbracket}\\
    &-(s+2)\dt C\Big(\tp0\app{s+2}-\tpp0\ap{s+2}+ \ap1\tpp{s+1}- \app1\tp{s+1}\Big)-\dPta F\paren{\tp{}, \tpp{}}^\ym_s.  \nn
\end{empheq}
We present the proof of this result in appendix \ref{AppEYM:Jacobi} (which at the same time serves as a reminder of the equivalence between demanding the Jacobi identity and requiring the actions $\vartriangleright,\blacktriangleright, \blacktriangleleft$ to satisfy the derivation and homomorphism properties).
\medskip

Next we prove that the Jacobi identity for the $\sfST$-bracket holds if and only if $\delta$ is a morphism, namely a realization of the $\sfST$-bracket onto the phase space $\PS$. 
This is the usual result when one deals with an algebroid bracket of the type ``$\lbr\cdot\,,\cdot\rbr=[\cdot\,,\cdot]+\delta_\cdot\cdot-\,\delta_\cdot\cdot$''.
We had an illustration of that fact in \cite{Cresto:2025bfo} when treating the pure YM $\sfS$-bracket.
The subtlety here (and the reason why the proof is non-trivial) is that the $C$- and YM-brackets are not Lie brackets \emph{per se}, in the sense that they have the aforementioned \emph{Jacobi identity anomaly}.
Besides the anchor map $\delta$ is not a derivative operator onto the YM-bracket (nor the $C$-bracket) as it would be in the usual symmetry algebroid setup.
Instead it satisfies a \emph{Leibniz rule anomaly} precisely equal to the Jacobi identity anomaly, so that the two compensate one another.
The essence of the demonstration is thus to establish this cancellation.
We encountered the same mechanism when treating the pure GR case and we refer the reader to \cite{Cresto:2024mne} 
and to the upcoming proof for details.

\begin{tcolorbox}[colback=beige, colframe=argile]
\textbf{Lemma [Jacobi identity]} \label{LemmaJacEYM}\\
The $\sfST$-bracket $\lbr\cdot\,,\cdot\rbr$ satisfies Jacobi if and only if  $\delta$ is an algebroid action:
\begin{equation}
    \big\lbr\Pta,\lbr\Ptap,\Ptapp\rbr\big\rbr\cyc 0 \qquad\Leftrightarrow \qquad \big[\dPta,\dPtap\big]+\delta_{\lbr\Pta,\Ptap \rbr}=0. \label{JacobiMorphismEYM}
\end{equation}
\end{tcolorbox}

\paragraph{Proof:}
For any bracket $\lbr\Pta,\Ptap\rbr=[\Pta,\Ptap]+\dPtap\Pta-\dPta\Ptap$, we have that
\begin{align}
    \big\lbr\Pta,\lbr\Ptap,\Ptapp\rbr\big\rbr &=\big[\Pta,\lbr\Ptap,\Ptapp\rbr\big]+ \delta_{\lbr\Ptap,\Ptapp\rbr}\Pta-\dPta\lbr\Ptap,\Ptapp\rbr \nn\\
    &=\big[\Pta,[\Ptap,\Ptapp]\big]+ \big[\Pta,\dPtapp\Ptap\big]-\big[\Pta,\dPtap\Ptapp\big]+ \delta_{\lbr\Ptap,\Ptapp\rbr}\Pta-\dPta\big[\Ptap,\Ptapp\big]-\dPta\dPtapp\Ptap+\dPta\dPtap\Ptapp \nn\\
    &\cyc \big[\Pta,[\Ptap,\Ptapp]\big]-\cA\big([\Ptap,\Ptapp],\dPta\big)+ \delta_{\lbr\Pta,\Ptap\rbr}\Ptapp+\big[\dPta, \dPtap\big]\Ptapp. \label{computationJacobiEYM}
\end{align}
Hence if the Jacobi identity anomaly of the bracket $[\cdot\,,\cdot]$ equates the Leibniz rule anomaly of the anchor onto this bracket, we get the result \eqref{JacobiMorphismEYM}.
In our specific case where $[\Pta,\Ptap]$ reduces to the $C$- and YM-brackets \eqref{TbracketEYM} and \eqref{YMbracketCFdef}, it is clear that 
\begin{equation}\label{LeibnizAnchorEYM}
\cA^\gr\big([\Ptap,\Ptapp],\dPta\big)= \eqref{JacobiEYMCbracket}\qquad\textrm{and}\qquad \cA^\ym\big([\Ptap,\Ptapp],\dPta\big)= \eqref{JacobiEYMYMbracket}.
\end{equation}

The combination of the 3 lemmas \hyperref[LemmaClosureEYM]{[$\sfST$-bracket closure]}, \hyperref[lemmaAlgebroidAction]{[Algebroid action]} and \hyperref[LemmaJacEYM]{[Jacobi identity]} proves the theorem \hyperref[theoremSTAlgebroid]{[$\cST$-algebroid]}.

\subsection{Bicrossed product of Lie algebroids \label{secEYM:BicrossedMathy}}

In this subsection, we come back to the generalization of the bicrossed product to Lie algebroids.
In \eqref{bicrossedBracket} we emphasized the bicrossed product structure of the single staff bracket $[\Pta,\Ptap]$, to which we added by hand the contribution $\dPtap\Pta-\dPta\Ptap$ of the anchor map in order to form the double staff bracket $\lbr\Pta,\Ptap\rbr$. 
Doing so, the Jacobi identity for the $\sfST$-bracket split into 3 easily identifiable contributions, cf.\,\eqref{computationJacobiEYM}.
However, the equality of the Jacobi identity anomaly \eqref{JacobiEYMCbracket}-\eqref{JacobiEYMYMbracket} with the Leibniz rule anomaly $\cA\big(\lbr\Ptap,\Ptapp\rbr,\dPta\big)$ is revealing of the fact that one should be able to define algebroid actions $\dotvartriangleright, \dotblacktriangleright$ and $\dotblacktriangleleft$, denoted with a dot, which are no longer anomalous.

\begin{tcolorbox}[colback=beige, colframe=argile]
\textbf{Definition [$\dotvartriangleright,\dotblacktriangleright, \dotblacktriangleleft$ Actions]}\\
We define\footnote{For simplicity we do not add dots on the products appearing in \eqref{defActionDot} or \eqref{STisomorphismBis}.}
\begin{align} \label{defActionDot}
    \dotvartriangleright\,&:\bcT\to\textrm{Der}(\cS)&   \dotblacktriangleright\,&: \ocT\to\textrm{End}\big(\bcT\Wprod\cS\big) &  \dotblacktriangleleft\,&:\bcT\Wprod\cS\to\textrm{End}(\ocT) \\
    &:\bt\mapsto\dWt, &  &:\ot\mapsto\dBt, &  &: \bPta\mapsto\,\dotblacktriangleleft\,\bPta. \nn
\end{align}
such that
\bs \label{defActionDot2}
\begin{align} 
    \dWt\ap{} &:=\Wt\ap{}-\dbt\ap{}, \\
    \dBt\bPtap &:=\Bt\bPtap-\odt\bPtap, \\
    \drBt\bPtap &:=\rBt\bPtap+\dbtp\ot.
\end{align}
\es
\end{tcolorbox}

\begin{tcolorbox}[colback=beige, colframe=argile, breakable]
\textbf{Lemma [Properties of $\dotvartriangleright,\dotblacktriangleright, \dotblacktriangleleft$]}\\
The dotted white action is a semi-direct action and thus satisfies the following derivative and homomorphism properties:
\bs \label{dotWhiteProperties}
\begin{align}
    \dWt\big\lbr\ap{},\app{}\big\rbr^\g &=\big\lbr\dWt\ap{},\app{}\big\rbr^\g+\big\lbr\ap{}, \dWt\app{}\big\rbr^\g,\\
    \overline{\big\lbr\bt,\btp\big\rbr} \mkern5mu\dotvartriangleright\mkern5mu\app{} &=\big[\dWt,\dWtp\!\big]\app{}.
\end{align}
\es
The dotted black action and back-reaction are bicrossed product actions and thus satisfy the following derivative and homomorphism properties characteristic of bicrossed products \cite{michor1992knitproductsgradedlie} (where we use the notation \eqref{LeibnizRuleNotation}):
\bs \label{dotBlackProperties}
\begin{align}
    \cA\Big(\big\lbr\bPtap,\bPtapp \big\rbr, \dBt\!\Big) &=(\drBt\bPtap)\,\dotblacktriangleright\, \bPtapp-(\drBt\bPtapp)\,\dotblacktriangleright\, \bPtap, \\
    \big\lbr\ot,\otp\big\rbr^\circ \,\dotblacktriangleright\,\bPtapp &=\big[\dBt,\dBtp\!\big]\bPtapp, \\
    \cA\Big(\big\lbr\otp,\otpp \big\rbr^\circ, \dotblacktriangleleft\, \bPta\Big)  &=\drBtp\big(\dBtpp\bPta\big)-\drBtpp\big(\dBtp\bPta\big), \\
    \drBtpp\big\lbr\bPta,\bPta'\big\rbr &=\otpp\big[ \dotblacktriangleleft\,\bPta\,,\dotblacktriangleleft\,\bPta'\,\big].
\end{align}
\es
\end{tcolorbox}

\ni We give the proofs for the actions $\dotvartriangleright$ and $\dotblacktriangleleft$ in App.\,\ref{AppEYM:DotWhiteLeibniz} and \ref{AppEYM:DottauHom}.
The other demonstrations go along the same lines.
We can then state again the theorem \hyperref[STisomorphism]{[Bicrossed product structure]}, this time in a full algebroid fashion.

\begin{tcolorbox}[colback=beige, colframe=argile]
\textbf{Theorem [Bicrossed product structure]}\\
\begin{equation} \label{STisomorphismBis}
\cST\simeq\ocT\Bprod\big(\bcT\Wprod\cS\big).
\end{equation}
Hence, using the definitions \eqref{defBracketobar} and \eqref{defActionDot2},
\begin{align}
    \big\lbr\Pta,\Ptap\big\rbr &=\Big(\big\lbr \ot,\otp\big\rbr^\circ+\drBt\bPtap-\drBtp\bPta\,;\overline{\big\lbr\bt,\btp\big\rbr}+ \big(\dBt\bPtap-\dBtp\bPta\big)^\gr, \label{bicrossedBracketBis}\\
    &\qquad\big\lbr\a{},\ap{}\big\rbr^\g+\dWt\ap{}-\dWtp\a{}+ \big(\dBt\bPtap-\dBtp\bPta\big)^\ym\Big).\nn
\end{align}
\end{tcolorbox}

\section{Covariant wedge algebras \label{secEYM:CovWedge}}

From the analysis done in the previous section we see that the Jacobi and Leibniz anomalies  associated with the bracket $[\cdot\,,\cdot]=\big([\cdot\,,\cdot]^C,[\cdot\,,\cdot]^\ym\big)$ defined in (\ref{TbracketEYM}-\ref{YMbracketCFdef})
vanish when the symmetry parameters are such that $\dPta \Ptap=0$. 
This imposes constraints on $\Pta$ which define a wedge algebra.
In other words, picking $\Pta$ in the kernel of the anchor map in the $\cST$-algebroid is the definition of the associated algebra $\Wcal_{CA}(\scri)$.

\begin{tcolorbox}[colback=beige, colframe=argile]
\textbf{Definition/Corollary 
[Covariant wedge on $\scri$]}\\
The covariant wedge space $\W_{CA}(\scri)$ and algebra $\Wcal_{CA}(\scri)$ are given by
\bs \label{CovWedgeScri}
\begin{align}
    \W_{CA}(\scri) &:=\Big\{\,\Pta\in\sfST~ \big|~\dt h=0=\dPta A\,\Big\}, \\
    \Wcal_{CA}(\scri) &:=\big(\W_{CA}(\scri),[\cdot\,,\cdot]\big),
\end{align}
\es
where the $\sfST$-bracket \eqref{STbracket} reduces to (\ref{CbracketCompact}-\ref{YMbracketCompact}). $\Wcal_{CA}(\scri) $ is a \emph{Lie} algebra.
\end{tcolorbox}

\paragraph{Remark:}
The fact that $\Wcal_{CA}(\scri)$ is a Lie algebra follows from the fact that $\cST$ is a Lie algebroid.
Indeed, the closure of the bracket $[\cdot\,,\cdot]$, namely that $[\Pta,\Ptap]\in \W_{CA}(\scri)$ when $\Pta,\Ptap \in \W_{CA}(\scri)$, is obvious from \eqref{LeibnizAnchorEYM}.
Moreover the dual EOM are respected by construction.
The important point to emphasize is that since $\dPta(C,A)=0$ we can view the radiative data $(C,A)$ as \emph{background data} which are not varied and label the wedge algebra.
On top of that, we show in (\ref{CbracketCompact}-\ref{YMbracketCompact}) that in the wedge, the Lie algebra brackets $[\cdot\,,\cdot]^C$ and $[\cdot\,,\cdot]^\ym$ can be recast in a form which makes manifest that their structure constants are indeed \emph{constant}, i.e. independent of the radiative data.
The dependence in these background fields only remains as a constraint among the symmetry generators.\footnote{For instance that $D\t{-1}=C\t0$, but also that $D^2\t0=N\t0+2DC\t1+3CD\t1-3C^2\t2$, which follows from the time derivative of the previous condition, and so on so forth. See \cite{Cresto:2024mne} for complimentary remarks.}
\medskip

From \eqref{Leibnizgr} and a similar condition for $[\cdot\,,\cdot]^\ym$, one establishes that $\Dcal$ is a covariant derivative in the covariant wedge and thus satisfies the Leibniz rule.
In essence the corollary simply follows from the well-known fact that an algebroid reduces to an algebra when the anchor map is trivial.  
The Leibniz property is a bonus result which is a consequence of the fact that the algebroid satisfies the dual EOM as constraints.
To see it, just notice that the dual EOM can also be written more abstractly as the vanishing of the adjoint action $\lbr\Ham,\Pta\rbr=0$ for the Hamiltonian $\Ham$,\footnote{$\lbr\Ham,\Pta\rbr=\Dcal\Pta-\delta_\Ham\Pta=\Dcal\Pta-\pa_u\Pta$. 
\eqref{HamConstraintVSLeibniz} assumes $\Ham\in\Wcal_{CA}$, i.e. $\delta_\Ham\Pta=0$, so that it holds in the non-radiative case where $\Dcal=\big(\DGR,\DEYM\big)$, cf.\,\eqref{defDEYM}.} cf.\,\eqref{Qham}.
Hence,
\begin{equation} \label{HamConstraintVSLeibniz}
    \big\lbr\Ham,\lbr\Pta,\Ptap\rbr\big\rbr= \big\lbr\lbr\Ham,\Pta\rbr,\Ptap\big\rbr+ \big\lbr\Pta,\lbr\Ham,\Ptap\rbr\big\rbr \quad\ensurestackMath{\stackon[4pt]{\Longrightarrow}{\scriptstyle{\Wcal_{CA}}}}\quad \Dcal[\Pta,\Ptap]=\big[\Dcal\Pta,\Ptap\big]+ \big[\Pta,\Dcal\Ptap\big].
\end{equation}

We are especially interested in the projection of $\Wcal_{CA}(\scri)$ onto a non-radiative cut $S$ of $\scri$.
For simplicity, let us consider the cut $u=0$.
The celestial symmetry parameters $T$ and $\Aa{}$ are then defined as the initial condition of the dual EOM \eqref{DualEOMEYM}, i.e. $\Pta\big|_{u=0}=(T,\Aa{})$.
On $\scri$, $\dPta(h,A)=0$ trivially imposes $\pa_u^n\big(\dPta(h,A)\big)\big|_{u=0}=0,\,\forall \,n\in\N$.
At a cut all conditions labeled by $n$ are independent.
In order to write them we need to encode the radiative data in terms of a collection of fields $\Cc{n} \in \Ccel{n+1,2}$ and $\Ac{n} \in \Ccel{n+1,1}$. 
These background fields are defined as
\be
 \Cc{n}\equiv(\pa_u^n C)\big|_{u=0}\qquad \textrm{and}\qquad\Ac{n}\equiv(\pa_u^n A)\big|_{u=0}.
\ee 
Besides we denote the collections as $\bfCc=(\Cc0,\Cc1,\Cc2,\ldots)$ and $\bfAc=(\Ac0,\Ac1,\Ac2,\ldots)$. 
Evaluating $\dPta(h,A)=0$ and $\dPta(C,F)=0$ at $u=0$, we get\footnote{It is understood that $\DYMc\Aa{s}=D\Aa{s}+\big[\Ac0,\Aa{s}\big]_\g$ and  $(\DGR T)_s=D\T{s+1}-(s+3)\Cc0\T{s+2}$.}
\bs
\begin{align}
    \DYMc\Aa0 &=\Cc0\Aa1+\Ac1\T0, \\
    \DYMc^2\Aa1 &=3\Cc0D_{\beta_0}\Aa2+ 2D\Cc0\Aa2-3\Cc0^2\Aa3-[\Ac1,\Aa0]_\g+2D_{\beta_0}\Ac1\T1 \\
    &+3\Ac1 D\T1+\Cc1\Aa1-6\Cc0\Ac1\T2+\Ac2\T0, \nn\\
    \DYMc^3\Aa2 &= \cdots, \nn
\end{align}
together with
\begin{align}
    D\T{-1} &=\Cc0\T0, \\
    D^2\T0 &=\Cc1\T0+2D\Cc0\T1+3\Cc0 D\T1-3\Cc0^2\T2, \\
    D^3\T1 &=\Cc2\T0+2D\Cc1\T1+4\Cc1 D\T1+8D\Cc0 D\T2+6\Cc0 D^2\T2+3D^2\Cc0\T2+\cdots,
\end{align}
\es
where $\cdots$ contains higher order terms in $\bfCc$.
This set of relations thus takes the form of a constraint on the derivatives $\DYMc^{s+1}\Aa{s}$ and $D^{n+2}\T{n}$ for $s\geq 0$ and $n\geq -1$.
If we focus on a non-radiative (NR) cut, for which $\Cc{n}=0=\Ac{n}, \forall\,n\geq 1$, then the constraint takes a very compact form.
Indeed, simply notice that for $n\in\N$,
\bs \label{CovWedgeEYMdef}
\begin{align}
    \big(\pa_u^n(\dPta C)\big)\big|_{u=0} &\equiv\dT \Cc{n} \overset{\textsc{nr}}{=}\big(\DGR{}^{n+2}T \big)_{-2}=0, \\
    \big(\pa_u^n(\dPta A)\big)\big|_{u=0}&\equiv \delta_{(T,\Aa{})}\Ac{n} \overset{\textsc{nr}}{=}\big(\DEYM{}^{n+1}\Aa{} \big)_{-1}=0,
\end{align}
\es
where $\DEYM\a{}$ is the non-radiative part of $\DTOT\Pta$, namely\footnote{In \eqref{CovWedgeEYMdef}, we use its projection onto $S$.}
\begin{equation} \label{defDEYM}
    \big(\DEYM\a{}\big)_s:= \big(\DTOT\Pta \big)_s\big|_{F=0}=\DYM\a{s+1}-(s+2)C\a{s+2}.
\end{equation}
Therefore, in order to preserve the non-radiative conditions $\Cc{n}=0=\Ac{n}, n\geq 1$ of the cut, together with the initial values of the shear $\Cc0$ and of the gauge potential $\Ac0$, we must impose the RHS of \eqref{CovWedgeEYMdef} to be 0.

This discussion then motivates the definition of the covariant wedge $\Wcal_{\bfCc\bfAc}(S)$ onto the sphere, or more precisely a non-radiative section of $\scri$ for which $\bfCc=(\Cc0,0,0,\ldots)$ and $\bfAc=(\Ac0,0,0,\ldots)$.
In the following $\sfSTK(S)$ is the same as $\sfSTK$, cf.\,\eqref{defSTKspace}, but for celestial fields.

\begin{tcolorbox}[colback=beige, colframe=argile]
\textbf{Definition [Covariant wedge space \& algebra on $S$]}\\
The covariant wedge space $\W_{\bfCc\bfAc}(S)$ at a non-radiative cut $S$ of $\scri$ characterized by the condition $\Cc{n}=0=\Ac{n}, \forall\,n\geqslant 1$ is defined as
\begin{equation} \label{CovWedgeEYMdefBis}
    \W_{\bfCc\bfAc}(S):=\Big\{\,(T,\Aa{})\in\sfSTK(S)~ \big|~\big(\DGR{}^{s+1}T \big)_{-2}=0=\big(\DEYM{}^{s+1}\Aa{} \big)_{-1},\,\forall\,s\in\N\,\Big\}.
\end{equation}
$\Wcal_{\bfCc\bfAc}(S)\equiv\big(\W_{\bfCc\bfAc}(S),[\cdot\,,\cdot]\big)$ is the associated Lie algebra with the bracket given by (the projection on $S$) of (\ref{CbracketCompact}-\ref{YMbracketCompact}).
\end{tcolorbox}

\ni $\Wcal_{\bfCc\bfAc}(S)$ is a Lie algebra since it corresponds to a special case of the projection of $\Wcal_{CA}(\scri)$ at a cut.

\paragraph{Remark:}
We leave for future investigation a more refined study of the non-radiative structure of the covariant wedge.
Indeed, it is natural to wonder what happens if we relax the non-radiation condition $\Cc{n}=\Ac{n}=0,\,\forall\, n\geq 1$ into, for instance, $\Cc{n}=\Ac{n}=0,\,\forall\, n\geq 2$ while keeping $\Cc0,\Cc1,\Ac0$ and $\Ac1$ different than 0. 
Note that the parameters $(\Cc{n},\Ac{n})$ are the higher goldstone fields introduced in \cite{Freidel:2022skz} which allowed the construction of the discrete basis of states.

We now give a compact form for the $\sfST$-bracket when restricted to the covariant wedge.
At the same time, we point out how the 3 actions $\vartriangleright, \blacktriangleright$ and $\blacktriangleleft$ recombine into a single semi-direct action $\blackwhitetriangle$ when the symmetry parameters pertain to $\Wcal_{\bfCc\bfAc}(S)$.

\subsection{Semi-direct structure \label{secEYM:SemiDirectWedge}}

We have seen already in \eqref{anomalyTsubAlgebroid} that the gravitational $\cT$-algebroid is not a sub-algebroid of $\cST$.
However, the culprit term in \eqref{anomalyTsubAlgebroid} involves $F$.
This suggests that in the non-radiative sector, where $F=0$, a semi-direct structure should hold.
Let us show that indeed, the symmetry algebra $\Wcal_{\bfCc\bfAc}(S)$, where $S$ is non-radiative as in the previous subsection, can be recast in the form $\Wcal_{\bfCc}^\gr(S)\BWprod \Wcal_{\bfCc\bfAc}^\ym(S)$ for a certain action $\blackwhitetriangle$.
Here $\Wcal_{\bfCc}^\gr(S)$ denotes the GR covariant wedge algebra \cite{Cresto:2024fhd, Cresto:2024mne} and $\Wcal_{\bfCc\bfAc}^\ym(S)$ is the YM covariant wedge in an asymptotically flat background.\footnote{Namely characterized by $\big(\DEYM{}^{s+1}\Aa{}\big)_{-1}=0$.
We show in App.\,\ref{AppEYM:LeibnizDEYM} that $\cA\big([\Aa{},\Aap{}]^\g,\DEYM\big)=0$ if $\Aa{},\Aap{}\in\Wcal_{\bfCc\bfAc}^\ym(S)$ such that $\big(\DEYM{}^{s+1}[\Aa{},\Aap{}]^\g\big)_{-1}=0$ if $\Aa{},\Aap{}\in\Wcal_{\bfCc\bfAc}^\ym(S)$.
We proved already in \cite{Cresto:2024fhd} that $\cA\big([T,T']^{\sigma_0},\DGR\big)=0$ when $T,T'\in \Wcal^\gr_{\bfCc}(S)$.
This is also a consequence of \eqref{HamConstraintVSLeibniz}.} 
Their brackets are respectively the $C$-bracket and $[\cdot\,,\cdot]^\g$.

To avoid too many changes of notation, but also because the first part of the computation is valid on $\scri$, we keep using the Carrollian parameter $\Pta$ in the following.
The projection onto $S$ can be applied at any point.
For $\tau\in\Wcal_{C}^\gr(\scri)$, we know that $D\t{-1}=C\t0$, which allows us to recast the deformation along the shear of the $C$-bracket as a constraint on $\t{-1}$.
Therefore, for $\tau,\tau'\in\Wcal^\gr_C(\scri)$, the $C$-bracket \eqref{TbracketEYM} takes the form
\begin{equation} \label{CbracketCompact}
    \big[\tau,\tau'\big]^C_s=\sum_{n=0}^{s+2}(n+1)\big(\t{n}D\tp{s+1-n}-\tp{n}D\t{s+1-n}\big).
\end{equation}
Note that the sum now runs till $s+2$ and the term $n=s+2$ precisely reproduces the GR Dali-bracket.

The same phenomenon occurs with the YM-bracket \eqref{YMbracketCFdef} when imposing $\dPta A=0$, i.e. $\DYM\a0=C\a1+F\t0$, together with $\dt h=C\t0-D\t{-1}=0$.
Indeed, for $\Pta,\Ptap\in\Wcal_{CA}(\scri)$, we have that\footnote{We correctly recover the action of a diffeomorphism onto a gauge transformation $\a0$ studied in \cite{Barnich:2013sxa}.}
\begin{align} \label{YMbracketCompact}
   \big[\Pta,\Ptap\big]^\ym_s= \big[\a{},\ap{}\big]^\g_s +\bigg\{\sum_{n=0}^{s+1}(n+1)\Big(\t{n}\DYM\ap{s+1-n}-\ap{n+1}D\t{s-n}\Big)-\Pta\leftrightarrow\Ptap\bigg\},
\end{align}
where the sum now runs till $s+1$.
The equations \eqref{CbracketCompact} and \eqref{YMbracketCompact} represent a compact way of writing the \emph{algebra part} of the $\sfST$-bracket.
It also emphasizes the fact that the brackets associated to the covariant wedge algebras have field space independent structure constants, as they should.
We had noticed that feature in \cite{Cresto:2024fhd} already.

Defining the action $\blackwhitetriangle$ as 
\begin{align}
     \blackwhitetriangle&:\Wcal_C^\gr(\scri)\to \textrm{End}\big(\Wcal^\ym_{CA}(\scri)\big) \nn\\
     &\quad \tau\mapsto\BWt, \label{BlackWhiteAction}\\
     \big(\BWt\ap{}\big)_s &:=\sum_{n=0}^{s+1}(n+1)\Big(\t{n}\DYM\ap{s+1-n}-\ap{n+1}D\t{s-n}\Big), \nn
\end{align}
we get that 
\begin{equation}
    \boxed{\big[\Pta,\Ptap\big]^\ym= [\a{},\ap{}]^\g+\BWt\ap{}-\BWtp\a{}}.
\end{equation}
The reader can check\footnote{We do not report these demonstrations since they are similar to the ones presented in appendices \ref{AppEYM:tauAction}-\ref{AppEYM:tauHom}.} that in the wedge (supplemented by the non-radiative condition $F\equiv 0$), the action $\blackwhitetriangle$ is a derivative and a homomorphism of Lie algebras, i.e.
\bs
\begin{align}
    \cA_s\big([\ap{},\app{}]^\g,\BWt\big) &=\t0\big[\ap1\dt h-\hdap A,\app{s+1}\big]_\g-\ap{}\leftrightarrow\app{} \overset{\Wcal_{\bfCc\bfAc}}{=}0, \\
    \Big([\t{},\tp{}]^C \blackwhitetriangle \app{}-\big[\BWt,\BWtp\!\big]\app{} \Big)_s\! &=\Big\{ (s+3)\app{s+2}D\big(\tp0\dt h\big)+(s+3)\t0\tp{s+2}\DYM\big(\hdapp A\big) \\
    &\hspace{-3.3cm}\!+\dt h\big(\tp0\DYM\app{s+3}-(s+3)\tp{s+2}\DYM\app1-\app1 D\tp{s+2}\big)-(s+3)\app1\tp{s+2}D\big(\dt h\big) \nn\\
    &\hspace{-3.3cm}\!+\hdapp A \big(\t0D\tp{s+2}+(s+3)\tp{s+2}D\t0\big)\Big\}-\t{}\leftrightarrow\tp{}\overset{ \Wcal_{\bfCc\bfAc}}{=}0.\nn
\end{align}
\es
with $\dt h$ and $\hda A$ given by \eqref{dtahEYM} and \eqref{hatdaA}.
This means in particular that the covariant wedge algebra associated to the $\cST$-algebroid at a non-radiative cut $S$ of $\scri$ is given by the semi-direct product\footnote{This structure was pointed out in \cite{Peraza:2022glu} for the restriction to $\gbms$.}
\begin{equation} \label{WedgeAlgebraEYMsemidirect}
    \boxed{\Wcal_{\bfCc\bfAc}(S)\simeq \Wcal^\gr_{\bfCc}(S)\BWprod \Wcal^\ym_{\bfCc\bfAc}(S)}.
\end{equation}

\subsection{Relation with the celestial bracket \label{secEYM:celestialBracket}}

Since we considered Schwartz fall-off conditions for $C$ and $A$, the deformation parameters are zero on the celestial sphere $\scri^+_-$.
From the Carrollian YM-bracket \eqref{YMbracketCFdef}, one can define its projection onto the sphere $S=\scri^+_-$, to obtain the celestial bracket which reduces to\footnote{For the upcoming argument, we just focus on the part of the bracket that mixes $T$ and $\Aa{}$.}
\begin{equation} \label{CelestialBracket}
    \big[(T,\Aa{}),(\Tp{},\Aap{})\big]^{\mathsf{cel}}_s=\bigg\{ \sum_{n=0}^s(n+1)\big(\T{n}D\Aap{s+1-n}-\Aap{n+1}D\T{s-n}\big)\bigg\} -(T,\Aa{})\leftrightarrow(\Tp{},\Aap{}),
\end{equation}
where $T$ and $\Aa{}$ are formally defined at the cut $\scri^+_-\equiv u\to-\infty$.

\paragraph{Relation with the celestial modes $w^p_m$ and $S^{q,b}_n$:}
The celestial bracket \eqref{CelestialBracket} corresponds to a symmetry \textit{algebra} bracket when $\Aa{}\in\Wcal^\ym_{\bfCc=\bfAc=0}(S)$ and $T\in\Wcal^\gr_{\bfCc=0}(S)$, which means that $D^{s+1}\Aa{s}=0$ and $D^{s+2}\T{s}=0$, see \eqref{CovWedgeEYMdefBis}.
Notice that if we work onto the twice punctured sphere $S=\C^*$, then we can expand $\Aa{s}$ and $\T{s}$ as\footnote{$\{Z_b\}$ is the set of generators of $\g$.}
\bs
\begin{align}
    \Aa{s} &=\sum_{m=0}^s\Aa{(s,m)}^b z^m Z_b,\qquad \textrm{with}\quad \Aa{(s,m)}^b=\Aa{(s,m)}^b(\bz), \\
    \T{s} &=\sum_{m=0}^{s+1}\T{(s,m)}z^m ,\qquad\quad\! \textrm{with}\quad \T{(s,m)}=\T{(s,m)}(\bz).
\end{align}
\es
Setting $s=2p-3$ for GR and $s=2p-2$ for YM,\footnote{This leads in both cases to $p=1,\frac32,2,\ldots$.} we equivalently get
\bs
\begin{align}
    \Aa{s} &=\sum_{m=1-p}^{p-1}\Aa{(2p-2,p-1+m)}^b z^{p-1+m} Z_b, \\
    \T{s} &=\sum_{m=1-p}^{p-1}\T{(2p-3,p-1+m)} z^{p-1+m}.
\end{align}
\es
Besides, notice that if we take $\Aa{}=0=T'$ and $T=\iota(\T{s})$, $\Aap{}=\iota(\Aap{s'})$,\footnote{$\iota$ is the inclusion map as in \cite{Cresto:2024fhd}.} then the only non-zero term in \eqref{CelestialBracket} reduces to
\begin{equation}
    \big[\big(\iota(\T{s}),0\big),\big(0, \iota(\Aap{s'})\big) \big]^{\mathsf{cel}}_{s+s'-1}=(s+1)\T{s}D\Aap{s'}-s'\Aap{s'}D\T{s}.
\end{equation}
In particular, the bracket of the modes 
\begin{equation}
    w^p_m=-\iota\big(\tfrac12 z^{p-1+m}\big)\qquad\textrm{and}\qquad S^{q,b}_n=-\iota\big(\tfrac12 z^{q-1+n}Z_b\big)
\end{equation}
is given by
\begin{equation}
    \big[w^p_m,S^{q,b}_n\big]^{\mathsf{cel} }_{2(p+q-2)-2}=\big(m(q-1)-n(p-1)\big)S^{p+q-2,b}_{m+n}.
\end{equation}
This is the usual $sw_{1+\infty}$ bracket introduced in \cite{Strominger:2021mtt}.

\subsection{The return of the Schouten-Nijenhuis bracket}

In \cite{Cresto:2024fhd}, we started the analysis of the higher spin symmetry bracket by pointing out the relation to the Schouten-Nijenhuis (SN) bracket for symmetric tensors \cite{Schouten, SNbracket} on the 2d manifold $S$.
In our notation, as explained in \cite{Cresto:2024fhd}, the SN-bracket reads
\begin{equation}
    [\tau,\tau']^\sn_s :=\sum_{n=0}^{s+1} n\big(\t{n}D\tp{s+1-n}-\tp{n}D\t{s+1-n}\big)=\sum_{n=0}^{s} (n+1)\big(\t{n+1}D\tp{s-n}-\tp{n+1}D\t{s-n}\big). \label{SNbracketEYM}
\end{equation}
If we define the tilde variable $\tt{s+1}\equiv\t{s}$, then (still assuming $\Pta,\Ptap\in\Wcal_{CA}(\scri)$ so that we can use \eqref{CbracketCompact} and \eqref{YMbracketCompact})
\begin{align}
    [\tau,\tau']^C_s=\sum_{n=0}^{s+2}(n+1)\big(\tt{n+1}D\ttp{s+2-n}-\ttp{n+1}D\tt{s+2-n}\big).
\end{align}
Furthermore,
\begin{align}
    \big[\Pta,\Ptap\big]^\ym_s =\big[\a{},\ap{}\big]^\g_s+\sum_{n=0}^{s+1}(n+1)\Big(\tt{n+1} \DYM\ap{s+1-n} &-\ap{n+1}D\tt{s+1-n} \\
    &-\ttp{n+1}\DYM\a{s+1-n}+\a{n+1}D\ttp{s+1-n}\Big). \nn
\end{align}
Comparing with \eqref{SNbracketEYM}, we obtain the following lemma.

\begin{tcolorbox}[colback=beige, colframe=argile] \label{lemmaSNbracket}
\textbf{Lemma [SN-bracket]}\\
In the covariant wedge algebra $\Wcal_{CA}(\scri)$, the $C$-bracket and the action $\blackwhitetriangle$ (cf.\,\eqref{BlackWhiteAction}) at degree $s$ are SN-brackets at degree $s+2$ and $s+1$ respectively, over the shifted variables $\tt{}$ and $\ttp{}$,
\bs
\begin{align}
    [\tau,\tau']^C_s &=\big[\tt{},\ttp{}\big]^\sn_{s+2}, \\
    \big(\BWt\ap{}\big)_s &=\big[\tt{},\ap{}\big]^\sn_{s+1}, \label{CFasSN}
\end{align}
\es
where $D\to\DYM$ when the argument is Lie algebra valued.
\end{tcolorbox}

\ni Since the SN-bracket is known to be a Lie bracket, this is a complementary way to confirm that $\Wcal_{CA}(\scri)$ is indeed a Lie algebra\footnote{This lemma implies that the $C$- and YM-brackets when $\dPta\equiv 0$ are Lie \textit{algebra} brackets, hence Jacobi holds. 
This however does \textit{not} imply that Jacobi is satisfied for the Lie \textit{algebroid} bracket.
See \hyperref[LemmaJacEYM]{Lemma [Jacobi identity]} for details.}---the dual EOM being preserved by the brackets is a particular case of \eqref{dualEOMGRbracket}-\eqref{dualEOMEYMbracket}.

\section{Summary}

We now recapitulate the main features of the phase space action $\dPta$.
First of all, it acts on the radiative variables as follows, cf.~\eqref{deltaPtaCA},
\bs \label{deltaPtaCAbis}
\begin{align}
    \dPta C &=N\t0-D^2\t0+2DC\t1+3CD\t1-3C^2\t2, \\
    \dPta A &=F\t0-\DYM\a0+C\a1.
\end{align}
\es
Next, as proven in \eqref{AlgebroidActionEYM}, this action is the realization of the algebroid $\sfST$-bracket \eqref{STbracket}, namely
\begin{equation}
    \big[\dPta,\dPtap\big]\cdot=-\delta_{\lbr\Pta,\Ptap\rbr}\cdot.
\end{equation}

Importantly, the time dependency of the Carrollian symmetry parameters $\Pta_s$ is determined by a set of dual EOM \eqref{DualEOMEYM} and the $\sfST$-bracket satisfies the latter as well (see the \hyperref[LemmaClosureEYM]{Lemma [$\sfST$-bracket closure]}).
Since the dual evolution equations involve the radiative data $(C,A,F)$, a major consequence is the field dependency of $\Pta$, which explains the necessity of working with an algebroid framework.
On top of that, the phase space variables appear explicitly in the $\sfST$-bracket itself.
Although the action $\dPta$ is not an algebroid action in the sense that it is not a derivative on the $\sfST$-bracket (cf.~\eqref{LeibnizAnchorEYM}), we still show that the Jacobi identity \eqref{JacobiMorphismEYM} holds for this bracket.
This comes from the remarkable property that the Leibniz rule anomaly of $\dPta$ precisely equals and compensates for the Jacobi identity anomaly of the $C$- and YM-brackets.

Besides, we showed that one can start from a kinematical $\sfK$-bracket \eqref{Kbracket}, for which the time evolution of $\Pta$ is arbitrary, and then construct the $\sfST$-bracket and its associated anchor map $\delta$ as the unique bracket and action which preserve the dual time evolution \eqref{DualEOMEYM}, cf.~\hyperref[KbracketTheorem]{Theorem [on-shell $\sfK$-bracket]}.
The phase space action is then the same as the one deduced from the conservation of the master charge, that we recalled in \eqref{deltaPtaCAbis}.

While the $\cSTK$-\textit{algebra} associated to the $\sfK$-bracket admits a semi-direct structure between its GR and YM parts---see \hyperref[KbracketTheorem]{Theorem [on-shell $\sfK$-bracket]} for details---the $\cST$-\textit{algebroid} splits into a more refined bicrossed product structure analyzed in details in section \ref{secEYM:BicrossedJacobi}, cf.~in particular \eqref{STisomorphismBis}:
\begin{equation}
\cST\simeq\ocT\Bprod\big(\bcT\Wprod\cS\big).
\end{equation}

Moreover, by restricting the anchor map to its kernel, namely
\begin{equation}
    \Big\{\Pta\in\sfST~ \big|~\dt h=0=\dPta A\Big\},
\end{equation}
the $\cST$-algebroid reduces to the covariant wedge Lie algebra $\Wcal_{CA}(\scri)$.
Of particular relevance, its projection onto a non-radiative cut $S$ of null infinity is characterized by the following constraints on the celestial parameters $(\T{},\Aa{})$:
\begin{equation}
    \big(\DGR{}^{s+1}T \big)_{-2}=0=\big(\DEYM{}^{s+1}\Aa{} \big)_{-1},\,\forall\,s\in\N.
\end{equation}
One can prove, cf.~\eqref{WedgeAlgebraEYMsemidirect}, that the EYM covariant wedge algebra is the semi-direct product of its GR sub-algebra onto the YM sub-algebra (in an asymptotically flat background), where the 3 actions $\vartriangleright, \blacktriangleright,\blacktriangleleft$ recombine to form the action $\blackwhitetriangle$.

Finally, let us mention that the algebroid symmetry acts on the dynamical phase space while the algebra symmetry obtained after imposing the wedge condition preserves the no-radiation boundary condition at $i^0$.

\section{Twistorial standpoint \label{secEYM:twistor}}

We now give one last perspective on the $\sfST$-bracket, by recasting it as a Poisson bracket in the $(q,u)$ plane, where $q$ is a spin 1 coordinate in the Newman space.
The latter is defined as the homogeneous bundle\footnote{Recall that a Carrollian field of weights $(\delta,s)$ can be matched to a homogeneous section of the bundle
$\Ocal\big(-(\delta+s),-(\delta-s)\big)$ once one introduces homogeneous coordinates onto the sphere.} $\TN=\Ocal(-1,1)\to\scri_{\C}$ over complexified $\scri$. Since $\scri_{\C}$ is a $\Ocal(1,1)$ bundle over  $\CP1$  \cite{Eastwood_Tod_1982} we have  equivalently that
$\TN=\Ocal(-1,1)\oplus \Ocal(1,1) \to\CP1$ when fibered over the Riemann sphere.
One can supplement the holomorphic coordinates $(q,u)$ of the fibers over $\CP1$ by their anti-holomorphic counterpart $(\bar q,\bar u)$.
The variables $q,\bar q$ have a remarkable geometric interpretation since they parametrize the notion of horizontality on $\scri_\C$.
In other words, they define an Ehresmann connection on $\scri_\C$.
To see this, recall that on top of the Carrollian structure $(\ell^a,q_{ab})$ on $\scri$, where $\ell=\pa_u$ is the null generator and $q_{ab}$ the degenerate metric, we need to specify a one-form $k_a$ to get a notion of connection \cite{Ashtekar:1981bq, Riello:2024uvs,Freidel:2022vjq,Freidel:2024emv}.
It satisfies $k_a\ell^a=1$ by definition.
Using a congruence of null geodesics transverse to $\scri$, labeled by a null vector $k^a$, the Ehresmann one-form can arise from a so-called null rigging structure \cite{Mars:1993mj}.
Upon a choice of reference one-form,\footnote{This means that $\TN$ is an affine bundle, since a reference one-form defines an origin for the coordinate $q$.} for instance $k_a\rd x^a=\rd u$, a generic Ehresmann connection, or equivalently the associated ruling vector, is parametrized as follows \cite{Cresto:2024mne}:
\begin{equation}
    k_{(q,\bar q)}=k+q\bm+\bar q m-q\bar q\ell.
\end{equation}
The sphere dyad normal to $k_{(q,\bar q)}$ is given by 
\begin{equation}
    m_q=m-q\ell,\qquad\bm_{\bar q}=\bm-\bar q\ell.
\end{equation}
As such, $\big(\ell,k_{(q,\bar q)},m_q,\bm_{\bar q} \big)$ forms a null tetrad.
The coordinate $q$ appeared in \cite{Adamo:2009vu, Adamo:2010ey} and recently in \cite{Kmec:2024nmu} as a natural way to parametrize the fibers of twistor space over $\CP1$.
\medskip

We can straightforwardly convert the expressions written in terms of $\tau$ and $\a{}$ as formulas involving the functions $\ft$ and $\fa$, cf. \cite{Cresto:2024mne, Cresto:2025bfo}.
For this we define the vector\footnote{The
graded vector $\Pta$ and the function $\fPta$ are two different representations of the
same abstract vector in $\sfSTK$.} 
\begin{equation}
\fPta(q,u,z,\bz)\equiv(\ft,\fa):=\left(\sum_{s=-1}^\infty\t{s}q^{s+1},\sum_{s=0}^\infty\a{s}q^{s}\right),
\end{equation}
where $q\in\Ccar{0,1}$, $\ft\in\Ccar{-1,1}$, and $\fa\in\Ccarg{0,0}$.
Besides we define the differential operators
\bs \label{defNabla}
\begin{align}
\nablaGR &:=q\pa_u-D+C\pa_q, \\
\nablaYM &:=q\pa_u-D-\ada, \\
\nablaEYM &:=\nablaYM+C\pa_q.
\end{align}
\es
Denoting also $\E_{\ft}(q)=\sum_{s=-1}^\infty\E^\gr_s(\tau)q^{s+1}$ and $\E^\ym_{\fPta}(q)=\sum_{s=0}^\infty\E^\ym_s(\Pta)q^{s}$, we deduce that
\begin{equation}
\nabla\left(\begin{array}{l}
        \ft \\
        \fa
        \end{array}\right):=
        \left( \begin{array}{lr}
        \nablaGR & 0 \\
        F\pa_q & \nablaEYM
        \end{array}\right)
        \left(\begin{array}{c}
        \ft \\
        \fa
        \end{array}\right)=
        \left(\begin{array}{c}
        \dt h+q\E_{\ft} \\
        \dPta A+q\E^\ym_{\fPta}
        \end{array}\right),
\end{equation}
which shows that $\nabla(\ft,\fa)$ is an infinitesimal gauge transformation if $(\ft,\fa)\in\sfST$.

Finally, as in the gravitational case \cite{Cresto:2024mne}, the YM-bracket corresponds to the Poisson bracket in the $(q,u)$ plane (on-shell of the dual EOM).
Indeed, after some algebra, we get that
\begin{align} \label{CFasPoisson}
    \big[\fPta,\fPta'\big]^\ym(q) &:=\sum_{s=0}^\infty \big[\Pta,\Ptap\big]^\ym_sq^s \nn\\
    &=\big[\fa,\fa'\big]^\g(q)+ \Big(\Big(\poisson{\ft,\fa'}-\pa_q\ft \E^\ym_{\fPta'}+\pa_q\fa'\E_{\ft}\Big)(q)-\fPta\leftrightarrow\fPta'\Big),
\end{align}
where
\begin{equation}
    \poisson{\ft,\fa'}:=\pa_q\ft\pa_u\fa'-\pa_u\ft\pa_q\fa'.
\end{equation}
In other words,
\begin{align}
    \poisson{\fPta,\fPta'}=\left(\sum_{s=-1}^\infty \poisson{\tau,\tau'}_s q^{s+1},\sum_{s=0}^\infty \poisson{\Pta,\Ptap}^\ym_s q^s\right),
\end{align}
where we used the kinematical $\sfK$-brackets \eqref{CbracketScridef}-\eqref{YMbracketScridef}.
The Poisson bracket on the LHS (up to the pure $\g$ part $\big[\fa,\fa'\big]^\g$) is the canonical Poisson bracket on the twistor fibers, expressed using the coordinates $(q,u)$.
Indeed, taking $\bar\mu^\alpha = u n^\alpha + q \lambda^\alpha$ where $\lambda^\alpha$ parametrizes holomorphic homogeneous coordinates on $\CP1$ and $n^\alpha$ is a spinor such that $\epsilon_{\alpha \beta}= \lambda_\alpha n_\beta-\lambda_\beta n_\alpha$, we see that
\begin{equation}
    \{\cdot\,,\cdot\}=\epsilon^{\alpha\beta} \frac{\pa\,\cdot }{\pa \bar\mu^\alpha} \frac{\pa\, \cdot}{\pa \bar\mu^\beta}=\pa_q\cdot\pa_u\cdot-\,\pa_u\cdot\pa_q\cdot.
\end{equation}

Notice finally that if we define the potential $\hat h \in \Ccar{0,2}$ which depends on $q\in \Ccar{0,1}$ as
\begin{equation}
    \hat h(q,u,z,\bz):=-\frac{q^2}{2}+h(u,z,\bz),
\end{equation}
then we can understand the covariant derivatives as 
\begin{equation}
-\big(\nabla\fPta\big)^\gr=-\nablaGR\ft=D\ft+\poisson{\hat h,\ft},\qquad 
    -\big(\nabla\fPta\big)^\ym=\DYM\fa +\poisson{\hat h,\fa}+\poisson{A,\ft},
\end{equation}
while $-\nablaEYM \fa=\DYM\fa+\poisson{\hat h,\fa}$.
These operators thus represent a deformation by the asymptotic data of the original complex structure $D$ (or $\DYM$) \cite{Adamo:2021lrv, Kmec:2024nmu}.

\section*{Acknowledgments}
\addcontentsline{toc}{section}{Acknowledgments}

NC would like to thank Romain Ruzziconi, Lionel Mason and Adam Kmec, together with Daniele Pranzetti, Shreyansh Agrawal, Laura Donnay and Panagiotis Charalambous, for discussions during his visit respectively at the Mathematical Institute in Oxford and at SISSA in Trieste.
NC thanks as well Marc Geiller together with Tom Wetzstein and Laurent Baulieu for the fruitful interactions while visiting ENS Lyon and LPTHE Paris respectively.
The authors are grateful to Atul Sharma and Roland Bittleston for enlightening discussions about twistor space.
Research at Perimeter Institute is supported by the Government of Canada
through the Department of Innovation, Science and Economic Development and by the
Province of Ontario through the Ministry of Colleges and Universities. This work was
supported by the Simons Collaboration on Celestial Holography.

\newpage
\appendix

\section{Closure of the $\sfST$-bracket \label{App:dualEOMEYM}}

In this appendix, we prove that
\begin{align} \label{force1}
    \pa_u\big\lbr\Pta,\Ptap\big\rbr^\ym =\DTOT\big\lbr\Pta,\Ptap\big\rbr.
\end{align}
For this, we shall take the time derivative of the algebroid YM-bracket.
This is a brute force computation, so we simply expand most of the terms to then reconstruct the RHS of \eqref{force1}, i.e.
\begin{align}
    \DTOT\big\lbr\Pta,\Ptap\big\rbr
    =\DTOT\big[\Pta,\Ptap \big] +\DTOT\big(\dPtap\Pta-\dPta\Ptap\big),
\end{align}
where
\begin{align}
    \Big(\DTOT\big[\Pta,\Ptap\big]\Big)_s &=\DYM\big[ \Pta,\Ptap\big]^\ym_{s+1} -(s+2)C\big[ \Pta,\Ptap\big]^\ym_{s+2}-(s+2)F[\tau,\tau']^C_{s+1}.
\end{align}
Let us proceed. 
In the following, for shortness, it is understood that equalities are valid upon anti-symmetrization of the RHS. 
Using the form \eqref{YMbracketCFdef} of the YM-bracket, we get that
\begin{align}
    &\pa_u \bigg\{\bigg\{ \sum_{n=0}^s(n+1) \Big(\t{n}\DYM\ap{s+1-n}+\a{n+1}D\tp{s-n}\Big)-(s+2)C \big(\t0\ap{s+2}+\a1\tp{s+1}\big)\bigg\}-\Pta\leftrightarrow\Ptap\bigg\}= \nn\\
    &=\sum_{n=0}^s(n+1)\Big\{\big(\DGR\tau\big)_n\DYM\ap{s+1-n}+\t{n}\DYM\big(\DTOT\Ptap\big)_{s+1-n}+\t{n}\big[F,\ap{s+1-n}\big]_\g \nn\\
    &\qquad\qquad\qquad+\big(\DTOT\Pta \big)_{n+1} D\tp{s-n}+\a{n+1}D\big(\DGR \tp{}\big)_{s-n}\Big\} -(s+2)N\big(\t0 \ap{s+2}+\a1\tp{s+1}\big)\nn\\
    &-(s+2)C\Big\{\big(\DGR\tau\big)_0\ap{s+2} +\t0\big(\DTOT\Ptap\big)_{s+2}+\big(\DTOT \Pta\big)_1\tp{s+1}+\a1\big(\DGR\tp{}\big)_{s+1}\Big\} \nn\\
    &=\sum_{n=0}^s(n+1)\Big\{D\t{n+1}\DYM \ap{s+1-n}+\t{n}\DYM^2\ap{s+2-n} +\DYM\a{n+2}D\tp{s-n}+\a{n+1}D^2\tp{s+1-n} \nn\\
    &\qquad\qquad\qquad -(n+3)C\t{n+2}\DYM\ap{s+1-n}-(s+3-n)DC\t{n}\ap{s+3-n}-(s+3-n)C\t{n}\DYM\ap{s+3-n} \nn\\
    &\qquad\qquad\qquad -(n+3)C\a{n+3} D\tp{s-n}-(s+3-n)DC\a{n+1}\tp{s+2-n}-(s+3-n)C \a{n+1}D\tp{s+2-n} \nn\\
    &\qquad\qquad\qquad -(s+3-n)\t{n} \DYM\big(F\tp{s+2-n}\big) +\t{n}\big[F,\ap{s+1-n}\big]_\g -(n+3)F\t{n+2}D\tp{s-n}\Big\} \nn\\
    &-(s+2)C\Big\{\big(\DGR\tau\big)_0 \ap{s+2}+\t0\DYM\ap{s+3}-(s+4)C\t0\ap{s+4}-(s+4)F\t0\tp{s+3} \nn\\
    &\qquad\qquad\qquad +\DYM\a2\tp{s+1}-3C\a3\tp{s+1}-3F\t2\tp{s+1}+\a1D\tp{s+2}-(s+4)C\a1\tp{s+3}\Big\} \nn\\
    &-(s+2)N\big(\t0\ap{s+2}+\a1\tp{s+1}\big) \nn\\
    &=\DYM\left(\sum_{n=0}^{s+1}(n+1)\t{n}\DYM\ap{s+2-n}\right)-\sum_{n=0}^{s+1}(n+1)\underline{ D\t{n}\DYM\ap{s+2-n}}-(s+2)\dashuline{\t{s+1}\DYM^2\ap1} \nn\\
    &+\DYM\left(\sum_{n=0}^{s+1}(n+1)\a{n+1}D\tp{s+1-n}\right)-\sum_{n=0}^{s+1}(n+1)\underline{ \DYM\a{n+1}D\tp{s+1-n}}-(s+2)\uwave{\a{s+2}D^2\tp0} \nn\\
    &+\sum_{n=0}^s(n+1)\underline{ D\t{n+1}\DYM\ap{s+1-n}}+\sum_{n=0}^s(n+1)\underline{\DYM\a{n+2}D\tp{s-n}} \nn\\
    &-(s+3)\DYM\big(C\t0\ap{s+3}\big) +\aunderbrace[l1r]{C\t0\DYM\ap{s+3}}+(s+3)\aunderbrace[l1r]{DC\t0\ap{s+3}}+(s+3)\aunderbrace[l1r]{CD\t0\ap{s+3}}\nn\\
    &-(s+3)\DYM\big(C\a1\tp{s+2}\big) +\aunderbrace[l1r]{C\a1D\tp{s+2}} +(s+3)\aunderbrace[l1r]{C\tp{s+2}\DYM\a1}+(s+3)\aunderbrace[l1r]{DC\a1\tp{s+2}} \nn\\
    &-(s+3)\DYM\big(F\t0\tp{s+2}\big)+(s+3) \aunderbrace[l3*{4}{01}03r]{F\t0 D\tp{s+2}}+(s+3)\aunderbrace[l3*{5}{01}03r]{\DYM (F\t0)\tp{s+2}} \nn\\
    &-(s+2)C\sum_{n=0}^{s+2}(n+1)\t{n}\DYM\ap{s+3-n}+(s+2)C\sum_{n=0}^{s+2}(n+1)\uuline{\t{n}\DYM\ap{s+3-n}} \nn\\
    &-C\sum_{n=0}^s(n+1)(s+3-n)\uuline{\t{n}\DYM\ap{s+3-n}}-C\sum_{n=0}^s(n+1)(n+3)\uuline{\t{n+2}\DYM\ap{s+1-n}} \nn\\
    &-(s+2)C\sum_{n=0}^{s+2}(n+1)\a{n+1}D\tp{s+2-n}+(s+2)C\sum_{n=0}^{s+2}(n+1) \dotuline{\a{n+1}D\tp{s+2-n}} \nn\\
    &-C\sum_{n=0}^s(n+1)(n+3)\dotuline{\a{n+3}D\tp{s-n}}-C\sum_{n=0}^s(n+1)(s+3-n)\dotuline{\a{n+1}D\tp{s+2-n}} \nn\\
    &-\sum_{n=0}^s(n+1)(s+3-n)\underbrace{DC\t{n}\ap{s+3-n}}-\sum_{n=0}^s(n+1)(s+3-n)\underbrace{DC\a{n+1}\tp{s+2-n}} \nn\\
    &+(s+2)(s+4)C^2\big(\t0\ap{s+4}+\a1 \tp{s+3}\big)+(s+2)(s+4)CF\t0\tp{s+3} \nn\\
    &-(s+2)\uwave{C\ap{s+2}\big(\DGR\t{}} \big)_0-(s+2)\uwave{N\ap{s+2}\t0} \nn\\
    &-(s+2)C\dashuline{\tp{s+1}\DYM\a2}+ 3(s+2)\dashuline{C^2\a3\tp{s+1}}+3(s+2)\dashuline{CF\t2\tp{s+1}} -(s+2)\dashuline{N\a1\tp{s+1}} \nn\\
    &+\sum_{n=0}^s(n+1)\Big\{ \t{n}\big[F,\ap{s+1-n}\big]_\g-(s+3-n)\aunderbrace[l3*{5}{01}03r]{\t{n} \DYM\big(F\tp{s+2-n}}\big) -(n+3)\aunderbrace[l3*{5}{01}03r]{F\t{n+2}D\tp{s-n}}\Big\}. \nn
\end{align}
Moreover, the time derivative of the YM Dali-bracket is simply
\begin{align}
    -\pa_u\big(F\paren{\t{},\tp{}}^\ym_s\big)
    &=-(s+2)\dashuline{\dot F\t0\tp{s+1}} -(s+2)\dashuline{F\tp{s+1}\big(\DGR\tau}\big)_0-(s+2)\aunderbrace[l3*{5}{01}03r]{F\t0 D\tp{s+2}} \nn\\
    &+(s+2)(s+4)CF\t0\tp{s+3}.
\end{align}
Next, it is not hard to check that the \underline{underlined} sums add up to 0 upon anti-symmetrization.
Moreover, the wavy underlined terms sum up to
\begin{equation}
    \uwave{~\cdots~}=-(s+2)\ap{s+2}\dt C+2(s+2)\aunderbrace[l1r]{DC\ap{s+2}\t1}+2(s+2)\aunderbrace[l1r]{C\ap{s+2}D\t1},
\end{equation}
while for the dashed underlined terms, we get
\begin{equation}
    \dashuline{~\,\cdots\,~}=-(s+2)\tp{s+1}\dPta F-(s+2)\tp{s+1}[F,\a0]_\g+2(s+2)\tp{s+1} \DYM\big(\aunderbrace[l1r]{C\a2}+\aunderbrace[l*{2}{01}0r]{F\t1}\big),
\end{equation}
where
\begin{align}
    \dPta F=\pa_u\big(\dPta A\big) &=-\DYM^2\a1+3C\DYM\a2+ 2DC\a2-3C^2\a3-[F,\a0]_\g+2\DYM F\t1 \nn\\
    &+3FD\t1+N\a1-6CF\t2+\dot F\t0.
\end{align}
Besides, the double underlined sums simplify to
\begin{equation}
    \uuline{~\cdots~}=(s+3)\aunderbrace[l1r]{C\t{s+2}\DYM\ap1}+ 2(s+2)\aunderbrace[l1r]{C\t{s+1}\DYM\ap2}-\aunderbrace[l1r]{C\t0\DYM\ap{s+3}},
\end{equation}
while the dotted ones reduce to
\begin{equation}
    \dotuline{~\cdots~}=2(s+2)\aunderbrace[l1r]{C\a{s+2}D\tp1}+(s+3)\aunderbrace[l1r]{C\a{s+3}D\tp0}-\aunderbrace[l1r]{C\a1 D\tp{s+2}}.
\end{equation}
Furthermore, the under-braced sums contribute as
\begin{equation}
    \underbrace{~\cdots~}=(s+3)DC\big(\aunderbrace[l1r]{\a{s+3}\tp0}-\aunderbrace[l1r]{\a1\tp{s+2}}\big)+2(s+2)DC\big( \aunderbrace[l1r]{\a{s+2} \tp1}-\aunderbrace[l1r]{\a2\tp{s+1}}\big).
\end{equation}
Finally, all the terms $\aunderbrace[l1r]{\,\cdots\,}$ cancel one another (upon anti-symmetrization) and the dashed under-braced contributions take the form
\begin{equation}
    \aunderbrace[l1*{2}{01}01r]{~\cdots~}=-(s+2)F\sum_{n=0}^{s+2}(n+1)\t{n}D\tp{s+2-n}.
\end{equation}
Therefore, gathering the factors that form the YM- and $C$-brackets, we obtain that
\begin{align} \label{force2}
    &\quad \,\pa_u\Big([\Pta,\Ptap]^\ym-[\a{},\ap{}]^\g\Big)_s \\
    &-\Big\{\DYM\Big([\Pta,\Ptap]^\ym-[\a{},\ap{}]^\g\Big)_{s+1} -(s+2)C\Big([\Pta,\Ptap]^\ym-[\a{},\ap{}]^\g\Big)_{s+2}-(s+2)F[\tau,\tau']^C_{s+1}\Big\} \nn\\
    &=\sum_{n=0}^s(n+1)\dashuline{\t{n} \big[F,\ap{s+1-n}}\big]_\g \!-(s+2)\dashuline{\tp{s+1}\mkern-2mu \big[F,\a0}\big]_\g\!-(s+2)\ap{s+2}\dt C\mkern-2mu-(s+2)\tp{s+1}\dPta F. \nn
\end{align}
On the other hand the time derivative of $[\a{},\ap{}]^\g$ is given by (using \eqref{LeibnizAnomalyDYM})
\begin{align} \label{force3}
    \pa_u[\a{},\ap{}]^\g_s -\DYM[\a{},\ap{}]^\g_{s+1} &=-\big[\dotuline{\DYM\a0,\ap{s+1}}\big]_\g-C\sum_{n=0}^s(n+2)\underline{ \big[\a{n+2},\ap{s-n}}\big]_\g \\
    &-\sum_{n=0}^s(n+2)\dashuline{\t{n+1} \big[F,\ap{s-n}}\big]_\g. \nn
\end{align}
The dashed underlined terms in \eqref{force2} and \eqref{force3} recombine into
\begin{equation} \label{force4}
    \dashuline{~\,\cdots\,~}=\t0 \big[\dotuline{F,\ap{s+1}}\big]_\g.
\end{equation}
Furthermore, writing the anti-symmetrization of the \underline{sum} proportional to $C$ explicitly, we notice that
\begin{align} 
    \underline{~\cdots~} &=-C\sum_{n=0}^s(n+2) \big[\a{n+2},\ap{s-n}\big]_\g+C \sum_{n=0}^s(n+2)\big[\ap{n+2},\a{s-n}\big]_\g \nn\\
    &=C\left\{-\!\sum_{n=-1}^s(n+2) \big[\a{n+2},\ap{s-n}\big]_\g+ \!\sum_{n=-1}^s(n+2)\big[\ap{n+2},\a{s-n}\big]_\g +\big[\a1,\ap{s+1}\big]_\g-\big[\ap1,\a{s+1}\big]_\g\right\} \nn\\
    &=C\left\{\sum_{n=-2}^{s-1}(s-n)\big[\ap{n+2},\a{s-n}\big]_\g+ \!\sum_{n=-1}^s(n+2)\big[\ap{n+2},\a{s-n}\big]_\g +\big[\a1,\ap{s+1}\big]_\g-\big[\ap1,\a{s+1}\big]_\g\right\} \nn\\
    &=C\left\{(s+2)\sum_{n=-2}^s \big[\ap{n+2},\a{s-n}\big]_\g +\big[\a1,\ap{s+1}\big]_\g-\big[\ap1,\a{s+1}\big]_\g\right\} \nn\\
    &=C\left\{-(s+2)[\a{},\ap{}]^\g_{s+2} +\big[\dotuline{\a1,\ap{s+1}}\big]_\g-\big[\dotuline{\ap1,\a{s+1}}\big]_\g \right\}. \label{force5}
\end{align}
The dotted underlined factors in \eqref{force3}, \eqref{force4} and \eqref{force5} simply add up to
\begin{equation}
    \dotuline{~\cdots~}=\big[\dPta A,\ap{s+1}\big]_\g.
\end{equation}
Hence
\begin{align}
    &\quad~\pa_u \big[\Pta,\Ptap\big]^\ym_s-\DYM \big[\Pta,\Ptap\big]^\ym_{s+1}+(s+2)C\big[\Pta,\Ptap\big]^\ym_{s+2}+(s+2)F[\t{},\tp{}]^C_{s+1} \nn\\
    &=\big[\dPta A,\ap{s+1}\big]_\g-(s+2)\ap{s+2}\dt C-(s+2)\tp{s+1}\dPta F. \label{force6}
\end{align}
Finally, the last contribution comes from the anchor map, for which we get
\begin{align} \label{force7}
    \pa_u\big(\dPtap\a{s}-\dPta\ap{s}\big)&=-\dPta\big(\DTOT\Ptap\big)_s \\
    &=-\big(\DTOT(\dPta\Ptap)\big)_s-\big[\dPta A,\ap{s+1}\big]_\g+(s+2)\ap{s+2}\dt C+(s+2)\tp{s+1}\dPta F. \nn
\end{align}
Summing \eqref{force6} and \eqref{force7}, we conclude that, cf. \eqref{dualEOMEYMbracket},
\begin{equation}
    \pa_u \big\lbr\Pta,\Ptap \big\rbr^\ym_s=\Big(\DTOT\big\lbr\Pta,\Ptap \big\rbr\Big)_s.
\end{equation}

\section{Proof of the algebroid action \eqref{AlgebroidActionEYM1} \label{AppEYM:AlgAct}}

Firstly, using the variation \eqref{deltaPtaCA} and the dual EOM \eqref{DualEOMEYM}, we compute
\bs
\begin{align}
    \dPta\dPtap A &=\dPta\Big(-\DYM\ap0+C\ap1+F\tp0\Big) \nn\\
    &=-\DYM\big(\dPta\ap0\big)-\big[-\DYM\a0+C\a1+F\t0,\ap0\big]_\g+C\dPta\ap1+F\dt\tp0 \nn\\
    &+\ap1\Big(N\t0-(\DGR{}^2\t{})_{-2}\Big) +\tp0\pa_u\Big(-\DYM\a0+C\a1+F\t0\Big) \nn\\
    &=-\DYM\big(\dPta\ap0\big)+\big[\ap0,-\DYM\a0\big]_\g +C\big(\dPta\ap1+[\ap0,\a1]_\g+\tp0(\DTOT\Pta)_1\big) \label{EYMline1}\\
    &-\t0[F,\ap0]_\g-\tp0[F,\a0]_\g+N\ap1\t0+N\a1\tp0+\dot F\t0\tp0 \label{EYMline2}\\
    &-\ap1\Big(D(\DGR\tau)_{-1}- C(\DGR\tau)_0\Big)-\tp0\DYM\big(\DTOT\Pta\big)_0 \label{EYMline3}\\
    &+F\big(\dt\tp0 +\tp0(\DGR\tau)_0\big). \label{EYMline4}
\end{align}
\es
Upon anti-symmetrization between $\Pta$ and $\Ptap$, the line \eqref{EYMline2} cancels out.
Moreover, we can easily identify the appearance of $[\ap{},\a{}]^\g$ and $\lbr\tp{},\t{}\rbr$.
Indeed, the anti-symmetrization of \eqref{EYMline1} reduces to
\begin{align}
    -\DYM\big([\ap{},\a{}]^\g_0+\dPta\ap0-\dPtap\a0\big)+C\Big( [\ap{},\a{}]^\g_1+\dPta\ap1-\dPtap\a1+\tp0(\DTOT\Pta)_1-\t0(\DTOT\Ptap)_1\Big), \label{EYMline5}
\end{align}
while the anti-symmetrization of \eqref{EYMline4} is simply
\begin{equation}
    F\Big(\lbr\tp{},\t{}\rbr_0-2\tp1(\DGR\t{})_{-1}+2\t1(\DGR\tp{})_{-1}\Big). \label{EYMline6}
\end{equation}
Finally, \eqref{EYMline3} can be recast as
\begin{align} \label{EYMline7}
    -\DYM\big(\ap1(\DGR\tau)_{-1}\big) +(\DYM\ap{})_0(\DGR\tau)_{-1}-\DYM\big(\tp0 (\DTOT\Pta)_0\big) +D\tp0(\DTOT\Pta)_0+C\ap1(\DGR\t{})_0.
\end{align}
Taking the anti-symmetrization of \eqref{EYMline7} and summing with \eqref{EYMline5} and \eqref{EYMline6}, we get that
\begin{align} \label{EYMline8}
    \big[\dPta,\dPtap\big]A &=-\DYM\lbr\Ptap,\Pta\rbr^\ym_0 +C[\ap{},\a{}]^\g_1+F\lbr\tp{},\t{}\rbr_0 +\bigg\{\bigg\{C\Big(\tp0( \DTOT\Pta)_1+\ap1(\DGR\t{})_0+\dPta\ap1 \Big) \nn\\
    &+2F\t1(\DGR\tp{})_{-1}+(\DYM\ap{})_0(\DGR\tau)_{-1} +D\tp0(\DTOT\Pta)_0\bigg\}-\Pta\leftrightarrow\Ptap\bigg\},
\end{align}
where we used the definition \eqref{YMbracketDTOTdef} of the YM-bracket.
We still have to massage the last line, which upon expanding each derivative turns out to be proportional to C:
\begin{align}
    &\quad \bigg\{2F\t1(\DGR\tp{})_{-1}+(\DYM\ap{})_0(\DGR\tau)_{-1} +D\tp0(\DTOT\Pta)_0\bigg\}-\Pta\leftrightarrow\Ptap \label{EYMline9}\\
    &= C\Big(2\tp1\DYM\a1+2\ap2 D\t0\Big)-\Pta\leftrightarrow\Ptap=C\Big(2\tp1(\DTOT\Pta )_0+2\ap2 (\DGR\tau)_{-1}\Big)-\Pta\leftrightarrow\Ptap. \nn
\end{align}
\vspace{-0.3cm}

\ni In the last step, we just added and subtracted the necessary terms to form $\DTOT$ and $\DGR$.
Using \eqref{EYMline9} in \eqref{EYMline8} we see that the terms proportional to the shear recombine to give the YM-bracket at degree 1.
We thus conclude that
\begin{align}
    \big[\dPta,\dPtap\big]A &=-\DYM\big\lbr\Ptap,\Pta\big\rbr^\ym_0+C \big\lbr\Ptap,\Pta\big\rbr^\ym_1+F \big\lbr\Ptap,\Pta\big\rbr^\gr_0=-\Big(\DTOT \big\lbr\Ptap,\Pta \big\rbr\Big)_{-1}=-\delta_{\lbr \Pta,\Ptap\rbr}A.
\end{align}

\section{Jacobi identity for the gravitational $\sfK$-bracket \label{AppEYM:JacobiScri}}

One uses that 
\be 
\left\{\t{},\tp{}\right\}_s=\sum_{a+b=s}(a+1)\big(\t{a}\pa_u\tp{b} -\tp{a}\pa_u\t{b}\big).
\ee 
The sum is over integers $-1\leq a,b\leq s+1$ such that $a+b=s$.
\begin{align}
    \big\{\t{},\{\tp{},\tpp{}\}\big\}_s &= \sum_{a+b=s}(a+1)\Big(\t{a}\pa_u\{\tp{},\tpp{}\}_b-\{\tp{}, \tpp{}\}_a\pa_u\t{b}\Big) \nn\\
    &=\sum_{a+b=s}\sum_{c+d=b}(a+1)(c+1)\t{a}\pa_u\big( \tp{c}\pa_u\tpp{d}-\tpp{c}\pa_u\tp{d}\big) \nn\\
    &-\sum_{a+b=s}\sum_{c+d=a}(a+1)(c+1)\pa_u\t{b}\big( \tp{c}\pa_u\tpp{d}-\tpp{c}\pa_u\tp{d}\big) \nn\\
    &=\sum_{a+b+c=s}(a+1)(c+1)\Big(\t{a} \pa_u\tp{c}\pa_u\tpp{b}+\t{a}\tp{c}\pa_u^2 \tpp{b}-\t{a}\pa_u\tpp{c}\pa_u\tp{b}-\t{a}\tpp{c} \pa_u^2\tp{b}\Big) \nn\\*
    &-\sum_{a+b+c=s}(a+c+1)(c+1)\Big(\tp{c}\pa_u\t{b} \pa_u\tpp{a}-\tpp{c}\pa_u\t{b}\pa_u\tp{a}\Big) \\
    &\cyc \sum_{a+b+c=s}\t{a}\pa_u\tpp{b}\pa_u\tp{c} (a+1)\Big((c+1)-(b+1)-(a+c+1)+(a+b+1)\Big) \nn\\
    &+\sum_{a+b+c=s}\t{a}\tp{c}\pa_u^2\tpp{b} \big((a+1)(c+1)-(c+1)(a+1)\big) \nn\\
    &\cyc 0. \nn
\end{align}
Therefore we have proven that the Jacobi identity holds, cf.\,\eqref{JacobiKGRbracket}.

\section{Properties of the action \includegraphics[width=0.40cm]{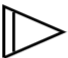}}

\subsection{Leibniz rule \label{AppEYM:tauBarAction}}

We first evaluate 
\begin{align}
\big(\bWt[\ap{},\app{}]^\g\big)_s &=\sum_{n=0}^s(n+1)\Big(\t{n}\pa_u\big[\ap{},\app{} \big]^\g_{s-n}-\big[\ap{},\app{}\big]^\g_{n+1} \pa_u\t{s-n-1} \Big) \\
&=\sum_{n=0}^s(n+1)\Big(\t{n}\big[\pa_u\ap{},\app{} \big]^\g_{s-n}+\t{n}\big[\ap{},\pa_u\app{} \big]^\g_{s-n}-\big[\ap{},\app{}\big]^\g_{n+1} \pa_u\t{s-n-1} \Big).\nn
\end{align}
We then focus on 
\begin{align}
    \big[\bWt\ap{},\app{}\big]^\g_s &=\sum_{k=0}^s\big[ (\bWt\ap{})_k,\app{s-k}\big]_\g \\
    &=\sum_{k=0}^s\sum_{n=0}^k(n+1)\t{n}\big[\pa_u \ap{k-n},\app{s-k}\big]_\g-\sum_{k=0}^s \sum_{n=0}^k(n+1)\pa_u\t{k-n-1}\big[\ap{n+1}, \app{s-k}\big]_\g \nn\\
    &=\sum_{n=0}^s\sum_{k=0}^{s-n}(n+1)\t{n} \big[\pa_u\ap{k},\app{s-k-n}\big]_\g-\sum_{k=0}^s \sum_{n=0}^{s-k}(n+1)\pa_u\t{s-n-k-1}\big[\ap{n+1},\app{k}\big]_\g, \nn
\end{align}
where we commuted the first double sum, namely $\sum_{k=0}^s\sum_{n=0}^k=\sum_{n=0}^s\sum_{k=n}^s$, and then changed $k-n\to n$; while we also changed $s-k\to k$ in the second double sum.
Taking $p=n+k$, we have that $\sum_{k=0}^s\sum_{n=0}^{s-k}=\sum_{p=0}^s \sum_{k=0}^p$.
Simply renaming $p\to n$, we then get
\begin{equation}
    \big[\bWt\ap{},\app{}\big]^\g_s =\sum_{n=0}^s(n+1) \t{n}\big[\pa_u\ap{},\app{}\big]^\g_{s-n}-\sum_{n=0}^s\sum_{k=0}^n(n-k+1)\pa_u\t{s-n-1} \big[\ap{n-k+1},\app{k}\big]_\g.
\end{equation}
The anti-symmetrization of this last double sum reduces to
\begin{align}
    &\quad -\sum_{n=0}^s\sum_{k=0}^n(n-k+1)\pa_u\t{s-n-1} \big[\ap{n-k+1},\app{k}\big]_\g\!-\ap{}\leftrightarrow\app{} \\
    &=- \sum_{n=0}^s\sum_{k=0}^n(n+1)\pa_u\t{s-n-1} \big[\ap{n-k+1},\app{k}\big]_\g =-\sum_{n=0}^s(n+1)\big[\ap{},\app{}\big]^\g_{n+1} \pa_u\t{s-n-1}. \nn
\end{align}
Therefore, the action $\!\barWhiteTriangle$ is a derivative, namely satisfies the Leibniz rule, cf.\,\eqref{barWhiteProperties}:
\begin{equation}
    \boxed{\bWt\big[\ap{},\app{}\big]^\g= \big[\bWt\ap{},\app{}\big]^\g+ \big[\ap{},\bWt\app{}\big]^\g}.
\end{equation}

\subsection{Homomorphism \label{AppEYM:HomtauBarAction}}

In this appendix, unless otherwise stated, we take the convention that equalities are valid upon anti-symmetrization of the RHS.
Firstly, we evaluate the action of the $\sfK$-bracket:
\begin{align}
    \big(\{\tau,\tp{}\}\barWhiteTriangle\app{}\big)_s &=\sum_{n=0}^s(n+1)\sum_{k=0}^{n+1}(k+1)\t{k} \pa_u\tp{n-k}\pa_u\app{s-n} \nn\\
    &-\sum_{n=0}^s(n+1) \app{n+1}\pa_u\left(\sum_{k=0}^{s-n}(k+1)\t{k}\pa_u\tp{s-n-k-1}\right). \label{intermHomBarAction}
\end{align}
Secondly, the commutator of the actions is given by
\begin{align}
    \Big(\big[\bWt,\bWtp\!\big]\app{}\Big)_s &=\sum_{ n=0}^s(n+1)\Big(\t{n}\pa_u\big(\bWtp\app{}\big)_{s-n}-\big(\bWtp\app{}\big)_{n+1}\pa_u\t{s-n-1}\Big) \nn\\
    &=\sum_{n=0}^s(n+1)\t{n}\sum_{k=0}^{s-n}(k+1) \Big(\dashuline{\pa_u\tp{k}\pa_u\app{s-n-k}}+\uline{\tp{k}\pa_u^2\app{s-n-k}} \\
    &\hspace{4.6cm}-\dashuline{\pa_u\app{k+1}\pa_u\tp{s-n-k-1}}-\uwave{\app{k+1}\pa_u^2\tp{s-n-k-1}}\Big) \nn\\
    &-\sum_{n=0}^s(n+1)\pa_u\t{s-n-1}\sum_{k=0}^{n+1}(k+1)\Big(\uuline{\tp{k}\pa_u\app{n+1-k}}-\dotuline{\app{k+1}\pa_u \tp{n-k}}\Big).\nn
\end{align}
The explicit anti-symmetrization of the underlined term vanishes,\footnote{Using that $\sum_{n=0}^s\sum_{k=0}^{s-n}A_nB_k =\sum_{n=0}^s\sum_{k=0}^{s-n}A_kB_n$.}
\begin{equation}
    \uline{~\cdots~}=\sum_{n=0}^s\sum_{k=0}^{s-n}(n+1)(k+1)\big(\t{n}\tp{k}-\tp{n}\t{k}\big)\pa_u^2 \app{s-n-k}=0.
\end{equation}
The wavy underlined term gives us
\begin{align}
    \uwave{~\cdots~} &=-\sum_{n=0}^s\sum_{k=0}^{s-n}(n+1)(k+1)\Big(\app{n+1}\pa_u\big(\t{k}\pa_u\tp{s-n-k-1}\big)-\dotuline{\app{k+1}\pa_u\t{n}\pa_u\tp{s-n-k-1}}\Big).
\end{align}
The dotted underlined terms reduce to\footnote{Changing $s-n-1\to n$ in the first sum.}
\begin{align}
    \dotuline{~\cdots~}&\!=\!\sum_{n=-1}^s \sum_{k=0}^{s-n}(s-n)(k+1)\app{k+1}\pa_u\t{n}\pa_u\tp{s-n-k-1}+\sum_{n=0}^s\sum_{k=0}^{s-n}(n+1)(k+1)\app{k+1}\pa_u\t{n}\pa_u\tp{s-n-k-1} \nn\\
    &=\sum_{n=-1}^s \sum_{k=0}^{s-n}(s+1)(k+1)\app{k+1}\pa_u\t{n}\pa_u\tp{s-n-k-1}=0,
\end{align}
which vanishes due to the anti-symmetrization.
To see it, make use of
\begin{equation}
    \sum_{n=-1}^s \sum_{k=0}^{s-n} A_k B_n C_{s-n-k-1} =\sum_{n=-1}^s \sum_{k=0}^{s-n} A_k B_{s-n-k-1} C_n.
\end{equation}
Next the double underlined contribution is (in the first step we already permuted the sums and used the anti-symmetry)
\begin{align}
    \uuline{~\cdots~}&=\sum_{k=0}^{s+1}\sum_{n=k-1}^s (n+1)(k+1)\t{k}\pa_u\tp{s-n-1}\pa_u\app{n+1-k} \nn\\
    &=\sum_{k=0}^{s+1}\sum_{n=k-1}^s(s+k-n)(k+1)\t{k} \pa_u\tp{n-k}\pa_u\app{s-n} \\
    &=\sum_{n=-1}^s\sum_{k=0}^{n+1}(s+k-n)(k+1)\t{k} \pa_u\tp{n-k}\pa_u\app{s-n}. \nn
\end{align}
Finally the dashed underlined factors gather as (where we already swapped $n\leftrightarrow k$ in the first step)
\begin{align}
     \dashuline{~\,\cdots\,~}&=\sum_{k=0}^s \sum_{n=-1}^{s-k}(n+1)(k+1)\t{k}\pa_u\tp{n}\pa_u\app{s-n-k}-\sum_{k=0}^s\sum_{n=-1}^{s-k}(n+1)(k+1)\t{k} \pa_u\tp{s-n-k-1}\pa_u\app{n+1} \nn\\
     &=\sum_{k=0}^s \sum_{n=-1}^{s-k}(s-2n-k-1)(k+1)\t{k} \pa_u\tp{s-n-k-1}\pa_u\app{n+1}\\
     &=\sum_{k=0}^s\sum_{n=k-1}^s(2n-k-s+1)(k+1)\t{k} \pa_u\tp{n-k}\pa_u\app{s-n}, \nn
\end{align}
where we changed $s-n-1\to n$ to get the last line. 
Moreover we can add the term $k=s+1$ for free since its contribution vanishes, which allows us to commute the sums.
Thus, we obtain that
\begin{equation}
    \uuline{~\cdots~}+\dashuline{~\,\cdots\,~}= \sum_{n=-1}^s\sum_{k=0}^{n+1}(n+1)(k+1)\t{k} \pa_u\tp{n-k}\pa_u\app{s-n}.
\end{equation}
Combining the latter with $\uwave{\,\cdots\,}$, we get \eqref{intermHomBarAction}.
Hence, cf.\,\eqref{barWhiteProperties}
\begin{equation}
    \boxed{\big\{\tau,\tp{}\big\}\barWhiteTriangle \app{}= \big[\bWt,\bWtp\!\big]\app{}}.
\end{equation}

\section{Properties of the action $\vartriangleright$}

\subsection{Leibniz rule anomaly \label{AppEYM:tauAction}}

We start by computing the action of $\bt$ onto the bracket $[\ap{},\app{}]$.
For this we use (as shown in \cite{Cresto:2025bfo}) that
\begin{equation}
    \cA_s\big([\ap{},\app{}]^\g,\DYM\big)=\big[ \DYM\ap0,\app{s+1}\big]_\g-\big[\DYM\app0, \ap{s+1}\big]_\g. \label{LeibnizAnomalyDYM}
\end{equation}
Hence
\begin{align}
    \!\big(\Wt[\ap{},\app{}]^\g\big)_s\! &=\sum_{n=1}^s\Big((n+1)\t{n}\DYM\big[\ap{},\app{}\big]^\g_{s+1-n}-n\big[\ap{},\app{}\big]^\g_n D\t{s+1-n}\Big)+(s+2)C\t{s+1}\big[\ap{},\app{}\big]^\g_1 \nn\\
    &=\sum_{n=1}^s\!\Big((n+1)\Big(\t{n}\big[ \DYM\ap{},\app{}\big]^\g_{s-n}\!+\t{n}\big[ \ap{},\DYM\app{}\big]^\g_{s-n}\Big)-n\big[\ap{},\app{}\big]^\g_{n}D\t{s+1-n}\Big) \label{tauAction2} \\
    &+(s+2)C\t{s+1}\big[\ap{},\app{}\big]^\g_1 +\sum_{n=1}^s(n+1)\Big(\t{n}\big[ \DYM\ap0,\app{s+1-n}\big]_\g-\t{n} \big[\DYM\app0, \ap{s+1-n}\big]_\g\Big).\nn
\end{align}
Next we calculate
\begin{align}
    \big[\Wt\ap{},\app{}\big]^\g_s &=\sum_{k=0}^s\big[(\Wt\ap{})_k,\app{s-k}\big]_\g \nn\\
    &=\sum_{k=1}^s\sum_{n=1}^k(n+1)\t{n} \big[\DYM\ap{k+1-n},\app{s-k}\big]_\g-\sum_{k=1}^s\sum_{n=1}^k n\big[\ap{n}, \app{s-k}\big]_\g D\t{k+1-n} \nn\\
    &+C\sum_{k=0}^s(k+2)\Big(\t{k+1}[\ap1,\app{s-k}]_\g\Big) \nn\\
    &=\sum_{n=1}^s\sum_{k=0}^{s-n}(n+1)\t{n} \big[\DYM\ap{k+1},\app{s-k-n}\big]_\g-\sum_{n=1}^{s}\sum_{k=1}^n k[\ap{k},\app{n-k}]_\g D\t{s+1-n} \label{tauAction1}\\
    &+C\sum_{k=0}^s(k+2)\t{k+1}[\ap1,\app{s-k}]_\g, \nn
\end{align}
where in the last step, for the term containing $\DYM\ap{}$, we commuted the sums and then changed $k-n\to k$; while for the term containing $D\t{}$, we first change $k-n\to n$, then commuted the sums, changed $k-n\to k$ and finally $s-n\to n$.
Still in that same double sum, we can add the term $k=0$ since its contribution is 0.
Anti-symmetrizing in $\ap{},\app{}$, we get in particular that the term involving $D\t{}$ reduces to
\begin{align}
    -\sum_{n=1}^s\sum_{k=0}^{n} D\t{s+1-n}\Big(k\big[\ap{k},\app{n-k}\big]_\g+k\big[\ap{n-k},\app{k}\big]_\g\Big) &=-\sum_{n=1}^s\sum_{k=0}^{n}n D\t{s+1-n}\big[\ap{k},\app{n-k}\big]_\g \nn\\
    &=-\sum_{n=1}^s n\big[\ap{},\app{}\big]^\g_nD\t{s+1-n}.
\end{align}
Besides, the anti-symmetrization of the term proportional to $C\ap1$ gives\footnote{The first step consists in extracting the term $n=s$ from the sum.}
\begin{align}
    C\sum_{n=0}^s(n+2)\t{n+1}&\Big(\big[\ap1, \app{s-n}\big]_\g+\big[\ap{s-n},\app1\big]_\g\Big)= \\
    &=(s+2)C\t{s+1}\big[\ap{},\app{}\big]^\g_1+C \sum_{n=1}^{s}(n+1)\t{n}\Big(\big[\ap1, \app{s+1-n}\big]_\g+\big[\ap{s+1-n},\app1\big]_\g\Big). \nn
\end{align}
Therefore, combining the last two equations with \eqref{tauAction1}, we infer that

\begin{align}
    &\quad \big[\Wt\ap{},\app{}\big]^\g_s+\big[ \ap{}, \Wt\app{}\big]^\g_s= -\sum_{n=1}^s n\big[\ap{},\app{}\big]^\g_n D\t{s+1-n}+(s+2)C\t{s+1}\big[\ap{},\app{}\big]^\g_1 \label{tauAction3}\\*
    &\!+\!\sum_{n=1}^s (n+1)\t{n}\Big(\big[\DYM \ap{},\app{} \big]^\g_{s-n}\!+\big[\ap{},\DYM \app{}\big]^\g_{s-n}\Big)\!+ C\sum_{n=1}^{s}(n+1)\t{n}\Big(\big[\ap1, \app{s+1-n}\big]_\g\!+\big[\ap{s+1-n},\app1\big]_\g\Big).\nn
\end{align}
Upon comparing \eqref{tauAction3} with \eqref{tauAction2}, we deduce that the Leibniz anomaly of the $\bt$-action takes the form \eqref{LeibnizAnomalyTauAction}
\begin{align}
    \boxed{\cA_s\big([\ap{},\app{}]^\g,\Wt\!\big) = \sum_{n=1}^s(n+1)\t{n}\big[ \DYM\ap0-C\ap1,\app{s+1-n}\big]_\g-\ap{}\leftrightarrow\app{}}.
\end{align}

\subsection{Homomorphism anomaly \label{AppEYM:tauHom}}

We start by computing the commutator of the actions of $\bt$ and $\btp$.
In the following, for shortness, it is understood that equalities are valid upon anti-symmetrization of the RHS.
\begin{align}
    &\big(\Wt(\Wtp\app{})\big)_s-\big(\Wtp(\Wt\app{})\big)_s =\nn\\
    &=\sum_{n=1}^s \Big((n+1)\t{n}\DYM(\Wtp\app{})_{s+1-n}-n(\Wtp\app{})_{n}D\t{s+1-n}\Big)+(s+2)C\t{s+1}(\Wtp\app{})_1 \nn\\
    &=\sum_{n=1}^s(n+1)\t{n}\DYM\bigg\{ \sum_{k=1}^{s+1-n}\Big((k+1)\tp{k}\DYM\app{s+2-n-k}-k\app{k}D\tp{s+2-n-k}\Big)+(s+3-n)C\app1\tp{s+2-n}\bigg\} \nn\\
    &-\sum_{n=1}^s nD\t{s+1-n}\bigg\{ \sum_{k=1}^n\Big((k+1)\tp{k}\DYM\app{n+1-k}-k\app{k}D\tp{n+1-k}\Big) +(n+2)C \app1\tp{n+1}\bigg\} \nn\\
    &+(s+2)C\t{s+1}\bigg\{2\tp1 \DYM\app1-\app1 D\tp1+3C\app1\tp2\bigg\} \nn\\
    &=\sum_{n=1}^s\sum_{k=1}^{s+1-n}(n+1)(k+1)\uline{\t{n}\tp{k}\DYM^2\app{s+2-n-k}}+\sum_{n=1}^s\sum_{k=1}^{s+1-n}(n+1)(k+1)\dotuline{\t{n}D\tp{k}\DYM\app{s+2-n-k}} \nn\\
    &-\sum_{n=1}^s\sum_{k=1}^{s+1-n}(n+1)k \dotuline{\t{n}D\tp{s+2-n-k}\DYM\app{k}}-\sum_{n=1}^s\sum_{k=1}^{s+1-n}(n+1)k \dashuline{\t{n}D^2\tp{s+2-n-k}\app{k}} \nn\\
    &+\sum_{n=1}^s(n+1)(s+3-n)\uwave{\t{n}D(C\tp{s+2-n}) \app1}+ \,C\sum_{n=1}^s(n+1)(s+3-n)\underline{\t{n}\tp{s+2-n}\DYM\app1}\nn\\
    &-\sum_{n=1}^s\sum_{k=1}^{n}n(k+1) \dotuline{\tp{k}D\t{s+1-n}\DYM\app{n+1-k}}+\sum_{n=1}^s \sum_{k=1}^{n}nk\dashuline{D\t{s+1-n}D\tp{n+1-k}\app{k}} \\
    &-C\sum_{n=1}^s n(n+2)D\t{s+1-n}\uwave{\app1\tp{n+1}}+(s+2)C\t{s+1} \bigg\{\underline{2\tp1 \DYM\app1}-\uwave{\app1 D\tp1}+3C\app1\tp2\bigg\}. \nn
\end{align}
Writing the anti-symmetrization of the underlined term explicitly, we immediately notice that\footnote{In the following, we often make use of the fact that $\sum_{n=1}^s\sum_{k=1}^{s+1-n}A_nB_k=\sum_{n=1}^s\sum_{k=1}^{s+1-n}A_kB_n$. \label{footEYM:sum}}
\begin{align}
    \underline{~\cdots~}=0.
\end{align}
We then treat the dotted underlined terms. 
In the first step, we do the appropriate changes of variables to get $\app{s+2-n-k}$ to appear in each sum and we leverage the fact that the RHS is understood to be anti-symmetrized to swap $\tau$ and $\tau'$ in the second double sum.
\begin{align}
    \dotuline{~\cdots~}\!&=\sum_{n=1}^s \sum_{k=1}^{s+1-n}(n+1)(k+1)\t{n}D\tp{k}\DYM\app{s+2-n-k}+\sum_{n=1}^s\sum_{k=1}^{s-n+1}(s-n+1)(k+1) \t{k}D\tp{n}\DYM\app{s-n+2-k} \nn\\
    &-\sum_{n=1}^s\sum_{k=1}^{s+1-n}(n+1)(s+2-n-k) \t{n}D\tp{k}\DYM\app{s+2-n-k}. \label{dottedUnderline}
\end{align}
Next, still in the second double sum, we use the property emphasized in the footnote \ref{footEYM:sum}, such that each term in \eqref{dottedUnderline} now involves $\t{n}D\tp{k}\DYM\app{s+2-n-k}$.
We thus obtain
\begin{align}
    \dotuline{~\cdots~}=\sum_{n=1}^s \sum_{k=1}^{s+1-n}(n+1)(n+k)\t{n}D\tp{k}\DYM\app{s+2-n-k}.
\end{align}
For later purposes, let us define the variable $p=n+k$ such that 
\begin{align}
    \dotuline{~\cdots~}=\sum_{p=2}^{s+1} \sum_{n=1}^{p-1} p(n+1)\t{n}D\tp{p-n}\DYM\app{s+2-p}=\sum_{n=1}^s \sum_{k=1}^n(n+1)(k+1)\aunderbrace[l1r]{\t{k}D\tp{n+1-k}\DYM\app{s+1-n}},
\end{align}
where in the last step, we renamed $p-1\to n$ and $n\to k$.
Concerning the dashed underlined contributions, we get
\vspace{-0.4cm}

\begin{adjustwidth}{-0.2cm}{0cm}
\begin{align}
    &\dashuline{\,~\cdots~\,}\!=-\sum_{n=1}^s \sum_{k=1}^{s+1-n}(n+1)kD\big(\t{n}D \tp{s+2-n-k}\big)\app{k}+\sum_{n=1}^s \sum_{k=1}^{s+1-n}(s+2)kD\t{n}D\tp{s+2-n-k}\app{k} \\
    &=-\sum_{n=1}^s \sum_{k=1}^{s+1-n}(k+1)nD\big(\t{k}D\tp{s+2-n-k}\big)\app{n}+\sum_{n=1}^s \sum_{k=1}^{s+1-n}(s+2)(s+2-n-k)D\t{n}D\tp{k}\app{s+2-n-k}. \nn
\end{align}
\end{adjustwidth}
The last double sum vanishes because of the anti-symmetrization. Hence,
\begin{align}
    \dashuline{\,~\cdots~\,}\!&=
    -\sum_{n=1}^s n\app{n}D\bigg\{ \sum_{k=1}^{s+1-n}(k+1)\aunderbrace[l1r]{\t{k}D\tp{s+2-n-k}}\bigg\}.\nn
\end{align}
We also get that the wavy underlined terms take the form
\begin{align}
    \uwave{~\cdots~}&=\app1\sum_{n=1}^s (n+1)(s+3-n)DC\t{n}\tp{s+2-n}+C\app1\sum_{n=1}^s(n+1)(s+3-n)\t{n}D\tp{s+2-n} \nn\\
    &+C\app1\sum_{n=2}^{s+1}(n-1)(n+1)\t{n} D\tp{s+2-n}-(s+2)C\app1\t{s+1} D\tp1 \\
    &=-2(s+2)DC\app1\t{s+1}\tp1-3(s+2)C\app1 \t{s+1}D\tp1+(s+2)C\app1\sum_{n=1}^{s+1}(n+1)\aunderbrace[l1r]{\t{n} D\tp{s+2-n}}. \nn
\end{align}
It is now straightforward to notice that the terms emphasized with the under-brace recombine to form the projection of the $C$-bracket onto the $\bcT$-sub-algebra as follows:\footnote{Of course here the anti-symmetrization of the RHS is explicit.}
\begin{equation}
    \aunderbrace[l1r]{~\cdots~}=\sum_{n=1}^s \Big((n+1)\overline{[\bt{},\btp{}]}_n\DYM\app{s+1-n}-n\app{n}D\overline{[\bt{} ,\btp{}]}_{s+1-n}\Big)+(s+2)C\app1\overline{[\bt{},\btp{}]}_{s+1}.
\end{equation}
We recognize the latter as the action of the bar-bracket onto $\app{}$, namely
\begin{equation}
    \aunderbrace[l1r]{~\cdots~}=\big(\overline{[\bt{},\btp{}]} \vartriangleright \app{}\big)_s.
\end{equation}
Therefore, just gathering the various contributions, we get that
\begin{align}
    &\big(\Wt(\Wtp\app{})\big)_s-\big(\Wtp(\Wt\app{})\big)_s -\big(\overline{[\bt{},\btp{}]} \vartriangleright \app{}\big)_s= \\
    &=-2(s+2)DC\app1\tp1\t{s+1} -3(s+2)C\app1 \t{s+1}D\tp1+3(s+2)C^2\app1\t{s+1}\tp2. \nn
\end{align}
The final result takes the form (cf. \eqref{HomAnomalyTauAction})
\begin{equation}
    \boxed{\Big(\overline{[\bt{},\btp{}]} \vartriangleright \app{}-\big[\Wt,\Wtp\!\big]\app{} \Big)_s=(s+2)\app1\Big(\t{s+1}\dbtp C-\tp{s+1}\dbt C\Big)},
\end{equation}
where $\dbt C$ is the variation $\dt C$ restricted to the $\bsfT$-subspace, i.e.
\begin{equation}
    \dbt C=2DC\t1+3CD\t1-3C^2\t2.
\end{equation}

\section{Properties of the action $\blacktriangleright$ \label{AppEYM:BlackTauAction}}

\subsection{Leibniz rule anomaly: GR projection \label{AppEYM:BlackTauActionLeibnizGR}}

We first compute the GR projection of the action of $\ot$ onto the bracket $\big[\bPtap,\bPtapp\big]$.
\begin{align}
    \big(\Bt\big[\bPtap,\bPtapp\big] \big)^\gr_s &=\t0 D\overline{\big[\btp,\btpp\big]}_{s+1}-(s+2)\overline{\big[\btp,\btpp\big]}_{s+1} D\t0-(s+3)C\t0\overline{\big[\btp, \btpp\big]}_{s+2} \nn\\
    &=\bigg\{\t0\sum_{n=1}^{s+1}(n+1)\big(D\tp{n}D\tpp{s+2-n}+\tp{n}D^2\tpp{s+2-n}\big)-(s+2)D\t0\sum_{n=1}^{s+1}(n+1)\tp{n}D\tpp{s+2-n} \nn\\
    &-(s+3)C\t0\sum_{n=1}^{s+2}(n+1)\tp{n}D\tpp{s+3-n}\bigg\}-\btp\leftrightarrow\btpp. \label{eq101}
\end{align}
Next,
\begin{align}
    \big[\Bt\bPtap,\bPtapp\big]^\gr_s &=\overline{\big[(\Bt\bPtap)^\gr,\btpp \big]}_s \nn\\*
    &=\sum_{n=1}^s(n+1)\Big(\big(\Bt\bPtap \big)^\gr_n D\tpp{s+1-n}-\tpp{n}D\big(\Bt \bPtap\big)^\gr_{s+1-n}\Big) \nn\\*
    &=\sum_{n=1}^s(n+1)D\tpp{s+1-n}\Big\{\t0 D\tp{n+1}-(n+2)\tp{n+1}D\t0-(n+3)C\t0\tp{n+2}\Big\} \nn\\
    &-\sum_{n=1}^s(n+1)\tpp{n}D\Big\{\t0 D\tp{s+2-n}-(s+3-n)\tp{s+2-n}D\t0-(s+4-n)C\t0\tp{s+3-n}\Big\} \nn\\
    &=\t0\sum_{n=1}^s(n+1)\uuline{D\tp{n+1} D\tpp{s+1-n}}-D\t0\sum_{n=1}^s(n+1)(n+2)\underline{\tp{n+1}D\tpp{s+1-n}} \nn\\
    &-C\t0\sum_{n=1}^s(n+1)(n+3)\uwave{\tp{n+2}D\tpp{s+1-n}} -D\t0\sum_{n=1}^s(n+1)\underline{\tpp{n}D\tp{s+2-n}} \nn\\
    &-\t0\sum_{n=1}^{s+1}(n+1)\aunderbrace[l1r]{\tpp{n}D^2\tp{s+2-n}}+(s+2)\t0\tpp{s+1}D^2\tp1  \\
    &+D\t0\sum_{n=1}^s(n+1)(s+3-n)\underline{\tpp{n}D\tp{s+2-n}}+D^2\t0\sum_{n=1}^s(n+1)(s+3-n)\dotuline{\tp{s+2-n}\tpp{n}} \nn\\
    &+D(C\t0)\!\sum_{n=1}^s(n+1)(s+4-n)\dashuline{\tp{s+3-n}\tpp{n}}+C\t0\!\sum_{n=1}^s(n+1)(s+4-n)\uwave{\tpp{n}D\tp{s+3-n}}. \nn
\end{align}
Anti-symmetrizing the straight underlined terms leads to
\begin{equation}
    \underline{~\cdots~}=D\t0\bigg\{(s+2)\sum_{n=1}^{s+1}(n+1)\aunderbrace[l1r]{\tpp{n}D\tp{s+2-n}}-2\tpp1D\tp{s+1}-(s+2)\tpp{s+1}D\tp1\bigg\}-\btp\leftrightarrow\btpp.
\end{equation}
Similarly for the wavy underlined terms, we get
\begin{align}
    \uwave{~\cdots~} &=C\t0\bigg\{(s+3) \sum_{n=1}^{s+2}(n+1)\aunderbrace[l1r]{\tpp{n}D\tp{s+3-n}}-3(s+2)\tpp{s+1}D\tp2-2(s+3)\tpp{s+2}D\tp1 \nn\\
    &\qquad\quad -3\tpp2D\tp{s+1}\bigg\}-\btp\leftrightarrow\btpp.
\end{align}
Also notice that the anti-symmetrization of the dashed underlined term reduces to
\begin{equation}
    \dashuline{\,~\cdots~\,}\!=-D(C\t0)\Big\{ 3(s+2)\tp2\tpp{s+1}+2(s+3)\tp1\tpp{s+2}\Big\}-\btp\leftrightarrow\btpp,
\end{equation}
and in a similar way
\begin{equation}
    \dotuline{~\cdots~} =-D^2\t0\Big\{2(s+2)\tp1\tpp{s+1}\Big\}-\btp\leftrightarrow\btpp.
\end{equation}
In order to identify the double underlined term with its counterpart in \eqref{eq101}, we conveniently write that
\begin{equation}
    \uuline{~\cdots~} =\t0\bigg\{\sum_{n=1}^{s+1}(n+1)\aunderbrace[l1r]{D\tp{n}D\tpp{s+2-n}}-D\tp1D\tpp{s+1}\bigg\}-\btp\leftrightarrow\btpp.
\end{equation}
Therefore, using the fact that the under-braced factors equate \eqref{eq101}, we find that
\begin{align}
    &\quad \big[\Bt\bPtap,\bPtapp\big]^\gr_s +\big[\bPtap,\Bt\bPtapp\big]^\gr_s =\big(\Bt\big[\bPtap,\bPtapp\big] \big)^\gr_s+ \nn\\
    &+\bigg\{\bigg\{(s+2)\t0\tpp{s+1}D^2\tp1 +D\t0\Big( 2\tp1D\tpp{s+1}-(s+2)\tpp{s+1}D\tp1\Big) -2(s+2)D^2\t0\tp1\tpp{s+1} \nn\\
    &-C\t0\Big(3(s+2)\tpp{s+1}D\tp2+ 2(s+3)\tpp{s+2}D\tp1-3\tp2D\tpp{s+1}\Big) \label{BlackLeibnizInterm1}\\
    &-D(C\t0)\Big( 3(s+2)\tp2\tpp{s+1}+2(s+3) \tp1\tpp{s+2}\Big)-\t0 D\tp1D\tpp{s+1}\bigg\}-\btp\leftrightarrow\btpp\bigg\}. \nn
\end{align}
Let us now compare the terms that did not recombine into $\big(\Bt\big[\bPtap,\bPtapp\big] \big)^\gr$ with
\begin{align}
    \big((\rBt\bPtap)\blacktriangleright \bPtapp\big)^\gr_s &=\big(\rBt\bPtap \big)_0D\tpp{s+1}-(s+2)\tpp{s+1}D\big(\rBt\bPtap \big)_0 -(s+3)C\tpp{s+2}\big(\rBt\bPtap \big)_0 \nn\\
    &=D\tpp{s+1}\Big(\t0D\tp1-2\tp1D\t0-3C\tp2\t0\Big) \label{BlackLeibnizInterm2}\\
    &-(s+2)\tpp{s+1}\Big(-D\t0D\tp1+\t0D^2\tp1 -2\tp1D^2\t0-3D(C\t0)\tp2-3C\t0D\tp2\Big) \nn\\
    &-(s+3)C\tpp{s+2}\Big(\t0D\tp1-2\tp1D\t0-3C\tp2\t0\Big). \nn
\end{align}
All the factors proportional to $D\tpp{s+1}$ and to $\tpp{s+1}$ (i.e. the first two lines of \eqref{BlackLeibnizInterm2}) appear as such in \eqref{BlackLeibnizInterm1}, up to a sign.
It is then not hard to see that the terms proportional to $\tpp{s+2}$ add up to form the variation $\dbtp C$.
Thence the Leibniz rule anomaly for the action $\blacktriangleright$ projected onto $\bsfT$ is given by \eqref{LeibnizAnomalyBlackTauActionGR}
\begin{equation}
    \boxed{\cA_s^\gr\Big(\big[\bPtap,\bPtapp \big], \Bt\!\Big)=\Big\{ \big((\rBt\bPtap)\blacktriangleright \bPtapp\big)^\gr_s+(s+3)\t0\tpp{s+2} \dbtp C\Big\}-\bPtap\leftrightarrow\bPtapp}.
\end{equation}

\subsection{Leibniz rule anomaly: YM projection \label{AppEYM:BlackTauActionLeibnizYM}}

The YM projection of the action of $\ot$ onto $\big[\bPtap,\bPtapp\big]$ takes the form
\begin{align}
    \big(\Bt\big[\bPtap,\bPtapp\big] \big)^\ym_s &=\t0\DYM\big[ \bPtap,\bPtapp\big]^\ym_{s+1}-(s+1)\big[\bPtap,\bPtapp\big]^\ym_{s+1}D\t0-(s+2)C\t0\big[\bPtap,\bPtapp \big]^\ym_{s+2} \nn\\
    &-(s+2)F\t0\overline{\big[\btp,\btpp\big] }_{s+1} \nn\\
    &=\t0\DYM\big[\ap{},\app{}\big]^\g_{s+1}-(s+1)\big[\ap{},\app{}\big]^\g_{s+1}D\t0-(s+2)C\t0\big[\ap{},\app{}\big]^\g_{s+2} \nn\\
    &+\bigg\{\bigg\{\t0\DYM\big(\Wtp\app{} \big)_{s+1}-(s+1)\big(\Wtp\app{} \big)_{s+1}D\t0-(s+2)C\t0\big(\Wtp\app{} \big)_{s+2} \nn\\
    &-(s+2)F\t0\sum_{n=1}^{s+1}(n+1)\tp{n}D\tpp{s+2-n}\bigg\}-\bPtap\leftrightarrow\bPtapp\bigg\}. \label{LeibnizYMinterm}
\end{align}
On the other hand,
\begin{align}
&~\quad \big[\Bt\bPtap,\bPtapp\big]^\ym_s= \nn\\ &=\big[(\Bt\bPtap)^\ym,\app{}\big]^\g_s+\big((\Bt \bPtap)^\gr\vartriangleright\app{}\big)_s-\big( \Wtpp(\Bt\bPtap)^\ym\big)_s \nn\\
&=\sum_{n=0}^s\big[(\Bt\bPtap)^\ym_n,\app{s-n}\big]_\g+\sum_{n=1}^s\Big((n+1)\big(\Bt\bPtap\big)^\gr_n\,\DYM\app{s+1-n}-n\app{n}D\big(\Bt\bPtap\big)^\gr_{s+1-n}\Big) \nn\\*
&+(s+2)C\app1\big(\Bt\bPtap\big)^\gr_{s+1}-\sum_{n=1}^s\Big((n+1)\tpp{n}\DYM\big(\Bt\bPtap\big)^\ym_{s+1-n}-n\big(\Bt\bPtap\big)^\ym_n D\tpp{s+1-n}\Big) \nn\\
&-(s+2)C\big(\Bt\bPtap\big)^\ym_1\tpp{s+1}\nn\\
&=\sum_{n=0}^s\Big[\underline{\t0\DYM\ap{n+1}}-(n+1)\underline{\ap{n+1}D\t0}-(n+2)\underline{C\t0\ap{n+2}}-(n+2)F\t0\tp{n+1},\app{s-n}\Big]_\g \nn\\
&+\sum_{n=1}^s(n+1)\DYM\app{s+1-n}\Big\{\uuline{\t0 D\tp{n+1}}-(n+2)\dashuline{\tp{n+1}D\t0}-(n+3)\uwave{C\t0\tp{n+2}}\Big\} \nn\\
&-\sum_{n=1}^sn\app{n}\Big\{\dashuline{D \t0D\tp{s+2-n}}+\uuline{\t0D^2\tp{s+2-n}}-(s+3-n)\dashuline{D\tp{s+2-n}D\t0}-\aunderbrace[l2*{8}{10}3r]{(s+3-n)\tp{s+2-n}D^2\t0} \nn\\
&\quad\qquad\qquad -(s+4-n)\aunderbrace[l1r]{D(C\t0)\tp{s+3-n}}-(s+4-n)\uwave{C\t0D\tp{s+3-n}}\Big\} \nn\\
&+(s+2)C\app1\Big\{\underbrace{\t0D\tp{s+2}}-(s+3)\underbrace{\tp{s+2}D\t0}-(s+4)\underbrace{C\t0\tp{s+3}}\Big\} \nn\\
&-\sum_{n=1}^s(n+1)\tpp{n}\Big\{ \dashuline{D\t0\DYM\ap{s+2-n}}+\uuline{\t0\DYM^2\ap{s+2-n}}-(s+2-n)\dashuline{\DYM\ap{s+2-n}D\t0} \nn\\
&\qquad\qquad -\aunderbrace[l2*{8}{10}3r]{(s+2-n)\ap{s+2-n}D^2\t0}-(s+3-n)\aunderbrace[l1r]{D(C\t0)\ap{s+3-n}}-(s+3-n)\uwave{C\t0\DYM\ap{s+3-n}} \nn\\
&\qquad\qquad -(s+3-n)\DYM(F\t0)\tp{s+2-n}-(s+3-n)\aunderbrace[l1*{3}{01}01D10*{3}{10}1r]{F\t0D\tp{s+2-n}}\Big\} \\
&+\sum_{n=1}^snD\tpp{s+1-n}\Big\{\uuline{\t0\DYM\ap{n+1}}-(n+1)\dashuline{\ap{n+1}D\t0}-(n+2)\uwave{C\t0\ap{n+2}}-\aunderbrace[l1*{3}{01}01D10*{3}{10}1r]{(n+2)F\t0\tp{n+1}}\Big\} \nn\\
&-(s+2)C\tpp{s+1}\Big\{\t0\DYM\ap2-\underbrace{2\ap2D\t0}-3C\t0\ap3-3F\t0\tp2\Big\}. \nn
\end{align}
After anti-symmetrization in $\bPtap,\bPtapp$, the underlined terms contribute as\footnote{We use that
\vspace{-0.2cm}
\begin{equation*}
    \sum_{n=0}^s(n+1)\big[\ap{n+1},\app{s-n}\big]_\g-\ap{}\leftrightarrow\app{}= \sum_{n=0}^{s+1}\Big(n\big[\ap{n},\app{s+1-n}\big]_\g+n\big[\ap{s+1-n},\app{n}\big]_\g \Big)=\sum_{n=0}^{s+1}(s+1)\big[\ap{n},\app{s+1-n}\big]_\g.
\end{equation*}
\vspace{-0.3cm}}
\begin{align}
    \underline{~\cdots~} &=\t0\big[\DYM\ap{}, \app{}\big]^\g_s+\t0\big[\ap{},\DYM\app{}\big]^\g_s-(s+1)D\t0\big[\ap{},\app{}\big]^\g_{s+1}-(s+2)C\t0\big[\ap{},\app{}\big]^\g_{s+2} \nn\\
    &+C\t0\Big(\big[\ap1,\app{s+1}\big]_\g +\big[\ap{s+1},\app1\big]_\g \Big) \nn\\
    &=\t0\DYM\big[\ap{},\app{}\big]^\g_{s+1}+\t0 \big[\hdap A,\app{s+1}\big]_\g-\t0 \big[\hdapp A,\ap{s+1}\big]_\g \\
    &-(s+1)D\t0\big[\ap{},\app{}\big]^\g_{s+1}-(s+2)C\t0\big[\ap{},\app{}\big]^\g_{s+2}, \nn
\end{align}
where in the last step, we used the property \eqref{LeibnizAnomalyDYM}.
Besides, the double underlined terms partially recombine (leveraging the anti-symmetrization) to give $\t0\DYM(\Wtp\app{})_{s+1}$:

\begin{align}
    \uuline{~\cdots~} &=\t0\bigg\{\sum_{n=2}^{s+1}n D\tp{n}\DYM\app{s+2-n}+\sum_{n=1}^s(n+1)\tp{n}\DYM^2\app{s+2-n}-\sum_{n=2}^{s+1}(n-1)\DYM\app{n}D\tp{s+2-n} \nn\\*
    &\qquad -\sum_{n=1}^s n\app{n}D^2\tp{s+2-n}\bigg\}-\bPtap\leftrightarrow\bPtapp \nn\\
    &=\t0\Big\{\DYM\big(\Wtp\app{}\big)_{s+1}-D\tp1\DYM\app{s+1}-(s+2)\tp{s+1}\DYM^2\app1+(s+1)\app{s+1}D^2\tp1 \\
    &\qquad -(s+3)\underbrace{\DYM\big(C\app1\tp{s+2} \big)}\Big\}-\bPtap\leftrightarrow\bPtapp. \nn
\end{align}
Concerning the dash-underlined terms, they partially recombine into $D\t0(\Wtp\app{})_{s+1}$:
\begin{align}
    \dashuline{~\,\cdots~\,}\! &=D\t0\bigg\{ \sum_{n=1}^sn(s+2-n)\app{n}D\tp{s+2-n}+\sum_{n=2}^{s+1}n(n-1)\app{n}D\tp{s+2-n} \nn\\
    &\qquad~~-\sum_{n=1}^s(n+1)(s+1-n)\tp{n}\DYM\app{s+2-n}-\sum_{n=2}^{s+1}n(n+1)\tp{n}\DYM\app{s+2-n}\bigg\}-\bPtap\leftrightarrow\bPtapp \nn\\
    &=-D\t0\Big\{ (s+1)\big(\Wtp\app{}\big)_{s+1}+(s+1)\app{s+1}D\tp1-2\tp1\DYM\app{s+1} \\
    &\qquad\quad~ -(s+1)(s+3)\underbrace{C\app1\tp{s+2}}\Big\}-\bPtap\leftrightarrow\bPtapp. \nn
\end{align}
We can rearrange the wavy underlined factors  as
\begin{align}
    \uwave{~\cdots~} &=-C\t0\bigg\{ \sum_{n=3}^{s+2}(n-1)(n+1)\tp{n}\DYM\app{s+3-n}-\sum_{n=1}^sn(s+4-n) \app{n}D\tp{s+3-n} \\
    &\qquad\quad~ +\sum_{n=1}^s(n+1)(s+3-n)\tp{n}\DYM\app{s+3-n}-\sum_{n=3}^{s+2}n(n-2)\app{n}D\tp{s+3-n} \bigg\}-\bPtap\leftrightarrow\bPtapp \nn\\
    &=-C\t0\Big\{(s+2)\big(\Wtp\app{}\big)_{s+2}-3\tp2\DYM\app{s+1}-2(s+2)\tp{s+1}\DYM \app2-(s+3)\underbrace{\tp{s+2}\DYM\app1}\nn\\
    &-\underbrace{\app1D\tp{s+2}}+3(s+1)\app{s+1} D\tp2+2(s+2)\app{s+2}D\tp1-(s+2)(s+4)\underbrace{C\app1\tp{s+3}}\Big\}-\bPtap\leftrightarrow\bPtapp. \nn
\end{align}
The straight under-braced contributions sum up to
\begin{align}
    \aunderbrace[l1r]{~\cdots~}&=D(C\t0)\bigg\{ \sum_{n=1}^sn(s+4-n)\app{n}\tp{s+3-n}-\sum_{n=3}^{s+2}n(s+4-n)\tp{s+3-n}\app{n} \bigg\}-\bPtap\leftrightarrow\bPtapp\nn\\
    &=CD\t0\Big\{(s+3)\underbrace{\app1\tp{s+2}}+2(s+2) \underbrace{\app2\tp{s+1}}- 3(s+1)\app{s+1}\tp2-2(s+2)\app{s+2}\tp1 \Big\} \\
    &+DC\t0\Big\{(s+3)\underbrace{\app1\tp{s+2}}+2(s+2) \app2\tp{s+1}- 3(s+1)\app{s+1}\tp2-2(s+2)\app{s+2}\tp1 \Big\}-\bPtap\leftrightarrow\bPtapp.\nn
\end{align}
The curly under-braced terms simply add up to 0,
\begin{equation}
    \underbrace{~\cdots~}=0,
\end{equation}
while the dash-under-braced factors reduce to
\begin{equation}
    \aunderbrace[l1*{4}{10}1r]{~\,\cdots\,~}= D^2\t0\Big\{-2(s+1)\app{s+1}\tp1+(s+2)\tp{s+1}\app{1}\Big\}-\bPtap\leftrightarrow\bPtapp.
\end{equation}
Finally the curly dash-under-braced terms simplify to
\begin{align}
    \aunderbrace[l0*{2}{02}00D00*{2}{20}0r]{~\,\cdots\cdots\,~}=-(s+2)F\t0\bigg\{\sum_{n=1}^{s+1}(n+1)\tp{n}D\tpp{s+2-n}-2\tp{s+1}D\tpp1\bigg\}-\bPtap\leftrightarrow\bPtapp.
\end{align}
Using the fact that
\begin{equation}
    \bigg\{\DYM(F\t0)\sum_{n=1}^s(n+1)(s+3-n)\tpp{n}\tp{s+2-n}\bigg\}-\btp\leftrightarrow\btpp=2(s+2)\DYM(F\t0) \tp{s+1}\tpp1 -\btp\leftrightarrow\btpp,
\end{equation}
we gather all these contributions and get that
\begin{align}
    &~\quad \Big(\big[\Bt\bPtap,\bPtapp\big]+\big[\bPtap,\Bt\bPtapp\big]\Big)^\ym_s= \nn\\ 
    &=\t0\DYM\big[\ap{},\app{}]_{s+1}-(s+1)D\t0\big[\ap{},\app{}\big]_{s+1}-(s+2)C\t0\big[\ap{},\app{}\big]_{s+2}-(s+2)F\t0\overline{\big[\btp, \btpp\big]}_{s+1} \nn\\
    &+\bigg(\!\bigg(\!\t0 \big[\hdap A,\app{s+1}\big]_\g+\t0 \DYM\big(\Wtp\app{}\big)_{s+1}-(s+1)D\t0\big(\Wtp\app{}\big)_{s+1}-(s+2)C\t0\big(\Wtp\app{}\big)_{s+2} \nn\\
    &+\t0\Big\{-D\tp1\DYM\app{s+1}-(s+2)\dotuline{\tp{s+1}\DYM^2\app1}+(s+1)\app{s+1}D^2\tp1\Big\} \nn\\
    &-C\t0\Big\{-3\tp2\DYM\app{s+1}-2(s+2)\dotuline{\tp{s+1}\DYM \app2}+3(s+1)\app{s+1} D\tp2+2(s+2)\app{s+2}D\tp1\Big\} \nn\\
    &-D\t0\Big\{(s+1)\app{s+1}D\tp1-2\tp1\DYM\app{s+1}\Big\}+CD\t0\Big\{- 3(s+1)\app{s+1}\tp2-2(s+2)\app{s+2}\tp1 \Big\} \nn\\
    &+DC\t0\Big\{2(s+2)\dotuline{\app2\tp{s+1}}- 3(s+1)\app{s+1}\tp2-2(s+2)\app{s+2}\tp1 \Big\} \label{LeibnizYMBig}\\
    &+D^2\t0\Big\{-2(s+1)\app{s+1}\tp1+(s+2)\tp{s+1}\app{1}\Big\}+(s+2)C\tp{s+1}\Big\{\dotuline{ \t0\DYM\app2}-3\dotuline{C\t0\app3}-\dotuline{3F\t0 \tpp2}\Big\} \nn\\
    &+2(s+2)\dotuline{F\t0\tp{s+1}D\tpp1} +2(s+2)\dotuline{\DYM F\t0 \tp{s+1}\tpp1}+2(s+2)FD\t0 \tp{s+1}\tpp1 \nn\\
    &-(s+2)\dotuline{\t0\tp{s+1}[F,\app0}]_\g-\sum_{n=0}^{s-1}(n+2)\t0\tp{n+1}\big[F, \app{s-n}\big]_\g\bigg)-\bPtap\leftrightarrow\bPtapp\bigg). \nn
\end{align}
In order to simplify this last expression, we compute (using the dual EOM \eqref{DualEOMEYM})
\begin{align}
    \dPta F=\pa_u\big(\dPta A\big) &=-\DYM^2\a1+3C\DYM\a2+ 2DC\a2-3C^2\a3-[F,\a0]_\g+2\DYM F\t1 \nn\\
    &+3FD\t1+N\a1-6CF\t2+\dot F\t0.
\end{align}
The dot-underlined terms then form the following combination:
\begin{equation}
    \dotuline{~\cdots~}=(s+2)\t0\tp{s+1}\Big(\dbPtapp F-N\app1-FD\tpp1+3CF\tpp2\Big).
\end{equation}
Hence, comparing with \eqref{LeibnizYMinterm} and grouping terms proportional respectively to $\app{s+1}, \DYM\app{s+1}$, $\app{s+2}$ and $\tpp{s+1}$, we deduce that

\begin{align}
    &\quad\Big(\big[\Bt\bPtap,\bPtapp\big]+ \big[\bPtap,\Bt\bPtapp\big]\Big)^\ym_s=\big(\Bt\big[\bPtap,\bPtapp\big]\big)^\ym_s+ \nn\\*
    &+\!\bigg(\!\bigg(\!\!-\!\Big(\underbrace{\t0D\tp1-2\tp1D\t0-3C\t0\tp2}_{\textrm{$\big(\rBt\bPtap\big)_0$}}\Big)\DYM\app{s+1}+(s+2)\app{s+2}\Big(\!\underbrace{-2C\t0D \tp1-2C\tp1D\t0-2DC\t0\tp1}_{\textrm{$-\t0\big(3CD\tp1+2DC\tp1-3C^2\tp2\big)+C\big(\rBt\bPtap\big)_0$}}\Big) \nn\\
    &+(s+1)\app{s+1}\Big\{\underbrace{\t0D^2\tp1-2\tp1D^2\t0-D\t0D\tp1-3C\t0D\tp2-3D(C\t0)\tp2}_{\textrm{$D\big(\rBt\bPtap\big)_0$}}\Big\} \nn\\
    &+(s+2)\tpp{s+1}\Big\{\underbrace{N\t0\ap1 +F\t0D\tp1-3CF\t0\tp2- \ap1D^2\t0-2F\tp1 D\t0}_{\textrm{$\ap1\big(N\t0-D^2\t0\big)+F\big(\rBt\bPtap\big)_0$}} \Big\}\\
    &-\sum_{n=0}^{s-1}(n+2)\t0\tp{n+1}\big[F, \app{s-n}\big]_\g+\t0 \big[\hdap A,\app{s+1}\big]_\g+(s+2)\t0\tp{s+1}\dbPtapp F\bigg)-\bPtap\leftrightarrow\bPtapp\bigg). \nn
\end{align}
As expected, we recognize the combination of actions
\begin{align}
    \big((\rBt\bPtap)\blacktriangleright \bPtapp\big)^\ym_s &=\big(\rBt\bPtap\big)_0 \DYM\app{s+1}-(s+1)\app{s+1}D\big(\rBt \bPtap\big)_0 \nn\\
    &-(s+2)C\big(\rBt\bPtap\big)_0 \app{s+2}-(s+2)F\big(\rBt\bPtap\big)_0 \tpp{s+1}.
\end{align}
Therefore the Leibniz rule anomaly for the action $\blacktriangleright$ projected onto $\sfS$ is given by \eqref{LeibnizAnomalyBlackTauActionYM}
\begin{empheq}[box=\fbox]{align}
    \cA_s^\ym\Big(\big[\bPtap,\bPtapp \big], \Bt\!\Big) &=\bigg\{\big((\rBt\bPtap)\blacktriangleright \bPtapp\big)^\ym_s+(s+2)\t0\app{s+2}\dbtp C-(s+2)\ap1\tpp{s+1}\odt C \nn\\
    &\quad -\t0 \big[\hdap A,\app{s+1}\big]_\g+(s+2)\t0\tpp{s+1}\dbPtap F \\
    &\quad +\sum_{n=1}^{s}(n+1)\tp{n}\big[\odt A, \app{s+1-n}\big]_\g\bigg\}-\bPtap\leftrightarrow\bPtapp, \nn
\end{empheq}
where we used that $\odt A=F\t0$.

\subsection{Homomorphism anomaly \label{AppEYM:BlackTauHom}}

From the definitions \eqref{defBlackActionGR}-\eqref{defBlackActionYM} of the action $\blacktriangleright$, we immediately get that the action of the commutator $\big[\ot,\otp\big]^\circ$ vanishes,
\begin{equation}
    \big[\ot,\otp\big]^\circ \blacktriangleright \bPtapp=(0,0).
\end{equation}
We then compute the commutator of the actions.
We give all the details for the YM sub-space part, since the GR part is analogous.
\begin{align}
    \big(\Bt\, &(\Btp\bPtapp)\big)^\ym_s =\nn\\
    &=\t0\DYM \big(\Btp\bPtapp\big)^\ym_{s+1}-(s+1)\big(\Btp\bPtapp\big)^\ym_{s+1}D\t0-(s+2)C\t0\big(\Btp\bPtapp\big)^\ym_{s+2} \nn\\
    &-(s+2)F\t0 \big(\Btp\bPtapp\big)^\gr_{s+1} \nn\\
    &=\t0\DYM\Big(\tp0\DYM\app{s+2}-(s+2)\app{s+2}D\tp0-(s+3)C\tp0\app{s+3}-(s+3)F\tp0\tpp{s+2}\Big) \nn\\*
    &-(s+1)D\t0\Big(\tp0\DYM\app{s+2}-(s+2)\app{s+2}D\tp0-(s+3)C\tp0\app{s+3}-(s+3)F\tp0\tpp{s+2}\Big) \nn\\*
    &-(s+2)C\t0\Big(\tp0\DYM\app{s+3}-(s+3)\app{s+3}D\tp0-(s+4)C\tp0\app{s+4}-(s+4)F\tp0\tpp{s+3}\Big) \nn\\*
    &-(s+2)F\t0\Big(\tp0D\tpp{s+2}-(s+3)\tpp{s+2}D\tp0-(s+4)C\tp0\tpp{s+3}\Big).
\end{align}
After anti-symmetrization in $\ot,\otp$, we obtain that
\begin{align}
    &\quad\Big(\Bt\big(\Btp\bPtapp\big)-\Btp\big(\Bt\bPtapp\big)\Big)^\ym_s =-\big(\ot\leftrightarrow\otp\big)\,+\nn\\
    &+\Big\{\underline{\t0D\tp0\DYM\app{s+2}}-(s+2)\underline{\t0D\tp0\DYM\app{s+2}}-(s+2)\app{s+2}\t0D^2\tp0-(s+3)C\uwave{\app{s+3}\t0D\tp0} \nn\\
    &-(s+3)F\dashuline{\tpp{s+2}\t0D\tp0}-(s+1)\underline{\tp0D\t0\DYM\app{s+2}}+(s+1)(s+3)C\uwave{\tp0 D\t0\app{s+3}} \\
    &+(s+1)(s+3)F\dashuline{\tpp{s+2}\tp0D\t0}+(s+2)(s+3)C\uwave{\app{s+3}\t0D\tp0}+(s+2)(s+3)F\dashuline{\tpp{s+2}\t0D\tp0}\Big\}. \nn
\end{align}
The various underlined terms cancel each other, so that we are left with\footnote{The term $N\t0$ in $\odt C$ does not contribute because of the anti-symmetrization. We can thus include it for free.}
\begin{equation}
    \boxed{\Big(\big[\ot,\otp\big]^\circ \blacktriangleright\bPtapp-\big[\Bt,\Btp\!\big]\bPtapp\Big)^\ym_s=(s+2)\app{s+2}\Big(\tp0\odt C-\t0\odtp C\Big)}.
\end{equation}
Similarly, the projection onto the $\bsfT$-subspace results in \eqref{HomAnomalyBlackTauActionGR}
\begin{equation}
    \boxed{\Big(\big[\ot,\otp\big]^\circ \blacktriangleright\bPtapp-\big[\Bt,\Btp\!\big]\bPtapp\Big)^\gr_s=(s+3)\tpp{s+2}\Big(\tp0\odt C-\t0\odtp C\Big)}.
\end{equation}

\section{Properties of the action $\blacktriangleleft$}

\subsection{Leibniz rule anomaly \label{AppEYM:BackBlackTauLeibniz}}

Since $\big[\otp,\otpp\big]^\circ$ is in the center of $\cST$, i.e. is non-zero only at degree $-1$, we readily see from the definitions \eqref{defReverseBlackAction-1}-\eqref{defReverseBlackAction0} that
\begin{equation}
    \big[\otp,\otpp\big]^\circ \blacktriangleleft\bPta =(0,0).
\end{equation}
Moreover, since $\big[\rBtp\bPta,\otpp\big]_0^\circ=0$, we trivially get that the Leibniz rule anomaly at degree 0 vanishes:
\begin{equation}
    \boxed{\cA_0\Big(\big[\otp,\otpp \big]^\circ, \blacktriangleleft\bPta \Big)=0}.
\end{equation}
In the following, we will also need to calculate
\begin{align}
    \big(\rBtp(\Btpp\bPta)\big)_0 &=\tp0D\big(\Btpp\bPta\big)^\gr_1-2\big(\Btpp\bPta\big)^\gr_1 D\tp0-3C\tp0 \big(\Btpp\bPta\big)^\gr_2 \nn\\
    &=\tp0D\big(\tpp0D\t2-3\t2D\tpp0-4C\tpp0\t3\big)-2D\tp0\big(\tpp0D\t2-3\t2D\tpp0-4C\tpp0\t3\big) \nn\\*
    &-3C\tp0\big(\tpp0D\t3-4\t3D\tpp0-5C\tpp0\t4\big),
\end{align}
which upon anti-symmetrization in $\otp,\otpp$ simplifies to\footnote{In the last step we added $N\tp0\tpp0-N\tpp0\tp0=0$.}
\begin{align}
    \big(\rBtp(\Btpp\bPta)\big)_0-\big(\rBtpp(\Btp\bPta)\big)_0 &=3\t2\tpp0D^2\tp0-3\t2\tp0D^2\tpp0 \nn\\
    &=3\t2\big(\tp0\odtpp C-\tpp0\odtp C\big). \label{IntermImportant}
\end{align}

Concerning the degree $-1$, we compute
\begin{align}
    &\quad~\big[\rBtp\bPta,\otpp\big]_{- 1}^\circ+ \big[\otp,\rBtpp\bPta\big]_{- 1}^\circ =\Big(\big(\rBtp\bPta\big)_0 D\tpp0-\tpp0 D\big(\rBtp\bPta\big)_0\Big)-\otp\leftrightarrow\otpp \nn\\
    &=\Big(\big(\tp0D\t1-2\t1D\tp0-3C\tp0\t2\big)D\tpp0+\tp0D\big(\tpp0D\t1-2\t1D\tpp0-3C\tpp0\t2\big)\Big)-\otp\leftrightarrow\otpp \nn\\
    &=\Big(-2\tp0\t1D^2\tpp0-6C\tp0\t2D\tpp0\Big)-\otp\leftrightarrow\otpp.
\end{align}
Now notice that
\begin{align}
    \big(\rBtp(\Btpp\bPta)\big)_{-1}=-2C\tp0 \big(\Btpp\bPta\big)_1^\gr=-2C\tp0\big( \tpp0D\t2-3\t2D\tpp0-4C\tpp0\t3\big),
\end{align}
which upon anti-symmetrization reduces to
\begin{equation}
    \big(\rBtp(\Btpp\bPta)\big)_{-1}-\big(\rBtpp(\Btp\bPta)\big)_{-1}=6C\tp0\t2 D\tpp0-6C\tpp0\t2 D\tp0.
\end{equation}
Consequently,
\begin{equation}
    \big[\rBtp\bPta,\otpp\big]_{-1}^\circ+ \big[\otp,\rBtpp\bPta\big]_{-1}^\circ =\Big(2\tp0\t1\odtpp C-\big(\rBtp(\Btpp\bPta)\big)_{-1}\Big)-\otp\leftrightarrow\otpp,
\end{equation}
where $\odt C=-D^2\t0+N\t0$.
The Leibniz rule anomaly at degree $-1$ is thus \eqref{LeibnizAnomalyrBlackTauAction}
\begin{equation}
    \boxed{\cA_{-1}\Big(\big[\otp,\otpp \big]^\circ, \blacktriangleleft\bPta \Big)=\Big\{ \big(\rBtp(\Btpp\bPta)\big)_{-1}+2\t1\tpp0 \odtp C\Big\}-\otp\leftrightarrow\otpp}.
\end{equation}

\subsection{Homomorphism anomaly \label{AppEYM:BackBlackTauHom}}

To check that the back reaction $\blacktriangleleft$ is a homomorphism of Lie algebras, we first calculate
\begin{align}
    \big((\rBtpp\bPta)\blacktriangleleft\bPtap \big)_{-1}-\big((\rBtpp\bPtap) \blacktriangleleft\bPta \big)_{-1} &=-2C\tp1\big(\rBtpp\bPta\big)_0-\bt\leftrightarrow\btp \nn\\
    &=-2C\tp1\big(\tpp0D\t1-2\t1D\tpp0-3C\tpp0\t2\big)-\bt\leftrightarrow\btp \nn\\
    &=-2\tpp0\tp1\big(CD\t1-3C^2\t2\big)-\bt\leftrightarrow\btp.
\end{align}
On the other hand,
\begin{equation}
    \big(\rBtpp[\bPta,\bPtap]\big)_{-1}= -2C\tpp0\overline{[\bt,\btp]}_1-\bt\leftrightarrow\btp=-4C\tpp0\t1D\tp1-\bt\leftrightarrow\btp.
\end{equation}
Therefore,\footnote{We can include the term $2DC\t1$ appearing in $\dbt C$ since the anti-symmetrization cancels it.}
\begin{equation} \label{HomAnomaly-1interm}
    \boxed{\Big(\rBtpp[\bPta,\bPtap]-\otpp\big[\!\blacktriangleleft\bPta, \blacktriangleleft\bPtap\big] \Big)_{-1}= 2\tpp0\big(\tp1\dbt C-\t1\dbtp C\big)}.
\end{equation}

At degree 0, we have that
\begin{align}
    \big((\rBtpp\bPta)\blacktriangleleft\bPtap \big)_0=\big(\rBtpp\bPta\big)_0D\tp1-2\tp1 D\big(\rBtpp\bPta\big)_0-3C\tp2\big(\rBtpp\bPta\big)_0.
\end{align}
Anti-symmetrizing between $\bPta$ and $\bPtap$ and using that
\begin{equation}
    \big(\rBtpp\bPta\big)_0=\tpp0D\t1-2\t1D\tpp0-3C\t2\tpp0,
\end{equation}
we get after some simple algebra that
\begin{equation}
    \big(\otpp\big[\!\blacktriangleleft\bPta, \blacktriangleleft\bPtap\big]\big)_0=\Big(-2\tpp0\tp1D^2\t1+4\tp1D\t1D\tpp0+6DC\tp1\t2\tpp0+6C\tp1D\t2\tpp0\Big)-\bt\leftrightarrow\btp.
\end{equation}
Since
\begin{align}
    \big(\rBtpp[\bPta,\bPtap]\big)_0 &=\tpp0 D\overline{[\bt,\btp]}_1-2\overline{[\bt,\btp]}_1 D\tpp0-3C\tpp0\overline{[\bt,\btp]}_2 \\
    &=\Big(2\tpp0\t1D^2\tp1-4\t1D\tp1D\tpp0-3C\tpp0\big(2\t1D\tp2+3\t2D\tp1\big)\Big)-\bt\leftrightarrow\btp, \nn
\end{align}
we infer that
\begin{equation}
    \Big(\rBtpp[\bPta,\bPtap]-\otpp\big[\!\blacktriangleleft\bPta, \blacktriangleleft\bPtap\big] \Big)_0= -3\tpp0\t2\big(2DC\tp1+3CD\tp1\big)-\bt\leftrightarrow \btp.
\end{equation}
The latter recombines with \eqref{HomAnomaly-1interm} such that for $s=-1,0$, one gets \eqref{HomAnomalyrBlackTauAction}
\begin{equation} 
    \boxed{\Big(\rBtpp[\bPta,\bPtap]-\otpp\big[\!\blacktriangleleft\bPta, \blacktriangleleft\bPtap\big] \Big)_{s}= (s+3)\tpp0\big(\tp{s+2}\dbt C-\t{s+2}\dbtp C\big)}.
\end{equation}

\section{Proof of the Jacobi identity anomaly \label{AppEYM:Jacobi}}

We compute the Jacobi identity anomaly for the YM-bracket. 
We first evaluate $\big[\Pta,[\Ptap,\Ptapp]\big]$ by splitting $\Pta$ as $(\ot,\bPta)$:
\begin{align}
    \big[\Pta,[\Ptap,\Ptapp]\big] &=\Big[ (\ot,\bPta),\big[(\otp,\bPtap),(\otpp,\bPtapp)\big]\Big] \nn\\
    &=\bigg(\Big[\ot,\big[\otp,\otpp \big]^\circ+\rBtp\bPtapp-\rBtpp\bPtap\Big]^\circ +\rBt\Big(\big[ \bPtap,\bPtapp\big]+\Btp\bPtapp-\Btpp\bPtap\Big) \nn\\
    &\qquad -\Big(\big[\otp,\otpp \big]^\circ+\rBtp\bPtapp-\rBtpp\bPtap\Big) \blacktriangleleft\bPta\,, \\
    &\qquad\Big[\bPta,\big[\bPtap,\bPtapp \big]+ \Btp\bPtapp-\Btpp\bPtap\Big]+\Bt \Big(\big[\bPtap,\bPtapp \big]+ \Btp\bPtapp-\Btpp\bPtap\Big) \nn\\
    &\qquad -\Big(\big[\otp,\otpp \big]^\circ+\rBtp\bPtapp-\rBtpp\bPtap\Big) \blacktriangleright\bPta\bigg). \nn
\end{align}
Using the symbol $\cyc$ to denote an equality valid upon cyclic permutation of both the LHS and the RHS, we conveniently rearrange the terms as follows:
\begin{align}
    \big[\Pta,[\Ptap,\Ptapp]\big] &\cyc \bigg(\cancel{\big[\ot,[\otp,\otpp]^\circ\big]^\circ}+\big[\otp, \rBtpp\bPta\big]^\circ+\big[ \rBtp\bPta,\otpp\big]^\circ+\rBtpp\big[\bPta,\bPtap\big]+\rBtp\big(\Btpp\bPta\big) \nn\\
    &\qquad -\rBtpp\big(\Btp\bPta\big)-[\otp,\otpp]^\circ\blacktriangleleft\bPta-\big(\rBtpp\bPta\big)\blacktriangleleft \bPtap+\big(\rBtpp\bPtap\big)\blacktriangleleft \bPta\,, \nn\\
    &\qquad\big[\bPta,[\bPtap,\bPtapp]\big]-\big[\Bt\bPtap,\bPtapp\big]-\big[\bPtap,\Bt\bPtapp\big]+\Bt\big[\bPtap,\bPtapp\big]+\Bt\big(\Btp\bPtapp\big) \nn\\
    &\qquad -\Btp\big(\Bt\bPtapp\big)-\big[\ot,\otp\big]^\circ\blacktriangleright\bPtapp-\big(\rBt\bPtap\big) \blacktriangleright \bPtapp+ \big(\rBt\bPtapp\big)\blacktriangleright \bPtap\bigg) \\
    &\hspace{-2cm}\cyc\bigg(\!-\cA\Big(\big[\otp,\otpp \big]^\circ, \blacktriangleleft\bPta\Big) +\rBtp\big(\Btpp\bPta\big)-\rBtpp\big(\Btp\bPta\big)+\otpp\Big(\! \blacktriangleleft\big[\bPta,\bPtap\big]-\big[\!\blacktriangleleft\bPta, \blacktriangleleft \bPtap\big]\Big)\,,\nn\\
    &\hspace{-2cm} \big[\bPta,[\bPtap,\bPtapp]\big]+\cA\Big(\big[\bPtap,\bPtapp\big],\Bt\!\!\Big)-\big(\rBt\bPtap\big) \blacktriangleright \bPtapp+ \big(\rBt\bPtapp\big)\blacktriangleright \bPtap-\Big(\big[\ot,\otp\big]^\circ\! \blacktriangleright-\big[\Bt,\Btp\!\big]\Big)\bPtapp\bigg). \nn
\end{align}
Since the cyclic permutation of $\big[\bPta,[\bPtap,\bPtapp]\big]$ is similarly given by
\begin{align}
    \big[\bPta,[\bPtap,\bPtapp]\big] &=\Big[(\bt,\a{}),\big[(\btp,\ap{}),(\btpp,\app{})\big]\Big] \\
    &\cyc \bigg(\overline{\big[\bt,\overline{[\btp,\btpp]}\big]},\big[\a{},[\ap{},\app{}]^\g\big]^\g+ \cA\Big(\big[\ap{},\app{}\big]^\g,\Wt\!\Big)-\Big(\overline{[\bt,\btp]} \vartriangleright-\big[\Wt,\Wtp\!\big]\Big)\app{}\bigg),\nn
\end{align}
we can gather all the properties of the actions $\vartriangleright, \blacktriangleright$ and $\blacktriangleleft$ to deduce the Jacobi identity anomaly of the $\sfST$-bracket.
More precisely, using (\ref{LeibnizAnomalyBlackTauActionGR}-\ref{LeibnizAnomalyrBlackTauAction}-\ref{HomAnomalyBlackTauActionGR}-\ref{HomAnomalyrBlackTauAction}) and \eqref{IntermImportant}, we get that\footnote{The reader can check that $\overline{\big[\bt,\overline{[\btp,\btpp]}\big]}\cyc 0$. 
Of course, since we already computed the Jacobi identity anomaly of the $C$-bracket in \cite{Cresto:2024fhd} and also in \cite{Cresto:2024mne}, we know that its restriction to the $\bsfT$-sub-space vanishes (the anomaly involving the degree 0 elements $\t0,\tp0,\tpp0$, cf. \eqref{JacobiAnomalyCbracketEYMinterm}).}
\begin{align}
    \big[\Pta,[\Ptap,\Ptapp]\big]^\gr_{-1} &\cyc 2\tpp0\big(\tp1\dbt C-\t1\dbtp C\big)-2\t1\big(\tpp0\odtp C-\tp0\odtpp C\big) \nn\\
    \big[\Pta,[\Ptap,\Ptapp]\big]^\gr_0 &\cyc 3\tpp0\big(\tp2\dbt C-\t2\dbtp C\big)+3\t2\big(\tp0\odtpp C-\tpp0\odtp C\big) \label{JacobiAnomalyCbracketEYMinterm}\\
    \big[\Pta,[\Ptap,\Ptapp]\big]^\gr_s &\cyc (s+3)\t0\big(\tpp{s+2}\dbtp C-\tp{s+2}\dbtpp C\big)-(s+3)\tpp{s+2}\big(\tp0\odt C-\t0\odtp C\big), \qquad  s\geqslant 1, \nn
\end{align}
which thanks to the cyclic permutation is more compactly written for all $s\geq -1$ as
\begin{equation}
    \boxed{\big[\Pta,[\Ptap,\Ptapp]\big]^\gr_s \cyc-\dt C\paren{\tp{},\tpp{}}^\gr_s}.
\end{equation}
Besides, collecting (\ref{LeibnizAnomalyTauAction}-\ref{LeibnizAnomalyBlackTauActionYM}-\ref{HomAnomalyTauAction}-\ref{HomAnomalyBlackTauActionYM}) we obtain that the YM projection of the Jacobi anomaly of the $\sfST$-bracket takes the form\footnote{Using that $\big[\a{},[\ap{},\app{}]^\g\big]^\g\cyc 0$, cf. \cite{Cresto:2025bfo}.}
\begin{align}
    \big[\Pta,[\Ptap,\Ptapp]\big]^\ym_s&\cyc\bigg\{\!\bigg\{ (s+2)\t0\app{s+2}\dbtp C-(s+2)\ap1\tpp{s+1}\odt C -\t0 \big[\hdap A,\app{s+1}\big]_\g\!+(s+2)\t0\tpp{s+1}\dbPtap F \nn\\
    &\quad +\sum_{n=1}^{s}(n+1)\tp{n}\big[\odt A, \app{s+1-n}\big]_\g-\sum_{n=1}^s(n+1)\t{n}\big[\hdap A,\app{s+1-n}\big]_\g\bigg\}-\bPtap\leftrightarrow\bPtapp\bigg\} \nn\\
    &\quad -(s+2)\app{s+2}\Big(\tp0\odt C-\t0\odtp C\Big)-(s+2)\app1\Big(\t{s+1} \dbtp C-\tp{s+1}\dbt C\Big).
\end{align}
Simplifying and permuting cyclically some of the terms, we get
\begin{empheq}[box=\fbox]{align}
    \big[\Pta,[\Ptap,\Ptapp]\big]^\ym_s &\cyc \sum_{n=0}^s(n+1)\Big(\tp{n}\big[\dPta A,\app{s+1-n}\big]_\g-\tpp{n}\big[\dPta A,\ap{s+1-n}\big]_\g\Big) \\
    &-(s+2)\dt C\Big(\tp0\app{s+2}-\tpp0\ap{s+2}+ \ap1\tpp{s+1}- \app1\tp{s+1}\Big)-\dbPta F\paren{\tp{}, \tpp{}}^\ym_s,  \nn
\end{empheq}
where we used the fact that
\begin{align}
    &\bigg\{\!-\sum_{n=1}^{s}(n+1)\tpp{n}\big[\odt A, \ap{s+1-n}\big]_\g-\sum_{n=1}^s(n+1)\t{n}\big[\hdap A,\app{s+1-n}\big]_\g-\t0 \big[\hdap A,\app{s+1}\big]_\g\bigg\}-\bPtap\leftrightarrow\bPtapp \nn\\
    &\cyc -\sum_{n=0}^s(n+1) \t{n}\big[\dPtap A,\app{s+1-n}\big]_\g-\Ptap\leftrightarrow\Ptapp.
\end{align}
Finally, notice that we can swap $\dbPta F$ for $\dPta F$ since $\t0\paren{\tp{},\tpp{}}^\ym\cyc 0$, thence getting \eqref{JacobiEYMYMbracket}.

\section{Properties of the algebroid actions}

\subsection{Leibniz rule \label{AppEYM:DotWhiteLeibniz}}

\paragraph{Action} \!\!$\dotvartriangleright\,$:
First notice that
\begin{equation}
    \big(\Wt(\hdapp\ap{})\big)_s=\hdapp\big( \Wt\ap{}\big)_s-\sum_{n=1}^s (n+1)\t{n}\big[\hdapp A,\ap{s+1-n}\big]_\g.
\end{equation}
Moreover we have that
\begin{equation}
    \big\lbr\Wt\ap{},\app{}\big\rbr^\g =\big[\Wt\ap{} ,\app{}\big]^\g +\hdapp\big(\Wt\ap{}\big)-\hat\delta_{\bt\mkern1mu \vartriangleright\mkern1mu\ap{}}\app{}.
\end{equation}
Using \eqref{LeibnizAnomalyTauAction}, this directly leads to
\begin{equation}
    \cA\big(\lbr\ap{},\app{}\rbr^\g ,\Wt\!\big)=\cA\big([ \ap{},\app{}]^\g ,\Wt\!\big)+\cA\big(\hdapp\ap{}-\hdap\app{},\Wt\!\big)= \hat\delta_{\bt\mkern1mu\vartriangleright \mkern1mu\ap{}}\app{}-\hat\delta_{\bt\mkern1mu \vartriangleright\mkern1mu\app{}}\ap{}.
\end{equation}
From \eqref{defActionDot2}, we also know that
\begin{equation}
    \cA\big(\lbr\ap{},\app{}\rbr^\g,\dWt\!\big)= \cA\big(\lbr\ap{},\app{}\rbr^\g,\Wt\!\big)- \cA\big(\lbr\ap{},\app{}\rbr^\g,\dbt\big),
\end{equation}
where
\begin{align}
    \cA\big(\lbr\ap{},\app{}\rbr^\g,\dbt\big) &=\Big(\dbt\hdapp\ap{}-\hdapp\dbt\ap{}+\hat\delta_{\dbt\ap{}}\app{}\Big)-\ap{}\leftrightarrow\app{} \nn\\
    &=\Big(-\hat\delta_{\bt\mkern1mu\vartriangleright \mkern1mu\app{}-\dbt\app{}}\ap{} +\hat\delta_{\dbt\ap{}}\app{}\Big)-\ap{}\leftrightarrow\app{} \\
    &=\hat\delta_{\bt\mkern1mu\vartriangleright \mkern1mu\ap{}}\app{}-\hat\delta_{\bt\mkern1mu\vartriangleright \mkern1mu\app{}}\ap{}. \nn
\end{align}
In the second line, we used the morphism property \eqref{AlgebroidActionEYM} of the anchor map, i.e.
\begin{equation}
    \big[\dbt,\hdapp\big]\ap{}=-\delta_{\lbr(0;\bt,0),(0;0,\app{})\rbr}\ap{},
\end{equation}
where the only non-vanishing part of $\lbr(0;\bt,0),(0;0,\app{})\rbr$ is in the YM subspace and equals
\begin{equation}
    \lbr(0;\bt,0),(0;0,\app{})\rbr^\ym=\dWt\app{}= \Wt\app{}-\dbt\app{}.
\end{equation}
Therefore, cf.\,\eqref{dotWhiteProperties},
\begin{equation}
    \boxed{\cA\big(\lbr\ap{},\app{}\rbr^\g ,\dWt\!\big) =0}.
\end{equation}

\paragraph{Action} \!\!$\dotblacktriangleleft\,$:
On the one hand,
\begin{align}
    \big\lbr\otp,\otpp\big\rbr^\circ\, \dotblacktriangleleft\,\bPta &=\big\lbr\otp,\otpp\big\rbr^\circ\!\blacktriangleleft \bPta+\dbt\big\lbr\otp,\otpp\big\rbr^\circ \\
    &=\big[\otp,\otpp\big]^\circ\! \blacktriangleleft\bPta +\big(\odtpp\otp-\odtp\otpp\big)\blacktriangleleft\bPta+ \dbt\big[\otp,\otpp\big]^\circ+\dbt\odtpp\otp-\dbt\odtp\otpp. \nn
\end{align}
On the other hand, 
\begin{align}
    \big\lbr\drBtp\bPta,\otpp\big\rbr^\circ&=\big[ \rBtp\bPta,\otpp\big]^\circ+\big[\dbt\otp,\otpp\big]^\circ+\odtpp\big(\rBtp\bPta\big)+\odtpp\dbt \otp-\delta_{\text{\scalebox{1.1}{$\ensurestackMath{\stackon[-1.9pt]{\tau'}{\mkern-4mu\scriptscriptstyle{\circ}}}$}} \scriptdotblacktriangleleft\,\bPta}\otpp.
\end{align}
We can now use the fact that, for $s=-1,0$,
\begin{equation}
    \odtpp\big(\rBtp\bPta\big)_s=\big((\odtpp\otp) \blacktriangleleft\bPta\big)_s+\big(\rBtp(\odtpp\bPta) \big)_s-(s+3)\tp0\t{s+2}\odtpp C,
\end{equation}
so that
\begin{align}
    \cA_s\Big(\big\lbr\otp,\otpp\big\rbr^\circ, \dotblacktriangleleft\,\bPta\Big) &=\cA_s\Big(\big[\otp,\otpp\big]^\circ, \blacktriangleleft\bPta\Big)+\cA_s\Big(\big[\otp,\otpp\big]^\circ, \dbt\Big)+\Big\{\Big\{\big[ \dbt, \odtpp\big]\otp_{\!\!s} \\
    & -\delta_{\text{\scalebox{1.1}{$\ensurestackMath{\stackon[-2pt]{\tau''}{\mkern-9mu\scriptscriptstyle{\circ}}}$}} \scriptdotblacktriangleleft\,\bPta}\otp_{\!\!s} -\big(\rBtp(\odtpp\bPta) \big)_s+(s+3)\tp0\t{s+2}\odtpp C\Big\}-\otp\leftrightarrow\otpp\Big\}. \nn
\end{align}
Note that for the $\circ$-bracket, $\cA\big([\otp,\otpp]^\circ,\dbt\big)=0$.
We then have that the $\sfST$-bracket resulting from $[\dbt,\odtpp]$ is
\begin{equation}
    \big\lbr(0,\bPta),(\otpp,0)\big\rbr=-\Big( \drBtpp\bPta,\dBtpp\bPta\Big),
\end{equation}
such that
\begin{equation}
    \big[\dbt,\odtpp\big]\otp= \delta_{\text{\scalebox{1.1}{$\ensurestackMath{\stackon[-2pt]{\tau''}{\mkern-9mu\scriptscriptstyle{\circ}}}$}} \scriptdotblacktriangleleft\,\bPta}\otp+ \delta_{(\text{\scalebox{1.1}{$\ensurestackMath{\stackon[-2pt]{\tau''}{\mkern-9mu\scriptscriptstyle{\circ}}}$}} \scriptdotblacktriangleright\,\bPta)^\gr}\otp.
\end{equation}
Next recall that $\cA_0\big([\otp,\otpp]^\circ, \blacktriangleleft\bPta\big)=0$.
Hence,
\begin{align}
    \cA_0\Big(\big\lbr\otp,\otpp\big\rbr^\circ, \dotblacktriangleleft\,\bPta\Big) &=\Big\{ \delta_{(\text{\scalebox{1.1}{$\ensurestackMath{\stackon[-2pt]{\tau''}{\mkern-9mu\scriptscriptstyle{\circ}}}$}} \scriptdotblacktriangleright\,\bPta)^\gr}\otp_{\!\!0}-\big(\rBtp(\odtpp\bPta) \big)_0+3\tp0\t{2}\odtpp C\Big\}-\otp\leftrightarrow\otpp.
\end{align}
Comparing with
\begin{equation}
    \drBtp\big(\dBtpp\bPta\big)=\rBtp\big(\Btpp \bPta-\odtpp\bPta\big)+ \delta_{(\text{\scalebox{1.1}{$\ensurestackMath{\stackon[-2pt]{\tau''}{\mkern-9mu\scriptscriptstyle{\circ}}}$}} \scriptdotblacktriangleright\,\bPta)^\gr}\otp
\end{equation}
and using \eqref{IntermImportant}, we see that
\begin{equation}
    \cA_0\Big(\big\lbr\otp,\otpp \big\rbr^\circ, \dotblacktriangleleft\, \bPta\Big)=\Big(\drBtp\big(\dBtpp\bPta\big)-\drBtpp\big(\dBtp\bPta\big) \Big)_0.
\end{equation}
Concerning the degree $-1$, we use the Leibniz rule anomaly \eqref{LeibnizAnomalyrBlackTauAction} and get
\begin{align}
    \cA_{-1}\Big(\big\lbr\otp,\otpp \big\rbr^\circ, \dotblacktriangleleft\, \bPta\Big) &=\Big\{ \big(\rBtp(\Btpp\bPta)\big)_{-1} -\cancel{2\tp0\t1\odtpp C} \\
    & +\delta_{(\text{\scalebox{1.1}{$\ensurestackMath{\stackon[-2pt]{\tau''}{\mkern-9mu\scriptscriptstyle{\circ}}}$}} \scriptdotblacktriangleright\,\bPta)^\gr}\otp_{\!\!-1}-\big(\rBtp(\odtpp\bPta) \big)_{-1}+\cancel{2\tp0\t1\odtpp C}\Big\}-\otp\leftrightarrow\otpp. \nn
\end{align}
We are therefore left with the same result both at degree $-1$ and 0, namely \eqref{dotBlackProperties}
\begin{equation}
    \boxed{\cA\Big(\big\lbr\otp,\otpp \big\rbr^\circ, \dotblacktriangleleft\, \bPta\Big)=\drBtp\big(\dBtpp\bPta\big)-\drBtpp\big(\dBtp\bPta\big)}.
\end{equation}

\subsection{Homomorphism \label{AppEYM:DottauHom}}

\paragraph{Action} \!\!$\dotvartriangleright\,$:
We first compute
\begin{equation}
    \overline{\lbr\bt,\btp\rbr} \,\dotvartriangleright\, \app{}=\overline{[\bt,\btp]} \vartriangleright \app{}-\delta_{\mkern3mu\overline{\lbr\bt,\btp\rbr} \mkern1mu} \app{}+\big(\dbtp\bt\big) \vartriangleright \app{}- \big(\dbt\btp\big)\vartriangleright \app{}.
\end{equation}
Next
\begin{align}
    \big[\dWt,\dWtp\!\big]\app{} =\Big\{\Wt \big(\Wtp\app{}\big)-\Wt\big(\dbtp\app{}\big) -\dbt\big(\Wtp\app{}\big)+\dbt\dbtp\app{}\Big\} -\bt\leftrightarrow\btp.
\end{align}
Using that\footnote{Notice that $\dbtp A=0$.}
\begin{equation}
    \big(\dbtp(\Wt\app{})\big)_s=\big((\dbtp\bt) \vartriangleright\app{}\big)_s+\big(\Wt(\dbtp \app{})\big)_s+(s+2)\dbtp C\t{s+1}\app1,
\end{equation}
and the fact that
\begin{equation}
    \big[\dbt,\dbtp\big]\app{}=-\delta_{\mkern3mu\overline{\lbr\bt,\btp\rbr} \mkern1mu}\app{},
\end{equation}
we infer that the commutator of the dotted action reduces to
\begin{align}
    \Big(\big[\dWt,\dWtp\!\big]\app{}\Big)_s &=\Big(\big[\Wt,\Wtp\!\big]\app{}\Big)_s-\delta_{\mkern3mu\overline{\lbr\bt,\btp\rbr} \mkern1mu}\app{s} \\
    &+\Big\{\Big\{\big((\dbtp\bt) \vartriangleright\app{}\big)_s+(s+2)\dbtp C\t{s+1}\app1\Big\}-\bt\leftrightarrow\btp\Big\}. \nn
\end{align}
From the result \eqref{HomAnomalyTauAction}, we conclude that
\begin{equation}
    \boxed{\overline{\lbr\bt,\btp\rbr} \,\dotvartriangleright \,\app{}-\big[\dWt,\dWtp\!\big]\app{}=0}.
\end{equation}

\paragraph{Action} \!\!$\dotblacktriangleleft\,$:
Firstly,
\begin{equation}
    \drBtpp\big\lbr\bPta,\bPta'\big\rbr=\rBtpp\Big( \big[\bPta,\bPta'\big]+\dbPtap\bPta-\dbPta\bPtap\Big)+ \delta_{\lbr\bt,\btp\rbr}\otpp.
\end{equation}
Secondly,
\begin{equation}
    \otpp\big[\dotblacktriangleleft\,\bPta\,, \dotblacktriangleleft\,\bPtap\,\big]=\Big\{ \big(\rBtpp\bPta\big)\blacktriangleleft\bPtap +\big(\dbt\otpp\big)\blacktriangleleft\bPtap +\dbtp\big(\rBtpp\bPta\big)+\dbtp\dbt\otpp\Big\} -\bPta\leftrightarrow\bPtap.
\end{equation}
Hence, for $s=-1,0$,
\begin{align}
    \Big(\otpp\big[\dotblacktriangleleft\,\bPta\,, \dotblacktriangleleft\,\bPtap\,\big]\Big)_s &=\Big(\otpp\big[\!\blacktriangleleft\bPta\,, \blacktriangleleft\bPtap\,\big]\Big)_s+ \delta_{\lbr\bt,\btp\rbr}\otpp_{\!\!\!s}+\Big\{\Big\{\cancel{\big((\dbt\otpp) \blacktriangleleft \bPtap}\big)_s \\
    &-\cancel{\big((\dbt\otpp)\blacktriangleleft \bPtap}\big)_s-\big(\rBtpp(\dbt\bPtap)\big)_s +(s+3)\tpp0\tp{s+2}\dbt C\Big\}-\bPta\leftrightarrow\bPtap\Big\}. \nn
\end{align}
Since $\rBtpp(\dbPta\bPtap)=\rBtpp(\dbt\bPtap)$, we just have to use \eqref{HomAnomalyrBlackTauAction} to infer that, cf.\,\eqref{dotBlackProperties}
\begin{equation}
    \boxed{\drBtpp\big\lbr\bPta,\bPta'\big\rbr -\otpp\big[ \dotblacktriangleleft\,\bPta\,,\dotblacktriangleleft\,\bPta'\,\big]=0}.
\end{equation}

\section{Covariant derivative in the wedge \label{AppEYM:LeibnizDEYM}}

Starting from $\big(\DEYM\a{}\big)_s=\DYM\a{s+1}-(s+2)C\a{s+2}$, we get that
\begin{align}
    \big(\DEYM[\a{},\ap{}]^\g\big)_s &=\DYM \big[\a{},\ap{}\big]^\g_{s+1}-(s+2)C\big[\a{},\ap{}\big]^\g_{s+2} \nn\\
    &=\big[\DYM\a{},\ap{}\big]^\g_s+ \big[\a{},\DYM\ap{}\big]^\g_s+\big[\DYM\a0,\ap{s+1}\big]_\g-\big[\DYM\ap0,\a{s+1}\big]_\g-(s+2)C\big[\a{},\ap{}\big]^\g_{s+2} \nn\\
    &=\sum_{n=0}^s\big[(\DEYM\a{})_n,\ap{s-n}\big]_\g+C\sum_{n=0}^s(n+2)\uline{\big[\a{n+2},\ap{s-n}\big]_\g}+\big[\DYM\a0, \ap{s+1}\big]_\g \nn\\
    &+\sum_{n=0}^s\big[\a{n},(\DEYM\ap{})_{s-n}\big]_\g+C\sum_{n=0}^s(s+2-n)\uline{\big[\a{n},\ap{s+2-n}\big]_\g}-\big[\DYM\ap0,\a{s+1}\big]_\g \nn\\
    &-(s+2)C\sum_{n=0}^{s+2} \uline{\big[\a{n},\ap{s+2-n}\big]_\g}.
\end{align}
The underlined terms simply give
\begin{equation}
    \uline{~\cdots~}=-C\big([\a1,\ap{s+1}]_\g- [\ap1,\a{s+1}]_\g\big).
\end{equation}
Hence
\begin{equation}
    \big(\DEYM[\a{},\ap{}]^\g\big)_s=\Big( \big[\DEYM \a{},\ap{}\big]^\g_s-\big[\hda A,\ap{s+1}\big]_\g\Big) -\a{}\leftrightarrow\ap{},
\end{equation}
where $\hda A=-\DYM\a0+C\a1=-\big(\DEYM \a{}\big)_{-1}$.
Projecting this equation onto a non-radiative cut, we infer that $\DEYM$ satisfies the Leibniz rule:
\begin{equation}
    \boxed{\cA\big([\Aa{},\Aap{}]^\g,\DEYM\big)=0,} \qquad \textrm{if}\quad\Aa{},\Aap{} \in\Wcal_{\bfCc\bfAc}^\ym(S).
\end{equation}

\newpage
\addcontentsline{toc}{section}{References}

\bibliographystyle{JHEP}
\bibliography{source_NoetherYM}

\end{document}